\documentclass{article}

\usepackage[noend]{algpseudocode}
\usepackage{algorithm}
\usepackage{amsthm}
\usepackage{amsmath}
\usepackage{amsfonts}
\usepackage{amssymb}
\usepackage[round]{natbib}
\usepackage{multirow}
\usepackage{stfloats}
\usepackage{float}
\usepackage{graphicx}
\usepackage{xcolor}
\usepackage{geometry}
\usepackage{afterpage}
\usepackage{pdflscape}
\usepackage{url}
\usepackage[
  bookmarks=false,
  pdfpagelabels=false,
  hyperfootnotes=false,
  hyperindex=false,
  pageanchor=false,
  colorlinks,
  citecolor=blue
]{hyperref}
\usepackage{cleveref}

\newcommand\NoDo{\renewcommand\algorithmicdo{}}     % remove "do" from algs
\newcommand{\E}{\mathbb{E}}             % Expectation
\newcommand{\R}{\mathbb{R}}             % Real numbers
\newcommand{\I}{\mathbb{I}}             % Indicator function
\newcommand{\D}{\mathcal{D}}            % Calligraphic D: Data
\newcommand{\Q}{\mathcal{Q}}            % Calligraphic Q: Variational Family
\renewcommand{\L}{\mathcal{L}}          % Calligraphic L: ELBO
\newcommand{\KL}{\text{KL}}             % KL divergence
\newcommand{\argmin}{{\arg\!\min}}      % arg min without space
\newcommand{\argmax}{{\arg\!\max}}      % arg max without space

\newcommand{\sidewaystable}[1]{
\afterpage{%
    \clearpage% Flush earlier floats (otherwise order might not be correct)
    \thispagestyle{empty}% empty page style (?)
    \newgeometry{left=2.5cm,right=2.5cm,bottom=0.1cm,top=0.1cm}
    \begin{landscape}% Landscape page
	\centering
	\vspace*{\fill} %
	#1
	\vspace*{\fill} %
    \end{landscape}
    \restoregeometry
    \clearpage% Flush page
}
}

\newcommand{\sidewaysfigure}[1]{
\afterpage{%
    \newgeometry{left=2.5cm,right=2.5cm,bottom=1.0cm,top=1.0cm}
    \begin{landscape}% Landscape page
	\centering
	#1
    \end{landscape}
    \restoregeometry
    \clearpage% Flush page
}
}

\graphicspath{{../figures/}}

\newcommand{\basedir}{./appendix}

\newcommand\blfootnote[1]{%
  \begingroup
  \renewcommand\thefootnote{}\footnote{#1}%
  \addtocounter{footnote}{-1}%
  \endgroup
}
\begin{document}

\setlength{\abovedisplayskip}{1.5em}
\setlength{\belowdisplayskip}{1.5em}
\hbadness=99999
\hfuzz=15pt

\title{Variational Bayes for high-dimensional proportional hazards models with applications within gene expression}

\author{Michael Komodromos$^{\text{1}}$, 
Eric Aboagye$^{\text{2}}$, 
Marina Evangelou$^{\text{1}, \dagger}$, \\
Sarah Filippi$^{\text{1}, \dagger}$, 
Kolyan Ray$^{\text{1}, \dagger}$ \\
\small{$^{\text{\sf1}}$Department of Mathematics, Imperial College London and,} \\
\small{$^\text{\sf2}$Department of Surgery and Cancer, Imperial College London}
}

\date{}

\maketitle

\begin{abstract}
    Few Bayesian methods for analyzing high-dimensional sparse survival data provide scalable variable selection, effect estimation and uncertainty quantification. Such methods often either sacrifice uncertainty quantification by computing maximum a posteriori estimates, or quantify the uncertainty at high (unscalable) computational expense. We bridge this gap and develop an interpretable and scalable Bayesian proportional hazards model for prediction and variable selection, referred to as SVB. Our method, based on a mean-field variational approximation, overcomes the high computational cost of MCMC whilst retaining useful features, providing a posterior distribution for the parameters %excellent point estimates 
    and offering a natural mechanism for variable selection via posterior inclusion probabilities. The performance of our proposed method is assessed via extensive simulations and compared against other state-of-the-art Bayesian variable selection methods, demonstrating comparable or better performance. Finally, we demonstrate how the proposed method can be used for variable selection on two transcriptomic datasets with censored survival outcomes, and how the uncertainty quantification offered by our method can be used to provide an interpretable assessment of patient risk.
    \\ \textbf{Availability and implementation:} our method has been implemented as a freely available R package \texttt{survival.svb} (\url{https://github.com/mkomod/survival.svb}).
    \\ \textbf{Contact:} \url{mk1019@ic.ac.uk} \blfootnote{$^\dagger$Equal contribution}
    \let\thefootnote\relax\footnotetext{Accepted in Bioinformatics \url{https://doi.org/10.1093/bioinformatics/btac416}}
\end{abstract}

\newpage

\section{Introduction}

The development of high-throughput sequencing technologies has led to the production of large-scale molecular profiling data, allowing us to gain insights into underlying biological processes \citep{Widlak2013}. One such technology is microarray sequencing, in which mRNA counts are used to describe gene expression. Such data, known as transcriptomics, are widely used in the biomedical domain and when analyzed alongside survival times have provided extraordinary opportunities for biomarker characterization and prognostic modelling  \citep{Bovelstad07, Lloyd15, Lightbody19, Lu21}. However, profiling data is often high-dimensional, which introduces several statistical challenges including: (i) variable selection, (ii) effect estimation of the features, (iii) uncertainty  quantification, and (iv) scalable computation. The task of variable selection is particularly important, as few genes typically have an effect on the outcome. Motivated by clinical applicability, we propose a state-of-the-art scalable (variational) Bayesian variable selection method for the proportional hazards model.

In recent years, several methods have been proposed to analyze sparse high-dimensional data, with one of the most popular being the LASSO \citep{Tibshirani96}. As biomedical studies are often concerned with clinical phenotypes, such as time to disease recurrence or overall survival time, these methods have been adapted to support survival analysis \citep{Antoniadis10, Witten10}. For instance, the LASSO, ridge and elastic-net penalties have all been extended to the proportional hazards model \citep{Tibshirani97, Gui05, Zou05, Simon11}. More recently, Bayesian shrinkage and variable selection methods have grown in popularity \citep{Park08, OHara09, Carvalho10, Li10, Bhadra19, Lewin19, Bai20}, with several methods being extended to survival data \citep{Tang17, Maity20, Nikooienejad20}.

Bayesian approaches to variable selection are popular, not least since the relevance of a covariate can be assessed simply by computing the posterior probability that it is included in a model. This recasts variable selection as a model selection problem \citep{Mitchell88, George93}, with every possible model assigned an individual posterior probability. One of the most popular such model selection priors is the spike-and-slab prior, see \citep{Banerjee21} for a recent survey. However, exact posterior computation involves summing over $2^p$ models, where $p$ is the number of covariates, which is intractable for even moderate $p$. Markov chain Monte Carlo (MCMC) methods avoid this problem, but are known to have difficulty efficiently exploring the model space for high-dimensional covariates \citep{Ormerod17}, especially for the problem sizes found in many modern omics studies which motivate our work. This high computational cost has led to several methods either making continuous relaxations, giving rise to \textit{continuous shrinkage priors} \citep{OHara09, Banerjee21}, or computing only \textit{maximum a posteriori} (MAP) estimates, thereby not offering the full Bayesian machinery. Since we wish to preserve certain interpretable features arising from the original discrete model selection approach, such as inclusion probabilities of particular covariates for variable selection, we instead turn to variational inference.

Variational inference (VI) is a popular scalable approximation technique, which has proven to be an effective tool for approximate Bayesian inference in many settings. VI involves minimizing the Kullback-Leibler divergence between a family of tractable distributions, called the \textit{variational family}, and the posterior distribution; thereby recasting conditional inference as an optimization problem. The resulting minimizer is then used for downstream Bayesian inference. Though the approximation does not provide exact Bayesian inference, computationally convenient variational families can dramatically increase scalability. A common choice being \textit{mean-field families}, under which the model parameters are independent. For a detailed review of VI, we direct the reader to \cite{Blei17} and \cite{Zhang19}.

We propose a scalable and interpretable Bayesian proportional hazards model using a sparsity-inducing spike-and-slab prior with Laplace slab and Dirac spike, referred to as \textit{sparse variational Bayes} (\textbf{SVB}). Since the posterior is computationally intractable, we use a mean-field variational approximation based on a factorizable family of spike-and-slab distributions, thereby preserving certain desirable discrete model selection aspects while providing scalable approximate Bayesian inference. We derive a coordinate-ascent algorithm for our implementation and investigate its performance in extensive simulations, comparing it against the posterior obtained via MCMC and demonstrating that the variational Bayes posterior can be used as a viable alternative, whilst being orders of magnitude faster to compute. We further compare with other state-of-the-art Bayesian variable selection methods, demonstrating comparable or better performance in many settings. Finally,  we analyze two transcriptomic datasets involving ovarian and breast cancer data with censored survival outcomes, yielding biologically interpretable results.

Various versions of this sparse variational family have been employed in linear and logistic regression models \citep{Logsdon10, Titsias2011, Carbonetto2012, Ormerod17, Ray20, Ray21} with some of these works specifically motivated by high-dimensional genomic applications. While most of these works use Gaussian distributions for the slab component, we instead follow \cite{Ray21} in using a Laplace prior slab since Gaussian prior slabs are known to cause excessive shrinkage leading to potentially poor performance, even when exact posterior computation is possible \citep{Castillo12}. Our work can thus be viewed as extending ideas from the sparse VI literature to the setting of survival analysis under censoring.

More generally, (not necessarily sparse) VI has proven to be an effective tool for approximate Bayesian inference and has seen wide use in several settings, including linear and logistic regression \citep{Jakkola97, Knowles11}, group factor analysis \citep{Klami15}, topic modelling \citep{Blei07}, clustering \citep{Teschendorff05} and Gaussian processes \citep{Opper09} amongst others, with many of these methods employed in genomic and transcriptomic studies \citep{Logsdon10, Papastamoulis2014, Zhang2017, Svensson2020}.

\section{Materials and Methods} \label{sec:methods}

\textbf{Notation:} Let $ \D = \{(t_i, \delta_i, x_i)\}_{i=1}^n $ denote the observed data, where $ t_i \in \R^+ $ is an observed (possibly right censored) survival time, $ \delta_i \in \{0, 1 \} $ is a censoring indicator with $ \delta_i = 0 $ if the observation is right censored and $ \delta_i = 1 $ if the observation is uncensored, and $ x_i = (x_{i1}, \dots, x_{ip})^\top \in \R^p $ is a vector of explanatory variables.

\subsection{Survival Analysis and the proportional hazards model}

Let $ \boldsymbol{T} $ denote a random variable for an event time with density $ f(t) $ and cumulative distribution function $ F(t) $. Then the \textit{survival function}, the probability a subject survives past time $t$, is given by
\begin{equation} \label{eq:survivor_function} 
    S(t) = 1 - F(t) = \exp \left( - \int_0^t h(s) ds \right) = \exp \left( - H(t) \right),
\end{equation}
where $ H(t) = \int_0^t h(s) ds $ is the cumulative hazard rate and $h(t) = f(t) / S(t)$ is the \textit{hazard rate}, the instantaneous rate of failure at time $t$. Importantly, expressing $S(t)$ in terms of the hazard function $h$ provides a natural mechanism for analysing survival times by assuming a form for $h$ \citep{Ibrahim01, Clark2003}.

One such form, used to quantify the effect of features collected alongside survival times, is the \textit{proportional hazards model} (PHM), wherein,
\begin{equation}
    h(t; x, \beta) = h_0(t) \exp \left( \beta^\top x \right),
\end{equation}
where $ h_0(t)$ is a baseline hazard rate and $\beta = (\beta_1, \dots, \beta_p)^\top \in \R^p$ are the model coefficients corresponding to the potential covariates of interest. Typically, estimating $ \beta $ is done by maximizing the \textit{partial likelihood},
\begin{equation} \label{eq:partial_likelihood}
    L_p(\D; \beta) = \prod_{\{i : \delta_i = 1 \}} \frac
        { \exp \left( \beta^\top x_i \right) }
        { \sum_{r \in R(t_i)} \exp \left (\beta^\top x_r \right) },
\end{equation}
where $ R(t_i) = \{ r : t_r \geq t_i \} $ \citep{Cox72, Cox75}. Under the partial likelihood, the baseline hazard rate $h_0(t)$ is treated as a nuisance parameter and not specified, meaning the survival function is not directly accessible without further assumptions on the hazard rate. This approach is commonly used when the main interest is on quantifying the effect of covariates on the survival time to understand the underlying mechanisms, rather than purely for predictive purposes. Since our focus is on variable selection and analysing effect sizes, we use the partial likelihood to compute the posterior.

The use of the partial likelihood \eqref{eq:partial_likelihood} is common in Bayesian survival analysis, and can be understood via multiple Bayesian and frequentist justifications \citep{Ibrahim01}. For the frequentist, the partial likelihood is the empirical likelihood with the maximum likelihood estimator (MLE) for the cumulative baseline hazard function $ H_0 $ plugged in, i.e. the profile likelihood \citep{Murphy00}. Using it in a Bayesian way thus means we are fitting a prior to our parameter of interest $\beta$ and an MLE on the nuisance parameter $H_0$. For the Bayesian, assigning a Gamma process prior to $H_0$, marginalizing the posterior over $H_0$ and taking the limit as the prior on $H_0$ becomes non-informative, gives a marginal posterior for $\beta$ exactly based on the partial likelihood \eqref{eq:partial_likelihood} \citep{Kalbfleisch78}. Thus using \eqref{eq:partial_likelihood} can be viewed as using a diffuse Gamma process prior on the nuisance parameter $H_0$.

\subsection{Prior and variational family}

We consider a spike-and-slab prior \citep{George93,Mitchell88} for the model coefficients $\beta$. Our choice of prior is conceptually natural for variable selection problems as it leads to interpretable inference regarding the inclusion probabilities of individual features. However, unlike the original formulation which uses Gaussian slabs, we use Laplace slabs, since Gaussian slabs are known to overly shrink the true large signals \citep{Castillo12, Ray21}. Formally, the prior distribution, $\Pi(\beta, z, w)$, has hierarchical representation,
\begin{equation} \label{eq:prior}
\begin{aligned}
    \beta_j | z_j  \ \overset{\text{ind}}{\sim} &\ \ z_j \text{Laplace}(\lambda) + (1-z_j) \delta_0  \\
    z_j | w_j      \ \overset{\text{ind}}{\sim} &\ \     \text{Bernoulli}(w_j)  \\
    w_j            \ \overset{\text{iid}}{\sim} &\ \     \text{Beta}(a_0, b_0),
\end{aligned}
\end{equation}
where $ \delta_0 $ is a Dirac mass at zero, Laplace($\lambda$) has density function $\tfrac{\lambda}{2} e^{-\lambda|x|}$ on $\R$ and $\lambda, a_0, b_0 > 0 $. Placing a hyperprior on $(w_j)$ allows mixing over the sparsity level and allows adaptation to the unknown sparsity. The posterior density is proportional to the partial likelihood $L_p$ in \eqref{eq:partial_likelihood} times the joint prior density, formally,
\begin{equation} \label{eq:posterior_phm}
    \pi(\beta, z, w | \D) \propto L_p(\D; \beta) \pi(\beta, z, w),
\end{equation} %
where $\beta = (\beta_1, \dots, \beta_p)^\top \in \R^p$, $z = (z_1, \dots, z_p)^\top \in \{0, 1\}^p$ and $w = (w_1, \dots, w_p)^\top \in [0, 1]^p$.

Since the posterior \eqref{eq:posterior_phm} is computationally intractable, we use a variational approximation. For the variational family, we choose a mean-field family given by the product of independent spike-and-slab distributions with normal slab and Dirac spike for each coefficient:
\begin{equation} \label{eq:variational_family}
    \Q = \left\{
        Q_{\mu, \sigma, \gamma} = \bigotimes_{j=1}^p
        \left[
            \gamma_j N(\mu_j, \sigma_j^2) + (1 - \gamma_j) \delta_0
        \right]
    \right\},
\end{equation}
where $\mu_j \in \R, \sigma_j \in \R^+, \gamma_j \in [0, 1] $. The notation $\otimes$ means a product measure implying coordinate independence, so that $\beta \sim Q_{\mu,\sigma,\gamma}$ means
$$\beta_j \overset{\text{ind}}{\sim} \gamma_j N(\mu_j, \sigma_j^2) + (1 - \gamma_j) \delta_0.$$
Our choice of $\Q$ thereby provides scalability and maintains the property of variable selection via the Dirac mass, since the quantities $\gamma_j = Q(\beta_j \neq 0)$ are the inclusion probabilities. The variational posterior is then given by finding an element $Q \in \Q$ minimizing the KL divergence between $Q$ and the posterior distribution $\Pi(\cdot | \D)$,
\begin{equation} \label{eq:variational_posterior} 
    \tilde{\Pi} = \underset{Q_{\mu, \sigma, \gamma } \in \Q}{\argmin} 
        \KL \left( Q_{\mu, \sigma, \gamma} \ \| \ \Pi( \cdot | \D ) \right),
\end{equation}
which is then used for inference. Note this approximation has $O(p)$ parameters compared to the full posterior dimension $O(2^p)$. As with all mean-field approximation, dependent information between the components of $\beta$ are lost, such as whether two coefficients $\beta_i$ and $\beta_j$ are likely to be selected simultaneously or not.

\subsection{Coordinate-ascent algorithm}

A convenient method for computing the mean-field variational posterior $ \tilde{\Pi}$ is coordinate-ascent variational inference (CAVI) \citep{Blei17}. In CAVI, the parameters $ \mu_j, \sigma_j, \gamma_j $ for $ j = 1, \dots, p $ are sequentially updated by finding the values that minimize the KL divergence between the variational family and the posterior, whilst all other parameters are kept fixed, iterating until convergence. This reduces the overall optimization problem to a sequence of one-dimensional optimization problems.

Minimizing the objective \eqref{eq:variational_posterior} is intractable for the Bayesian PHM due to the form of likelihood \eqref{eq:partial_likelihood} and so we instead minimize an upper bound for the KL divergence. Such surrogate type functionals are well-used in variational inference, for example in logistic regression \citep{Jakkola97, Knowles11, Depraetere17}, and can lead to an increase in accuracy.

The component-wise variational updates for $ \mu_j $ and $ \sigma_j $ are given by the minimizers of
\begin{equation} \label{eq:update_mu}
\begin{aligned}
    f(\mu_j; \mu_{-j},&\ \sigma, \gamma, z_j = 1) =\ 
    \sum_{\{i : \delta_i = 1 \}} \left(
        \log \sum_{r \in R(t_i)} 
            M(x_{rj}, \mu_j, \sigma_j) P_j(x_r, \mu, \sigma, \gamma)
        - \mu_j x_{ij}
    \right)  \\
    & + \lambda \sigma_j \sqrt{2/\pi} e^{-\mu_j^2/(2\sigma_j^2)}
    + \lambda \mu_j (1 - 2 \Phi( - \mu_j / \sigma_j )) \\
\end{aligned}
\end{equation}
and
\begin{equation} \label{eq:update_sigma}
\begin{aligned}
    g(\sigma_j; \mu,&\ \sigma_{-j}, \gamma, z_j = 1) =
    \sum_{\{i : \delta_i = 1 \}} \left(
        \log \sum_{r \in R(t_i)} 
            M(x_{rj}, \mu_j, \sigma_j) P_j(x_r, \mu, \sigma, \gamma)
    \right) \\
    & + \lambda \sigma_j \sqrt{2/\pi} e^{-\mu_j^2/(2\sigma_j^2)}
    + \lambda \mu_j (1 - 2 \Phi( - \mu_j / \sigma_j ))
    - \log \sigma_j
\end{aligned}
\end{equation}
where $ M(x_{rj}, \mu_j, \sigma_j) = \exp (\mu_j x_{rj} + \frac{1}{2} \sigma_j^2 x_{rj}^2 )$, $P_j(x_r, \mu, \sigma, \gamma) = \prod_{k \neq j} \left( \gamma_k M(x_{rk}, \mu_k, \sigma_k) + (1-\gamma_k) \right)$ and $ \Phi $ denotes the CDF of the standard normal distribution. The minimizers of these expressions do not have closed-form solutions, and therefore optimization routines are needed to find them, for instance via Brent's method \citep{Brent73}. Finally, the component-wise variational update for $ \gamma_j $ is given by solving,
\begin{equation} \label{eq:update_gamma}
\begin{aligned}
    \log & \frac{\gamma_j}{1 - \gamma_j} = 
    \frac{1}{2} + \log \frac{a_0}{b_0} - \Bigg(
	\lambda \sigma_j \sqrt{2/\pi} e^{-\frac{\mu_j^2}{2\sigma_j^2}}
	+ \lambda \mu_j (1 - 2 \Phi ( -\frac{ \mu_j}{\sigma_j} ) )
	+ \log \frac{\sqrt{2}}{\sqrt{\pi} \sigma_j \lambda } \\
	& + \sum_{\{i : \delta_i = 1 \}} \Bigg(
            \log \sum_{r \in R(t_i)} M(x_{rj}, \mu_j, \sigma_j) P_j(x_r, \mu, \sigma, \gamma)
	- \log \sum_{r \in R(t_i)}  P_j(x_r, \mu, \sigma, \gamma) 
            - \mu_j x_{ij} \Bigg) \Bigg)
\end{aligned}
\end{equation}
A full derivation of these expressions is provided in Section \ref{appendix:cavi_derivation} of the Supplement.

\Cref{alg:cavi_phm} summarizes the coordinate-ascent variational inference algorithm. We denote the RHS of \eqref{eq:update_gamma} by $ \zeta(\gamma_j; \mu, \sigma, \gamma_{-j})$, and assess convergence by computing the change in $ \mu, \sigma $ and $ \gamma $ after each iteration, stopping when the total absolute change is below a specified threshold (e.g. $10^{-3}$). While the evidence lower bound (ELBO) is often used to assess convergence, the ELBO is not analytically tractable in the present setting, which instead requires computationally expensive Monte Carlo integration to evaluate it. For this reason, we instead choose to assess convergence using the absolute change in $\mu, \sigma$ and $\gamma$.

Due to the non-convex objective in \eqref{eq:variational_posterior}, CAVI generally only guarantees convergence to a local optimum, and therefore can be sensitive to initialization \citep{Blei17}. We found this to be the case for our method, particularly for $\mu$ and $\gamma$, therefore providing good starting values is generally important (see Section \ref{appendix:sensitivity_analysis} of the Supplementary materials for more details). In turn, we initialized $\mu$ using the LASSO with a small regularization hyperparameter, since $\mu$ corresponds to the unshrunk means if the variables are included in the model, and $\gamma$ as $(0.5,\dots, 0.5)^\top$, since this corresponds to an initial inclusion probability of $0.5$ for each feature. We found the proposed method is less sensitive to initial value of $\sigma$, for example initializing $ \sigma $ as $ (0.05, \dots, 0.05)^\top $ is sufficient.

\begin{algorithm}[htp]
    \caption{CAVI for VB approximation to posterior \eqref{eq:posterior_phm}}
    \label{alg:cavi_phm}
    \begin{algorithmic}[1]
	\State \textbf{require} $ \D, \lambda, a_0, b_0$
	\State Initialize $ \mu, \sigma, \gamma $
	\NoDo \While {not converged} \NoDo
	    \NoDo \For {$j = 1,\dots, p $}
	    \State $ \mu_j \leftarrow {\argmin}_{\mu_j \in \R}\ \  f(\mu_j; \mu_{-j}, \sigma, \gamma, z_j = 1) $ \texttt{\hspace*{\fill} // \ \Cref{eq:update_mu}}
	    \State $ \sigma_j \leftarrow {\argmin}_{\sigma_j \in \R^{+}}\  g(\sigma_j; \mu, \sigma_{-j}, \gamma, z_j = 1) $ \texttt{\hspace*{\fill} // \ \Cref{eq:update_sigma}}
	    \State $ \gamma_j \leftarrow \text{sigmoid}\ \zeta(\gamma_j; \mu, \sigma, \gamma_{-j})$ \texttt{\hspace*{\fill} // \Cref{eq:update_gamma}}
	    \EndFor
	\EndWhile
	\State \Return $ \mu, \sigma, \gamma $.
    \end{algorithmic}
\end{algorithm}

\subsection{Parameter tuning}

The proposed method involves three prior parameters $\lambda, a_0$ and $b_0$ defined in \eqref{eq:prior}, where $\lambda$ controls the shrinkage imposed on $ \beta_j | z_j = 1 $, with large values imposing more shrinkage, and $a_0$ and $b_0$ control the shape of the Beta distribution, whose expectation $a_0/(a_0 + b_0)$ reflects the \textit{a priori} proportion of non-zero coefficients. Generally, our method is not particularly sensitive to the prior parameters (see Section \ref{appendix:sensitivity_analysis} of the Supplement for a numerical investigation) and in practice using sensible \textit{a priori} choices is appropriate for most settings. For example, if it is believed there are a small number of non-zero coefficients with moderate effect sizes, taking $a_0$ as a small constant (such as $1, 10, p/100$), $b_0 = p$ and $\lambda$ between $0.5$ and $2.0$ is appropriate.

If an \textit{a priori} choice is unavailable, the prior parameters can be tuned using the data. To do so, we suggest performing a grid search over a predefined set of values, selecting the element that maximizes a given goodness of fit measure, several options of which are presented in section \ref{appendix:goodness_of_fit} of the Supplement. Furthermore, when tuning $a_0$ and $b_0$, to limit computation we suggest fixing $ b_0 $ and searching across a set of values for $a_0$, thereby exploring different values of the \textit{a priori} inclusion probability.

% Table placed here so it can be displayed on page 5
\begin{table*}[t]
    \centering
    \resizebox{\textwidth}{!}{ %
{\setlength{\tabcolsep}{1.07em} 
\begin{tabular}{| c c l | c c c c c c |}
    \hline
    Setting & c & Method & $\ell_2$-error & $\ell_1$-error & TPR & FDR & AUC & runtime\\
    \hline
    % --- Indep
    \rule{0pt}{1\normalbaselineskip}
    \multirow{4}{*}{\textit{1}} &
    \multirow{2}{*}{0.25} & SVB &
 0.368 (0.21, 0.70)& 1.000 (0.52, 1.86)& 1.000 (0.90, 1.00)& 0.000 (0.00, 0.00)& 1.000 (1.00, 1.00)&  18.5s  (13.5s,25.6s)\\
    & & MCMC &
 0.412 (0.20, 0.75)& 1.017 (0.48, 2.01)& 1.000 (0.90, 1.00)& 0.000 (0.00, 0.00)& 1.000 (1.00, 1.00)&  1h 24m  (1h 7m,1h 50m)\\
    [.5em]
    & \multirow{2}{*}{0.4} & SVB &
 0.428 (0.23, 0.89)& 1.138 (0.63, 2.45)& 1.000 (0.90, 1.00)& 0.000 (0.00, 0.00)& 1.000 (0.95, 1.00)&  21.9s  (14.5s,30.5s)\\
    & & MCMC &
 0.506 (0.26, 0.98)& 1.300 (0.69, 2.74)& 1.000 (0.80, 1.00)& 0.000 (0.00, 0.00)& 1.000 (1.00, 1.00)&  1h 28m  (1h 25m,1h 30m)\\
    [.4em]
    \hline
    % --- Diag
    \rule{0pt}{1\normalbaselineskip}
    \multirow{4}{*}{\textit{2}} &
    \multirow{2}{*}{0.25} & SVB &
 0.376 (0.20, 0.73)& 1.031 (0.58, 2.07)& 1.000 (0.90, 1.00)& 0.000 (0.00, 0.00)& 1.000 (1.00, 1.00)&  18.9s  (14.4s,25.4s)\\
    & & MCMC &
 0.405 (0.21, 0.81)& 1.059 (0.58, 2.18)& 1.000 (0.90, 1.00)& 0.000 (0.00, 0.00)& 1.000 (1.00, 1.00)&  1h 14m  (1h 6m,1h 17m)\\
    [.5em]
    & \multirow{2}{*}{0.4} & SVB &
 0.472 (0.23, 1.08)& 1.176 (0.61, 2.96)& 1.000 (0.90, 1.00)& 0.000 (0.00, 0.00)& 1.000 (0.95, 1.00)&  24.0s  (17.3s,33.1s)\\
    & & MCMC &
 0.520 (0.25, 1.08)& 1.319 (0.62, 2.91)& 1.000 (0.90, 1.00)& 0.000 (0.00, 0.00)& 1.000 (1.00, 1.00)&  1h 38m  (1h 25m,2h 4m)\\
    [.4em]
    \hline
    % --- Block
    \rule{0pt}{1\normalbaselineskip}
    \multirow{4}{*}{\textit{3}} &
    \multirow{2}{*}{0.25} & SVB &
 0.392 (0.18, 1.40)& 1.079 (0.53, 3.28)& 1.000 (0.90, 1.00)& 0.000 (0.00, 0.09)& 1.000 (0.95, 1.00)&  29.2s  (16.9s,44.9s)\\
    & & MCMC &
 0.418 (0.21, 1.01)& 1.092 (0.54, 2.58)& 1.000 (0.90, 1.00)& 0.000 (0.00, 0.00)& 1.000 (1.00, 1.00)&  1h 45m  (1h 24m,1h 49m)\\
    [.5em]
    & \multirow{2}{*}{0.4} & SVB &
 0.470 (0.24, 1.57)& 1.263 (0.63, 4.16)& 1.000 (0.80, 1.00)& 0.000 (0.00, 0.10)& 1.000 (0.95, 1.00)&  21.7s  (13.7s,33.2s)\\
    & & MCMC &
 0.508 (0.23, 1.26)& 1.236 (0.61, 3.45)& 1.000 (0.80, 1.00)& 0.000 (0.00, 0.09)& 1.000 (1.00, 1.00)&  1h 36m  (1h 30m,1h 45m)\\
    [.4em]
    \hline 
    % --- Block
    \rule{0pt}{1\normalbaselineskip}
    \multirow{4}{*}{\textit{4}} &
    \multirow{2}{*}{0.25} & SVB &
 0.393 (0.18, 1.12)& 1.067 (0.50, 2.54)& 1.000 (0.90, 1.00)& 0.000 (0.00, 0.10)& 1.000 (0.95, 1.00)&  17.0s  (9.2s,24.9s)\\
    & & MCMC &
 0.382 (0.17, 0.95)& 1.007 (0.44, 2.47)& 1.000 (0.90, 1.00)& 0.000 (0.00, 0.10)& 1.000 (1.00, 1.00)&  1h 5m  (1h 3m,1h 8m)\\
    [.5em]
    & \multirow{2}{*}{0.4} & SVB &
 0.425 (0.18, 1.38)& 1.171 (0.50, 2.85)& 1.000 (0.90, 1.00)& 0.000 (0.00, 0.10)& 1.000 (0.95, 1.00)&  25.8s  (14.8s,39.9s)\\
    & & MCMC &
 0.486 (0.21, 1.13)& 1.158 (0.53, 3.17)& 1.000 (0.80, 1.00)& 0.000 (0.00, 0.00)& 1.000 (0.95, 1.00)&  1h 38m  (1h 14m,1h 46m)\\
    [.4em]
    \hline 
\end{tabular} %
}}

    \caption{Comparison of variational to MCMC posterior taking $(n, p, s)=(200, 1000, 10)$ and $c \in \{0.25, 0.4\}$, presented is the median and $(5\%, 95\%)$ quantiles. Simulations were ran on Intel$^\circledR$ Xeon$^\circledR$ E5-2680 v4 2.40GHz CPUs.}
    \label{tab:comparison_results}
\end{table*}

\subsection{Implementation}

A freely available implementation is available for the R programming language via the package \texttt{survival.svb}, with functions available for fitting and evaluating models.

\section{Simulation study} \label{sec:simulations}

We use simulations to validate the proposed method, referred to as \textbf{SVB}. Firstly, we compare the variational posterior to the posterior obtained via Markov-Chain Monte-Carlo (MCMC), assessing whether our approximation can be used as a viable alternative. Secondly, we compare against other state-of-the-art Bayesian variable selection methods for the proportional hazards model. R scripts to reproduce our results can be found at \url{https://github.com/mkomod/svb.exp}

\subsection{Simulation design} \label{sec:simulation_design}

Data is simulated for $i=1,\dots,n$ observations, each having a survival time $t_i$, censoring indicator $\delta_i$, and $p$ continuous predictors $x_i \in \R^p$. The survival time is sampled independently from $\boldsymbol{T} \ | \ x_i, \beta_0, h_0 $, which has density $ f(t; x, \beta_0, h_0) = h_0(t) \exp \left( \beta_0^\top x - e^{\beta_0^\top x} \int_0^t h_0(s) ds \right) $, where we have taken $ h_0(t) = 1 $ and where the coefficient vector $\beta_0 \in \R^p$ contains $s$ non-zero elements with values sampled iid. uniformly from $[-2.0, -0.5]\cup[0.5, 2.0]$ and indices chosen uniformly at random. To introduce censoring, we sample $d_i \overset{\text{iid}}{\sim} U(0, 1) $, letting $ \delta_i =  \mathbb{I}(d_i > c) $ where $c \in [0, 1]$ is the censoring proportion, and set $ t_i \leftarrow t'_i $ where $ t'_i \overset{\text{ind}}{\sim} U(0, t_i) $ if $\delta_i = 0 $, leaving $t_i$ unchanged otherwise. Finally, the predictors are generated from one of four different settings designed to examine the behaviour under varying degrees of difficulty:

\begin{itemize}
    \item \textit{Setting 1,} an independent setting where $x_i \overset{\text{iid}}{\sim} N(0_p, I_p)$.
    \item \textit{Setting 2,} a fairly challenging setting where predictors are moderately correlated within groups and independent between groups, formally $x_i \overset{\text{iid}}{\sim} N(0, \Sigma)$ with $\text{diag}(\Sigma) = 1 $, $ \Sigma_{ij} = 0.6 $ for $ i \neq j,\  i,j = 50k + 1, \dots, 50(k + 1),\ k = 0, \dots, p/50 - 1 $, $\Sigma_{ij} = 0 $ otherwise. The setting is similar to \cite{Tang17}.
    \item \textit{Setting 3,} a challenging setting where $x_i \overset{\text{iid}}{\sim} N(\mu, \Sigma)$ with $\mu, \Sigma$ estimated from the design of the TCGA dataset analyzed in \Cref{sec:tcga}. The $s$ causal variables are randomly selected to correspond to features with a variance of at least $1.0$.
    \item \textit{Setting 4,} a realistic setting where the first $p$ predictors are taken from the TCGA dataset analyzed in \Cref{sec:tcga} and the $s$ causal features are selected as in setting 3.
\end{itemize}

To evaluate the methods, we examine the accuracy of the corresponding point estimates, quality of the variables selected, and (if applicable) the uncertainty quantification. The point estimates are assessed via the $\ell_2\text{-error}$, $ \| \beta_0 - \widehat{\beta} \|$ and the $\ell_1\text{-error}$, $| \beta_0 - \widehat{\beta} |$, where $\widehat{\beta} $ is either a MAP estimate for $\beta$ or the posterior mean if a distribution is available. For the variables selected the: (i) true positive rate (TPR) (ii) false discovery rate (FDR) and (iii) area under the curve (AUC) of the receiver operator characteristic curve are computed. For the TPR and FDR a coefficient is considered to have been selected if the posterior inclusion probability is at least $0.5$. Finally, regarding uncertainty quantification, we evaluate the marginal credible sets by computing the: (i) empirical coverage, i.e. the proportion of times the true coefficient $\beta_{0, j}$ is contained in the credible set, and (ii) set size, given by the Lebesgue measure of the set. Details regarding the construction of the credible sets are presented where appropriate. For all metrics, we report the median, $5\%$ and $95\%$ quantiles across 100 replicates unless otherwise stated.

\subsection{Simulation results}

\subsubsection{Comparison to MCMC} \label{sec:comparison_to_mcmc} 

To assess how well the variational posterior matches the target (computationally challenging) posterior from \eqref{eq:posterior_phm}, we compare the performance of our approach against the approximate yet asymptotically exact posterior obtained via MCMC. To do so, data is generated as described in \Cref{sec:simulation_design}, taking $(n, p, s)=(200, 1000, 10)$ and $c \in \{0.25, 0.4\}$, where we have kept $n$ and $p$ small so we can run our MCMC sampler in a reasonable amount of time. The MCMC sampler (described in Section \ref{appendix:mcmc} of the Supplement) was run for $10,\!000$ iterations with a burn-in period of $1,\!000$ iterations. For both methods we used prior parameters $\lambda= 1, a_0=1$ and $b_0=p$. Results are presented in \Cref{tab:comparison_results}.

Regarding the point estimates, for both the MCMC and the variational posteriors we took $\widehat{\beta} = (\widehat{\beta}_1, \dots, \widehat{\beta}_p) \in \R^p$ as the posterior mean, which for the latter is given by $\widehat{\beta}_j = \gamma_j \mu_j$. Promisingly, both methods produce similar results, with near identical performance in all settings (\Cref{tab:comparison_results}). In particular, the similarity of the $\ell_2$-error and $\ell_1$-error suggests the posterior means are near identical. In terms of variable selection, both methods performed similarly. In particular, the TPR is comparable across the different settings, suggesting both methods are selecting a similar set of truly associated features. However, the upper quantile for the FDR is slightly larger for the variational posterior, meaning the MCMC posterior selects fewer spurious variables.

Finally, we examine the uncertainty quantification of each method via $95\%$ marginal credible sets $S_j, j=1,\dots,p$, which are given by: $S_j = I_j$ if the posterior inclusion probability is greater than $0.95$, $ S_j = \{0\} $ if the posterior inclusion probability is less than $0.05$, and  $S_j = I_j \cup \{0\}$ otherwise, where $I_j$ is the smallest interval from the continuous component of our posterior such that $S_j$ contains $95\%$ of the posterior mass. As expected, for the non-zero coefficients, the coverage of the MCMC posterior is slightly better than the coverage of the variational posterior (\Cref{tab:mcmc_coverage}), meaning the credible sets of the variational posterior are sometimes not large enough to capture the true non-zero coefficients. This is further reflected by the smaller set sizes, highlighting the well known fact that VI can underestimate the posterior variance \citep{Carbonetto2012, Blei17, Zhang19, Ray20}. Promisingly, the coverage of the zero coefficients is equal to one for both methods, meaning the credible sets contain zero, and typically, as reflected by the set size, contain only zero.

\begin{table}[t]
    \centering
    \resizebox{1\textwidth}{!}{ %
{\setlength{\tabcolsep}{2.2em} 
\begin{tabular}{| c c l | c c c c |}
    \hline
    Setting & c & Method & cov. $\beta_0 \neq 0 $ & set size  $\beta_0 \neq 0 $ & cov. $\beta_0=0$ & set size $\beta_0=0$ \\
    \hline
    % --- Indep
    \rule{0pt}{1\normalbaselineskip}
    \multirow{4}{*}{1} & 
    \multirow{2}{*}{0.25} & SVB &
    0.770 (0.202)& 0.320 (0.013)& 1.000 (0.000)& 0.000 (0.000)\\
    & & MCMC &
    0.928 (0.138)& 0.506 (0.039)& 1.000 (0.000)& 0.000 (0.000)\\
    [.5em]
    & \multirow{2}{*}{0.4} & SVB &
    0.774 (0.208)& 0.355 (0.021)& 1.000 (0.000)& 0.000 (0.000)\\
    &     & MCMC &
    0.914 (0.127)& 0.570 (0.054)& 1.000 (0.000)& 0.000 (0.000)\\
    [.4em]
    \hline
    % --- Block
    \rule{0pt}{1\normalbaselineskip}
    \multirow{4}{*}{2} & 
    \multirow{2}{*}{0.25} & SVB &
0.703 (0.227)& 0.306 (0.028)& 1.000 (0.001)& 0.000 (0.000)\\
    & & MCMC &
0.904 (0.161)& 0.522 (0.053)& 1.000 (0.000)& 0.000 (0.000)\\
    [.5em]
    & \multirow{2}{*}{0.4} & SVB &
0.683 (0.262)& 0.333 (0.039)& 1.000 (0.001)& 0.000 (0.000)\\
    & & MCMC &
0.845 (0.218)& 0.567 (0.101)& 1.000 (0.000)& 0.000 (0.000)\\
    [.4em]
    \hline 
    % --- 
    \rule{0pt}{1\normalbaselineskip}
    \multirow{4}{*}{3} & 
    \multirow{2}{*}{0.25} & SVB &
 0.626 (0.288)& 0.251 (0.020)& 1.000 (0.000)& 0.000 (0.000)\\
    & & MCMC &
 0.903 (0.140)& 0.482 (0.047)& 1.000 (0.000)& 0.000 (0.000)\\
    [.5em]
    & \multirow{2}{*}{0.4} & SVB &
 0.619 (0.278)& 0.276 (0.028)& 1.000 (0.000)& 0.000 (0.000)\\
    & & MCMC &
 0.873 (0.197)& 0.540 (0.078)& 1.000 (0.000)& 0.000 (0.000)\\
    [.4em]
    \hline
    % --- Block
    \rule{0pt}{1\normalbaselineskip}
    \multirow{4}{*}{4} & 
    \multirow{2}{*}{0.25} & SVB &
 0.672 (0.224)& 0.252 (0.021)& 1.000 (0.000)& 0.000 (0.000)\\
    & & MCMC &
 0.921 (0.144)& 0.483 (0.047)& 1.000 (0.000)& 0.000 (0.000)\\
    [.5em]
    & \multirow{2}{*}{0.4} & SVB &
 0.660 (0.249)& 0.277 (0.025)& 1.000 (0.001)& 0.000 (0.000)\\
    & & MCMC &
 0.906 (0.156)& 0.547 (0.059)& 1.000 (0.000)& 0.000 (0.000)\\
    [.4em]
    \hline 
\end{tabular} %
}}

    \caption{Coverage and set size for variational and MCMC posterior. Presented are means and std. dev.}
    \label{tab:mcmc_coverage}
\end{table}

Overall, the variational posterior displays similar performance to the MCMC posterior in key aspects for this setting with $p=1000$, and can be computed orders of magnitude faster (\Cref{tab:comparison_results}). Our results highlight that the variational posterior is particularly good at capturing the key features (posterior means and inclusion probabilities) and provides reasonable uncertainty quantification for individual features.

\subsubsection{Comparison to other methods} \label{sec:comparison_to_other_nethods}

We perform a large-scale simulation study to empirically compare the performance of our method to two Bayesian variable selection methods. Within our study, data is generated as described in \Cref{sec:simulation_design}, taking $(n, p, s)=(500, 5000, 30)$ and $c\in \{0.25, 0.4\}$ for all settings. Notably, under such a setting running MCMC would be computationally prohibitive, as highlighted in the previous section.

We compare against \textbf{BhGLM} \citep{Tang17}, a spike-and-slab LASSO method that uses a mixture of Laplace distributions with one acting as the spike and the other the slab, and \textbf{BVSNLP} \citep{Nikooienejad20}, which uses a mixture prior composed of a point mass at zero and an inverse moment prior. Notably, both \textbf{BhGLM} and \textbf{BVSNLP} use Cox's partial likelihood in the posterior and return a MAP estimate for $\beta$ as well as posterior inclusion probabilities for each feature. Finally, for each method we use the default hyperparameters and let $ \lambda = 1 $, $ a_0 = 1 $ and $ b_0 = p $ for \textbf{SVB}.

Generally, all methods produced excellent point estimates, with \textbf{SVB} obtaining the smallest median $\ell_2$-error and $\ell_1$-error in settings 1 and 2, and \textbf{BhGLM} in setting 4 (\Cref{tab:simulation_results}). Notably, \textbf{SVB} obtained the smallest lower ($5\%$) quantile for the $\ell_2$-error and $\ell_1$-error in settings 3 and 4, meaning the method can perform better than \textbf{BhGLM} but may be sensitive to the design matrix.

Regarding the variables selected, all methods performed exceptionally well achieving the ideal values for the TPR, FDR and AUC in Settings 1 and 2 (\Cref{tab:simulation_results}). Within settings 3, \textbf{BhGLM} obtained the best TPR, FDR and AUC closely followed by \textbf{SVB} and \textbf{BVSNLP}. Within setting 4, all three methods obtained the ideal values when the censoring was low ($c=0.25$) and \textbf{BhGLM} performed best under moderate censoring ($c=0.40$). Further, \textbf{BhGLM} best controlled the FDR in settings 3 and 4, obtaining the lowest upper ($95\%$) quantile, closely followed by \textbf{SVB}. Finally, we note, \textbf{SVB} is the only method that provides uncertainty quantification, a direct application of which is demonstrated in \Cref{sec:yau}.

\begin{table*}[ht]
    \centering
    \resizebox{\textwidth}{!}{ %
{\setlength{\tabcolsep}{1.07em} 
\begin{tabular}{| c c l | c c c c c |}
    \hline
    Setting & c & Method & $\ell_2$-error & $\ell_1$-error & TPR & FDR & AUC \\
    \hline
    % --- Indep
    \rule{0pt}{1\normalbaselineskip}
    \multirow{6}{*}{\textit{1}} &
    \multirow{3}{*}{0.25} & SVB &
 0.378 (0.26, 0.89)& 1.747 (1.16, 4.17)& 1.000 (1.00, 1.00)& 0.000 (0.00, 0.00)& 1.000 (1.00, 1.00)\\
    & & BhGLM &
 1.206 (0.79, 1.78)& 9.590 (7.22, 12.88)& 1.000 (1.00, 1.00)& 0.000 (0.00, 0.00)& 1.000 (1.00, 1.00)\\
    & & BVSNLP &
 0.456 (0.33, 0.96)& 2.007 (1.41, 4.57)& 1.000 (1.00, 1.00)& 0.000 (0.00, 0.03)& 1.000 (1.00, 1.00)\\
    [.5em]
    & \multirow{3}{*}{0.4} & SVB &
 0.449 (0.31, 0.99)& 2.056 (1.37, 4.87)& 1.000 (1.00, 1.00)& 0.000 (0.00, 0.00)& 1.000 (1.00, 1.00)\\
    & & BhGLM &
 0.807 (0.53, 1.35)& 6.458 (4.52, 9.31)& 1.000 (1.00, 1.00)& 0.000 (0.00, 0.00)& 1.000 (1.00, 1.00)\\
    & & BVSNLP &
 0.518 (0.35, 1.44)& 2.231 (1.52, 6.85)& 1.000 (0.96, 1.00)& 0.000 (0.00, 0.03)& 1.000 (0.99, 1.00)\\
    [.4em]
    \hline
    % --- Block
    \rule{0pt}{1\normalbaselineskip}
    \multirow{6}{*}{\textit{2}} &
    \multirow{3}{*}{0.25} & SVB &
 0.405 (0.29, 0.80)& 1.823 (1.28, 3.78)& 1.000 (1.00, 1.00)& 0.000 (0.00, 0.00)& 1.000 (1.00, 1.00)\\
    & & BhGLM & 
 0.596 (0.45, 1.04)& 4.494 (3.51, 6.89)& 1.000 (1.00, 1.00)& 0.000 (0.00, 0.00)& 1.000 (1.00, 1.00)\\
    & & BVSNLP & 
 0.475 (0.33, 0.90)& 2.130 (1.47, 4.01)& 1.000 (1.00, 1.00)& 0.000 (0.00, 0.00)& 1.000 (1.00, 1.00)\\
    [.5em]
    & \multirow{3}{*}{0.4} & SVB & 
 0.491 (0.33, 1.05)& 2.208 (1.45, 5.03)& 1.000 (0.97, 1.00)& 0.000 (0.00, 0.03)& 1.000 (1.00, 1.00)\\
    & & BhGLM & 
 0.551 (0.44, 0.86)& 3.716 (2.98, 5.36)& 1.000 (0.97, 1.00)& 0.000 (0.00, 0.00)& 1.000 (1.00, 1.00)\\
    & & BVSNLP &
 0.515 (0.37, 1.47)& 2.238 (1.54, 6.71)& 1.000 (1.00, 1.00)& 0.000 (0.00, 0.00)& 1.000 (1.00, 1.00)\\
    [.4em]
    \hline 
    % --- N(m, S)
    \rule{0pt}{1\normalbaselineskip}
    \multirow{6}{*}{\textit{3}} &
    \multirow{3}{*}{0.25} & SVB & 
 1.040 (0.30, 3.17)& 3.881 (1.36, 15.37)& 0.967 (0.83, 1.00)& 0.000 (0.00, 0.14)& 0.983 (0.92, 1.00)\\
    & & BhGLM & 
 0.590 (0.36, 1.57)& 3.279 (2.23, 6.73)& 1.000 (0.93, 1.00)& 0.000 (0.00, 0.03)& 1.000 (0.97, 1.00)\\
    & & BVSNLP & 
 3.107 (1.74, 9.73)& 12.262 (6.88, 47.67)& 0.967 (0.53, 1.00)& 0.000 (0.00, 0.53)& 0.983 (0.78, 1.00)\\
    [.5em]
    % --- 
    & \multirow{3}{*}{0.4} & SVB &
 1.379 (0.36, 3.47)& 5.728 (1.55, 17.47)& 0.933 (0.77, 1.00)& 0.033 (0.00, 0.13)& 0.967 (0.88, 1.00)\\
    & & BhGLM & 
 0.796 (0.41, 2.18)& 4.035 (2.25, 10.92)& 0.967 (0.87, 1.00)& 0.000 (0.00, 0.07)& 1.000 (0.95, 1.00)\\
    & & BVSNLP & 
 3.867 (1.98, 11.44)& 15.874 (7.99, 51.07)& 0.967 (0.20, 1.00)& 0.033 (0.00, 0.69)& 0.983 (0.65, 1.00)\\
    [.4em] 
    \hline 
    % --- Real
    \rule{0pt}{1\normalbaselineskip}
    \multirow{6}{*}{\textit{4}} &
    \multirow{3}{*}{0.25} & SVB & 
 0.603 (0.29, 2.02)& 2.298 (1.21, 8.84)& 1.000 (0.90, 1.00)& 0.000 (0.00, 0.08)& 1.000 (0.95, 1.00)\\
    & & BhGLM & 
 0.503 (0.35, 1.36)& 3.141 (2.25, 5.59)& 1.000 (0.93, 1.00)& 0.000 (0.00, 0.03)& 1.000 (0.97, 1.00)\\
    & & BVSNLP & 
 2.946 (1.96, 8.72)& 11.426 (6.98, 36.46)& 1.000 (0.90, 1.00)& 0.000 (0.00, 0.07)& 1.000 (0.95, 1.00)\\
    [.5em]
    % --- 
    & \multirow{3}{*}{0.4} & SVB &
 1.092 (0.32, 2.83)& 3.878 (1.40, 14.06)& 0.967 (0.83, 1.00)& 0.000 (0.00, 0.08)& 0.983 (0.92, 1.00)\\
    & & BhGLM & 
 0.674 (0.40, 1.64)& 3.610 (2.28, 7.72)& 1.000 (0.93, 1.00)& 0.000 (0.00, 0.04)& 1.000 (0.97, 1.00)\\
    & & BVSNLP & 
 3.163 (2.14, 10.53)& 12.227 (8.14, 45.64)& 1.000 (0.73, 1.00)& 0.000 (0.00, 0.32)& 1.000 (0.87, 1.00)\\
    [.4em] 
    \hline 
\end{tabular} %
}}

    \caption{Comparison of Bayesian variable selection methods, taking $(n, p, s) = (500, 5000, 30)$ and $c \in \{0.25, 0.4\}$, presented is the median and $(5\%, 95\%)$ quantiles.}
    \label{tab:simulation_results}
\end{table*}

\section{Application} \label{sec:real_world_data}

\subsection{TCGA ovarian cancer data} \label{sec:tcga}

The first dataset we analyzed is a transcriptomic dataset where the outcome of interest is overall survival. The dataset was collected from patients with ovarian cancer and has a sample size of $ n = 580 $, of which $ 229 $ samples are right censored and $ 351 $ samples are uncensored, corresponding to a censoring rate of $ 39.5\%$ \citep{TCGA}. Within the dataset there are $ p = 12,042 $ covariates, which we pre-processed by removing features with a coefficient of variation below the median value \citep{Mar11}, leaving 6,021 covariates which we centered before fitting our method.

When applying our method, we set $a_0 = p/100$ and $b_0 = p$, reflecting our prior belief that few genes are associated with the response. As we had no prior belief for $\lambda$, we performed 10-fold cross validation to select the value, exploring a grid of values $\Lambda = \{0.05,\ 0.1,\ 0.25, 0.5,\ 0.75,\ 1.0,\ 1.25,\ 1.5,\\ 1.75,\ 2.0,\ 2.5,\ 3.0,\ 4.0,\ 5.0\}$. To evaluate model fit we compute the: (i) ELBO$=\E_Q \left[ \log L_p \right] - \text{KL}(Q \| \Pi )$, (ii) expected log-likelihood under the variational posterior (ELL = $\E_{Q}[\log L_p(\D; \beta)]$), and (iii) c-index, reporting the mean and standard deviation across the 10 folds for the training and validation sets in \Cref{tab:tcga_models} of the Supplement. Notably, no single hyperparameter value stands out as being best, meaning the model is not particularly sensitive to the value of $\lambda$.

\begin{figure}[htp]
    \centering
    \includegraphics[width=.7\textwidth]{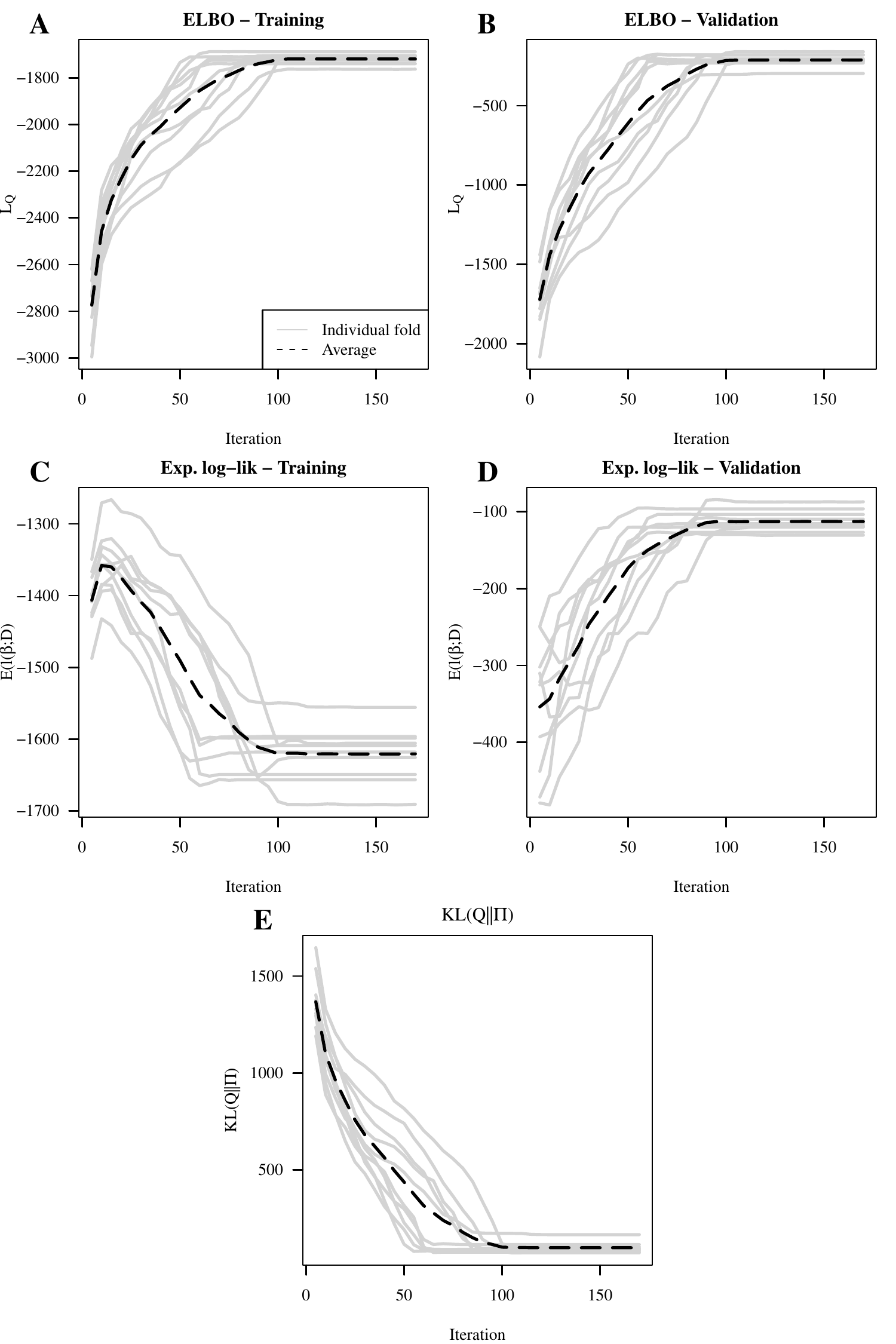}
    \caption{Ovarian cancer dataset model convergence diagnostics for $\lambda=1$.}
    \label{fig:tcga_convergence}
\end{figure}

To assess the model's convergence diagnostics we examine the fit for $\lambda=1$, and examine the change in: (i) ELBO, (ii) ELL,  and (iii) KL between the variational posterior and the prior, as we iterate our co-ordinate ascent algorithm (\Cref{fig:tcga_convergence}). Note the ELBO and ELL are computed for the training and validation sets, whereas $\KL(Q\|\Pi)$ need only be computed for the training set. Notably, across the different folds the ELBO is increasing as the co-ordinate ascent algorithm is iterated [Figures \ref{fig:tcga_convergence} (A) and (B)], suggesting that the model fit is improving. Interestingly, the training ELL is decreasing [\Cref{fig:tcga_convergence} (C)], whereas the inverse is true for the validation ELL [\Cref{fig:tcga_convergence} (D)], meaning, initially the model is overfitting to the training data, and as we iterate begins to fit better to the unseen validation set. Further, the $\KL(Q \| \Pi)$ is decreasing [\Cref{fig:tcga_convergence} (E)], therefore a greater degree of sparsity is enforced as we iterate, excluding more features and preserving the ones that best explain the variation in the response.

As we are using our model for variable selection, we examine the genes selected across the different values of $\lambda$ and folds. \Cref{tab:tcga_gene_frequencies} reports the names and selection proportion of genes, where the selection proportion is the number of times a particular gene has posterior inclusion probability greater than
\begin{equation}
    k^* = \underset{k \in [0, 1]}{\argmax} \left\{ \frac{\sum_{j=1}^p (1-\gamma_j) \I \{ \gamma_j > k \} }{\sum_{j=1}^p \I \{ \gamma_j > k \} } < \alpha \right\}
\end{equation}
Notably, $k^*$ is computed for each fit and is a threshold used to control the Bayesian false discovery rate at significance level $\alpha$, which we have set as $\alpha=0.10$ \citep{Newton2004}. Promisingly, the most frequently selected gene, \textit{PI3}, has a known, albeit limited, role in ovarian cancer, a disease characterised by copy number aberration. \cite{Clauss2010} reported the first link of \textit{PI3} gene product (elafin) to ovarian cancer \citep{Clauss2010}. Elafin, a serine proteinase inhibitor involved in inflammation and wound healing, is overexpressed in ovarian cancer and overexpression is associated with poor overall survival and is due, in part, to genomic gains on chromosome 20q13.12, a locus frequently amplified in ovarian carcinomas. There is less known about the gene encoding the alpha isoform of the calcineurin catalytic subunit (\textit{PPP3CA}) in ovarian cancer. A recent report indicates that higher expression of calcineurin predicts poor prognosis in ovarian cancer, particularly those of serous histology \citep{Xin19}. It is also plausible that other know functions of calcineurin/nuclear factor of activated T cells (NFAT), in controlling adaptive T-cell function or innate immunity \citep{Fric2012a}, in this cancer that warrants further investigation. Finally, \textit{CCR7}, the third most abundant gene was recently reported, in single cell RNA-seq analysis, to be emphasised in high-grade serous ovarian cancer \citep{Izar2020}.

\begin{table}[H]
    \centering
    \resizebox{\textwidth}{!}{ %
{\setlength{\tabcolsep}{2.2em} 
    \begin{tabular}{c c c c c c c c}
    \hline
    % PI3 & VSIG4 & PPP3CA & IL7R & SDF2L1 & D4S234E & DAP & CCR7 \\
 % 0.786 & 0.307 & 0.257 & 0.243 & 0.207 & 0.2 & 0.193 & 0.186 \\ \hline
% ACSL3 & PLA2G2D & ADORA3 & FLNA & SLAMF7 & UBD & CD14 & HABP2 \\
 % 0.157 & 0.157 & 0.121 & 0.121 & 0.107 & 0.107 & 0.086 & 0.086 \\ \hline
% LPXN & LCE2B & TBP & GALNT10 & NOTCH4 & RNF128 & C5orf28 & PPM2C \\
 % 0.086 & 0.079 & 0.079 & 0.071 & 0.071 & 0.071 & 0.064 & 0.064 \\ \hline
 % FJX1 & TSPAN13 & HSPB7 & TREML2 & & & & \\
 % 0.057 & 0.057 & 0.05 & 0.05 & & & & \\ \hline
    % PI3 & VSIG4 & PPP3CA & IL7R & SDF2L1 & D4S234E & CCR7 & ACSL3 \\
 % 0.714 & 0.307 & 0.214 & 0.207 & 0.2 & 0.193 & 0.157 & 0.143 \\ \hline
% DAP & FLNA & PLA2G2D & SLAMF7 & ADORA3 & CD14 & LPXN & UBD \\
 % 0.143 & 0.121 & 0.114 & 0.107 & 0.1 & 0.086 & 0.086 & 0.086 \\ \hline
% TBP & GALNT10 & RNF128 & LCE2B & TSPAN13 & FJX1 \\
 % 0.079 & 0.071 & 0.071 & 0.064 & 0.057 & 0.05 \\ \hline
   PI3 & PPP3CA & CCR7 & SDF2L1 & D4S234E & VSIG4 & DAP & IL7R \\
 0.7 & 0.379 & 0.293 & 0.286 & 0.229 & 0.171 & 0.136 & 0.136 \\ \hline
TBP & ACSL3 & SLAMF7 & UBD & IL2RG & GALNT10 & FLJ20323 & RNF128 \\
 0.121 & 0.114 & 0.1 & 0.1 & 0.064 & 0.057 & 0.05 & 0.05 \\ \hline
\end{tabular} %
}}

    \caption{Gene names and selection proportions for ovarian cancer dataset.}
    \label{tab:tcga_gene_frequencies}
\end{table}

\break
\subsection{Breast cancer data set} \label{sec:yau} 

The second dataset we analyzed is a transcriptomic dataset collected from patients with breast cancer, where the outcome of interest is overall survival \citep{Yau10}. The dataset consists of $n=682$ samples and $p=9,168$ features which we preprocessed as before, leaving $p=4,584$ features. Within the dataset $454$ observations are right censored, corresponding to a censoring rate of $66.5\%$.

As in the previous section, we set $a_0 = p / 100$ and $b_0 = p$, and tuned the prior parameter $\lambda$ via 10-fold cross validation using the same set $\Lambda$. \Cref{tab:yau_models} in the Supplement reports the ELBO, ELL and c-index averaged across the validation and training sets. We note that the model is not particularly sensitive to the value of $\lambda$. Furthermore, an assessment of the convergence diagnostics for $\lambda=2.5$, presented in \Cref{fig:yau_convergence} of the Supplement, carries a similar interpretation as with the TCGA data.

\Cref{tab:yau_gene_frequencies} reports the names and selection proportions of the genes within the dataset. The most frequently selected gene, Rho GTPase activating protein 28 (\textit{ARHGAP28}) is a negative regulator of RhoA. There is paucity of data on this gene in cancer generally, however, a report by \cite{Planche2011} identified the gene as downregulated in reactive stroma of prostate tumours. Further assessment of this gene in breast cancer is warranted. Notably, \textit{NEK2, GREM1} and \textit{ABCC5} have been examined in the biomedical literature, and have been associated with cancer cell proliferation and metastasis. More specially, overexpression of \textit{NEK2} induces epithelial-to-mesenchymal transition, a process which leads to functional changes in cell invasion, overexpression of \textit{GREM1} has been associated with metastasis and poor prognosis, and \textit{ABCC5} has been associated with breast cancer skeletal metastasis \citep{Rivera21, Park20, Mourskaia12}. As with the TCGA data, it is encouraging that genes with pre-existing biological interpretation have been selected by our model.

\begin{table}[htp]
    \centering
    \resizebox{\textwidth}{!}{ %
{\setlength{\tabcolsep}{2.2em} 
    \begin{tabular}{c c c c c c c c}
    \hline
ARHGAP28 & NEK2 & ABCC5 & GREM1 & DUSP4 & ITGA5 & CCL2 & IGFBP7 \\
 0.386 & 0.25 & 0.2 & 0.2 & 0.193 & 0.193 & 0.164 & 0.143 \\ \hline
NFE2L3 & TRPC1 & PKMYT1 & DDX31 & EMILIN1 & SSPN & ABO & HSPC072 \\
 0.114 & 0.114 & 0.1 & 0.086 & 0.086 & 0.086 & 0.079 & 0.079 \\ \hline
% EIF5A2 & ARL6IP5 & NUAK1 & RABEP1 & ASPM & COL4A1 & DPM2 & RACGAP1 \\
%  0.071 & 0.064 & 0.064 & 0.064 & 0.057 & 0.057 & 0.057 & 0.057 \\ \hline
% SNX5 & NEIL3 & NP & PGR & RRM2 \\
%  0.057 & 0.05 & 0.05 & 0.05 & 0.05 \\ \hline
\end{tabular} %
}}

    \caption{Gene names and selection proportions for the breast cancer dataset.}
    \label{tab:yau_gene_frequencies}
\end{table}

Finally, we want to highlight that our method, in contrast to the methods compared in \Cref{sec:comparison_to_other_nethods}, quantifies the uncertainty of $\beta$. Crucially, the availability of uncertainty serves as a powerful inferential tool for computing (variational) posterior probabilities with respect to risk scores ($\beta^\top x)$. Such probabilities can be useful in comparing patients between one another, or assessing the risk of patients against chosen benchmarks (depending on the aims of the practitioner).

\begin{figure}[H]
    \centering
    \includegraphics[width=.7\textwidth]{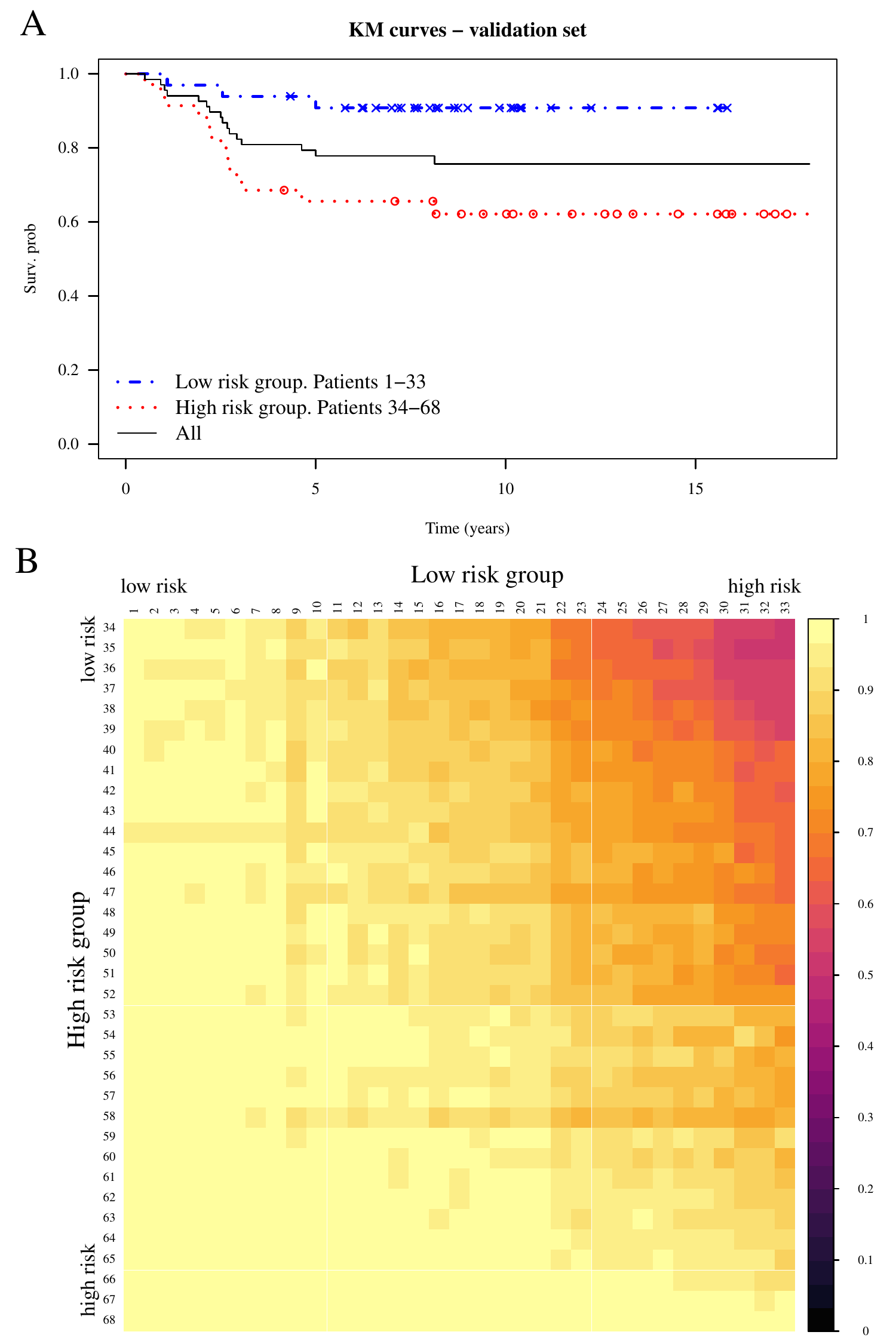}
    \caption{\textbf{(A)} Kaplan-Meier curves for patients in low and high risk groups. \textbf{(B)} Comparison of patients in the low and high risk groups (ordered by $\widehat{\eta}$) -- within each cell the (variational) posterior probability patient in row $i$ is at greater risk than patient in column $j$ is computed. Samples are taken from the second validation fold and the fit with $\lambda=2.5$ is used.
}
    \label{fig:yau_risk}
\end{figure}

To demonstrate, we opt to compute the posterior probability that one patient is at greater risk than another, formally, $\tilde{\Pi}( \beta^\top x_i\geq \beta^\top x_j)$, where $i \neq j$. To illustrate the application, we split patients into low and high risk groups based on the estimated prognostic index, $ \widehat{\eta}_i = \widehat{\beta}^\top x_i $, where $\widehat{\beta}$ is the posterior mean. Patients with prognostic index less than the median (computed for the training set) are considered low risk, whilst patients with prognostic index greater than or equal to the median are considered high risk. The Kaplan-Meier (KM) curves for these groups are shown in \Cref{fig:yau_risk} (A). Critically, Bayesian approaches that only compute the MAP of $\beta$ are only able to provide a point estimates for $\widehat{\eta}$. In turn, our method is able to provide uncertainty around this quantity and therefore with respect to the ranking of the patients. For instance, in \Cref{fig:yau_risk} (B) we present the posterior probabilities comparing the risks between patients. We observe that the highest risk patients in the low risk group are comparable to the lowest risk patients in the high risk group, and that the highest risk patients within the high risk group are with high probability more at risk than the patients within the low risk group.

\section{Discussion} \label{sec:discussion}

Variable selection and effect estimation for high-dimensional survival data has been an issue of great interest in recent years, particularly given the ever growing production of large scale omics data. However, the high-dimensionality and heterogeneity in the predictors, alongside the censoring in the response, makes the analysis a non-trivial task. While many recent methods have tackled these issues through a Bayesian approach, due to long compute times they often only produce point estimates rather than the full posterior and thereby fall short in providing the full Bayesian machinery.

We have bridged this gap and developed a scalable and interpretable mean-field variational approximation for Bayesian proportional hazards models with a spike-and-slab prior. We have demonstrated that the resulting variational posterior displays similar performance to the posterior obtained via MCMC whilst requiring a fraction of the compute time. Furthermore, we have demonstrated via an extensive simulation study that our proposed method performs comparably to state-of-the art Bayesian variable selection methods. 
% Although this is not an aspect we have considered, our results suggest that the variational posterior can also be useful for prediction when we are less interested in identifying which variables are included in the predictive model. 

Finally, we have demonstrated that our method can be used for variable selection on two real world transcriptomics datasets, giving rise to results with pre-existing biological interpretations, thereby validating the practical utility of our method. We have also shown that the risk of patients can be compared through (variational) posterior probabilities, highlighting that the availability of a posterior distribution can be a powerful inferential tool. For illustrative purposes we examined the pairwise probabilities of patients grouped based on the prognostic index, however, patients could have alternatively been compared to other baselines e.g. the feature vector corresponding to median prognostic index. Furthermore, although this is not an aspect we have considered, grouping based on: age, cancer status, stage etc. may yield insightful results for practitioners and bioinformaticians.

A natural extension of our work would be to develop approximations with relaxed independence assumptions by using a more flexible variational family \citep{Ning2021a}. Finally, we would like to highlight that improving the uncertainty quantification is an active area of research in the general VI community, see e.g. \citep{Jerfel21}.

\newpage
\subsection*{Acknowledgements}

We would like to thank the reviewers for their constructive suggestions and comments.

\subsection*{Funding}

MK gratefully acknowledges funding provided by EPSRC's StatML CDT, Imperial's CRUK centre and Imperial's Experimental Cancer Medicine centre.

\subsection*{Conflict of interest}

The authors declare they have no competing interests.

\bibliography{./references.bib}

\newpage
\appendix
\section{Variational algorithm} \label{appendix:cavi_derivation}

Recall that our mean-field variational family is given by
\begin{equation}
    \Q = \left\{
        Q_{\mu, \sigma, \gamma} = \bigotimes_{j=1}^p
        \left[
            \gamma_j N(\mu_j, \sigma_j^2) + (1 - \gamma_j) \delta_0
        \right]
        : \mu_j \in \R, \sigma_j \in \R^+, \gamma_j \in [0, 1]
    \right\},
\end{equation}
and the posterior is given by
\begin{equation}
   d \Pi(\beta| \D) = \Pi_D^{-1} e^{l_p(\D; \beta)} d \Pi(\beta),
\end{equation}
where $ l_p = \log{L_p} $ is the log (partial) likelihood and $\Pi_D $ is the normalising constant. Our aim is to evaluate the KL divergence between an element $ Q_{\mu, \sigma, \gamma} $ of the variational family and the posterior $ \Pi( \cdot | \D )$ as a function of $ \mu_j, \sigma_j $ and $ \gamma_j $, whilst keeping all other parameters fixed. Due to the discrete components of the prior \eqref{eq:prior}, one has to be careful with the different terms since they may not be absolutely continuous with respect to one another (as measures), and hence may not have densities.

\subsection{Update equations for $\mu_j$ and $ \sigma_j $}

We first compute the KL divergence between $ Q_{\mu, \sigma, \gamma} $ and the posterior $ \Pi(\cdot | \D ) $, conditional on $ z_j = 1$, as a function of $ \mu_j $ and $ \sigma_j $. We recall the notation $\mu_{-j}$, which refers to all the components of $(\mu_1,\dots,\mu_p)$ except $\mu_j$. Firstly, since both the prior \eqref{eq:prior} and variational family \eqref{eq:variational_family} consist of factorisable distributions, the Radon-Nikodym derivative between them also factorizes
\begin{equation}\label{eq:factorize}
\frac{dQ_{\mu,\sigma,\gamma}}{d\Pi}(\beta) = \prod_{j=1}^n \frac{dQ_j}{d\Pi_j}(\beta_j),
\end{equation}
where $Q_j = \gamma_j N(\mu_j,\sigma_j^2) + (1-\gamma_j) \delta_0$ and $\Pi_j = \bar{w} \text{Lap}(\lambda) + (1-\bar{w}) \delta_0$ with $\bar{w} = a_0 /(a_0+b_0)$. The latter expression for the prior follows from integrating out the hierarchical representation \eqref{eq:prior}, whereupon we have weights equal to the prior mean weight $\bar{w}$.

Since the variational probability distribution of $ \beta_j $ conditional on $ z_j = 1 $ (i.e. the slab component) is singular with respect to the Dirac measure $ \delta_0$, in this case it is sufficient to consider only the continuous part of the prior measure in the denominator of the Radon-Nikodym derivative, that is
$$\frac{dQ_{\mu_j,\sigma_j | z_j = 1 }}{d\Pi_j}(\beta_j) = \frac{dN(\mu_j,\sigma_j^2)}{\bar{w}d\text{Lap}(\lambda)}(\beta_j).$$
Using the above facts, $ \KL(Q_{\mu, \sigma, \gamma | z_j = 1} \ \| \ \Pi(\cdot | \D)) $ equals, as a function of $ \mu_j $ and $\sigma_j$,
{\allowdisplaybreaks        % allow breaks over pages
\begin{align}
   \E_{\mu, \sigma, \gamma | z_j = 1} & \left[ 
        \log \frac{dQ}{d\Pi(\cdot | \D)}
   \right] 
   = \ 
   \E_{\mu, \sigma, \gamma | z_j = 1} \left[ 
        \log \frac{ dQ_{\mu, \sigma, \gamma} }{ d\Pi } 
        - l_p(\D; \beta) + \log \Pi_D
   \right] \nonumber \\
   % ---
   =&\ 
   \E_{\mu, \sigma, \gamma | z_j = 1} \left[ 
       \log \left( \frac{dN(\mu_j,\sigma_j^2)}{\bar{w}d\text{Lap}(\lambda)} (\beta_j) \prod_{k\neq j} \frac{dQ_k}{d\Pi_k}(\beta_k) \right)
        - l_p(\D; \beta) + \log \Pi_D
   \right] \nonumber \\
   % ---
   =&\ 
   \E_{\mu, \sigma, \gamma | z_j = 1} \Bigg[ 
        \log \left( \prod_{k\neq j} \frac{dQ_k}{d\Pi_k}(\beta_k) \right)
        + \log \left( \frac{1}{\sqrt{2\pi\sigma_j^2}} e^{-\frac{(\beta_j-\mu_j)^2}{2\sigma_j^2}} \frac{2}{\lambda \bar{w}} e^{\lambda|\beta_j|} \right)
        - l_p(\D; \beta) + \log \Pi_D
   \Bigg] \nonumber \\
   % ---
   %\overset{\text{(i)}}{=} &\
   =&\
   \E_{\mu, \sigma, \gamma | z_j = 1} \left[ 
        \lambda | \beta_j |
        - \log \sigma_j
        - \frac{ (\beta_j - \mu_j)^2 }{2\sigma_j^2}
        - l_p(\D; \beta)
   \right] + C \nonumber \\
   % ---
   = &\ 
   \E_{\mu, \sigma, \gamma | z_j = 1} \left[ 
        \lambda | \beta_j |
        - \log \sigma_j
        - \frac{ (\beta_j - \mu_j)^2 }{2\sigma_j^2}
        - \sum_{\{i : \delta_i = 1 \}} \left(
            \beta^\top x_i
            - \log \sum_{r \in R(t_i)} \exp(\beta^\top x_r)
        \right)
   \right] + C \nonumber,
\end{align}
} % end of \allowdisplaybreaks 
where the constant $C$ is independent of $\mu_j$ and $\sigma_j$ and may vary from line to line. Since $| \beta_j | $ has a folded normal distribution, it has expectation $ \sigma_j \sqrt{2/\pi} e^{-\mu_j^2/(2\sigma_j^2)} + \mu_j (1 - 2 \Phi( - \mu_j / \sigma_j )) $, where $ \Phi $ is the CDF of the standard normal distribution. The previous display equals
{\allowdisplaybreaks        % allow breaks over pages
\begin{align}
   % ---
   %\overset{\text{(ii)}}{=} &\ 
&     \sum_{\{i : \delta_i = 1 \}} \left(
        \E_{\mu, \sigma, \gamma | z_j = 1} \left[ 
            \log \sum_{r \in R(t_i)} \exp(\beta^\top x_r)
        \right] 
        - \mu_j x_{ij}
    \right)  \nonumber \\
    & \quad + \lambda \sigma_j \sqrt{2/\pi} e^{-\mu_j^2/(2\sigma_j^2)}
    + \lambda \mu_j (1 - 2 \Phi( - \mu_j / \sigma_j ))
    - \log \sigma_j  + C \nonumber \\
   % ---
   %\overset{\text{(iii)}}{\leq} &\ 
   & \leq
    \sum_{\{i : \delta_i = 1 \}} \left(
        \log \sum_{r \in R(t_i)} \E_{\mu, \sigma, \gamma | z_j = 1} \left[ 
            \exp(\beta^\top x_r)
        \right] 
    -  \mu_j x_{ij} 
    \right) \nonumber \\
    & \quad + \lambda \sigma_j \sqrt{2/\pi} e^{-\mu_j^2/(2\sigma_j^2)}
    + \lambda \mu_j (1 - 2 \Phi( - \mu_j / \sigma_j ))
    - \log \sigma_j + C \nonumber \\
    % ---
    & =
    \sum_{\{i : \delta_i = 1 \}} \left(
        \log \sum_{r \in R(t_i)} 
            M(x_{rj}, \mu_j, \sigma_j) P_j(x_r, \mu, \sigma, \gamma)
        - \mu_j x_{ij}
    \right)  \nonumber \\
    & \quad + \lambda \sigma_j \sqrt{2/\pi} e^{-\mu_j^2/(2\sigma_j^2)}
    + \lambda \mu_j (1 - 2 \Phi( - \mu_j / \sigma_j ))
    - \log \sigma_j
    + C, \label{eq:expression_mu_sigma}
\end{align}
} % end of \allowdisplaybreaks 
where $ M(x_{rj}, \mu_j, \sigma_j) = \exp (\mu_j x_{rj} + \frac{1}{2} \sigma_j^2 x_{rj}^2 )$ and $P_j(x_r, \mu, \sigma, \gamma) = \prod_{k \neq j} \left( \gamma_k M(x_{rk}, \mu_k, \sigma_k) + (1-\gamma_k) \right)$ and we have used Jensen's inequality. Noting, to evaluate an expression for $ \E_{\mu, \sigma, \gamma | z_j = 1} \left[ e^{\beta^\top x_r}\right] $ we exploit the independence structure of $\beta$, i.e. $\beta_j \overset{\text{ind}}{\sim} \gamma_j N(\mu_j, \sigma^2) + (1-\gamma_j)\delta_0$, and hence $ \E_{\mu, \sigma, \gamma | z_j = 1} \left[ e^{\beta^\top x_r}\right] = \E_{\mu_j, \sigma_j, \gamma_j | z_j = 1} \left[ e^{\beta_j x_{rj}}\right] \prod_{k\neq j} \E_{\mu_k, \sigma_k, \gamma_k} \left[ e^{\beta_k x_{rk}}\right] $.

The last display thus provides an upper bound for the KL divergence and hence a surrogate objective function. Minimising this expression with respect to either $ \mu_j $ or $ \sigma_j $ gives the same minimisers as minimising the objective functions \eqref{eq:update_mu} or \eqref{eq:update_sigma}, respectively, and hence gives our update equations.

\subsection{Update equation for $ \gamma_j $}

In a similar way, the KL divergence between $ Q_{\mu, \sigma , \gamma} $ and $ \Pi(\cdot | \D )$ as of function of $ \gamma_j $ equals
\begin{align*}
   \E_{\mu, \sigma, \gamma} & \left[ 
        \log \left( \prod_{k\neq j} \frac{dQ_k}{d\Pi_k}(\beta_k) \right)
        + \log \frac
            {d[\gamma_jN(\mu_j, \sigma_j^2) + (1-\gamma_j) \delta_0 ]}
            {d[\bar{w} \text{Lap}(\lambda) + (1-\bar{w}) \delta_0 ]} (\beta_j)
        - l_p(\D; \beta)
   \right] + C.
\end{align*}
We now split this expression in terms of the events $z_j = 1$ or 0. Noting that with $Q_{\mu,\sigma,\gamma}$-probability one, $\beta_j = 0$ if and only if $z_j = 0$, the last display can rewritten as
{\allowdisplaybreaks
\begin{align}
   \E_{\mu, \sigma, \gamma} & \left[ 
        \mathbb{I}_{\{z_j =1 \}} \log \frac
            {\gamma_j d N(\mu_j, \sigma_j^2)}
            {\bar{w} d \text{Lap}(\lambda)} (\beta_j)
        + \mathbb{I}_{\{z_j = 0 \}} \log \frac
            {(1-\gamma_j)}
            {(1-\bar{w})}
        - l_p(\D; \beta)
   \right] + C \nonumber \\
   = &\
      \E_{\mu, \sigma, \gamma}  \left[ 
        \mathbb{I}_{\{z_j =1 \}} \left( \log \left( \frac{\gamma_j}{\bar{w}} 
        \frac{\sqrt{2}}{\sqrt{\pi}\sigma_j \lambda} \right) + \lambda |\beta_j| 
    - \frac{(\beta_j-\mu_j)^2}{2\sigma_j^2} \right)
        + \mathbb{I}_{\{z_j = 0 \}} \log \frac
            {(1-\gamma_j)}
            {(1-\bar{w})}
        - l_p(\D; \beta)
   \right] + C \nonumber \\
   % ---
   = &\
    \gamma_j \left\{ 
        \log \frac{\gamma_j}{\bar{w}}
        + \log \frac{\sqrt{2}}{\sqrt{\pi} \sigma_j \lambda }
        + \lambda \sigma_j \sqrt{2/\pi} e^{-\mu_j^2/(2\sigma_j^2)}
        + \lambda \mu_j (1 - 2 \Phi( - \mu_j / \sigma_j ))
        - \frac{1}{2}
    \right\}  \nonumber \\
    & + (1-\gamma_j) \log \frac
            {(1-\gamma_j)}
            {(1-\bar{w})} 
   + \sum_{\{i : \delta_i = 1 \}} \left(
        \E_{\mu, \sigma, \gamma} \left[ 
            \log \sum_{r \in R(t_i)} \exp(\beta^\top x_r)
       \right]
       - \gamma_j \mu_j x_{ij}
   \right)
   + C, \nonumber
\end{align}
where $C$ does not depend on $\gamma_j$. Upper bounding the expectation in the last display using Jensen's inequality,
\begin{align}
    & \E_{\mu, \sigma, \gamma} \left[ 
        \log \sum_{r \in R(t_i)} \exp(\beta^\top x_r)
   \right] \nonumber \\
   & = \sum_{\{i : \delta_i = 1 \}} \Bigg(
        \gamma_j \E_{N(\mu_j, \sigma_j^2) \otimes Q_{-j}} \left[ 
            \log \sum_{r \in R(t_i)} e^{\beta^\top x_r}
       \right]
   + (1- \gamma_j) \E_{\delta_0 \otimes Q_{-j}} \left[ 
            \log \sum_{r \in R(t_i)} e^{\beta^\top x_r}
       \right]
    \Bigg) \nonumber \\
   & \leq \sum_{\{i : \delta_i = 1 \}} \Bigg(
        \gamma_j 
        \log \sum_{r \in R(t_i)} \E_{N(\mu_j, \sigma_j^2) \otimes Q_{-j}} \left[ 
            e^{\beta^\top x_r}
       \right]
   + (1 - \gamma_j) \log \sum_{r \in R(t_i)} \E_{\delta_0 \otimes Q_{-j}} \left[ 
            e^{\beta^\top x_r}
       \right]
    \Bigg) \nonumber \\
    & = \sum_{\{i : \delta_i = 1 \}} \Bigg(
        \gamma_j 
        \log \sum_{r \in R(t_i)} 
        M(x_{rj}, \mu_j, \sigma_j) P_j(x_r, \mu, \sigma, \gamma)
   + (1 - \gamma_j) \log \sum_{r \in R(t_i)} 
        P_j(x_r, \mu, \sigma, \gamma) 
    \Bigg). \nonumber
\end{align}
} % end of \allowdisplaybreaks
Substituting this into the second to last display, we can upper bound the KL divergence as a function of $ \gamma_j $ by
\begin{equation} \label{eq:expression_gamma}
\begin{aligned}
   & \gamma_j \sum_{\{i : \delta_i = 1 \}} \Bigg(
        \log \sum_{r \in R(t_i)} M(x_{rj}, \mu_j, \sigma_j) P_j(x_r, \mu, \sigma, \gamma)
        - \log \sum_{r \in R(t_i)}  P_j(x_r, \mu, \sigma, \gamma) 
        - \mu_j x_{ij}
    \Bigg) \ + \\
    & \gamma_j \left( 
        \lambda \sigma_j \sqrt{2/\pi} e^{-\mu_j^2/(2\sigma_j^2)}
        + \lambda \mu_j (1 - 2 \Phi( - \mu_j / \sigma_j ))
        + \log \frac{\sqrt{2}}{\sqrt{\pi} \sigma_j \lambda }
        - \frac{1}{2}
        + \log \frac{\gamma_j}{1-\gamma_j}
        - \log \frac{a_0}{b_0}
    \right)  \\
    & + \log(1 - \gamma_j) + C
\end{aligned}
\end{equation}
where $ C $ does not depend on $ \gamma_j $ and we have used $\bar{w} = a_0/(a_0+b_0)$. Setting the derivative with respect to $ \gamma_j $ of the last equation equal to zero and rearranging gives the update equation \eqref{eq:update_gamma} for $\gamma_j$.

% --------------------------------------------------------------------------------
% --------------------------------------------------------------------------------
% -------------------------------------------------------------------------------- 
\section{Goodness of fit measures} \label{appendix:goodness_of_fit}

A commonly used measure of goodness of fit in the VI literature is the evidence lower bound (ELBO), which acts as a lower bound for the Bayesian marginal likelihood and is defined as
\begin{equation} \label{eq:elbo}
\L_Q = \E_Q \left[ \log L_p \right] - \text{KL}(Q \| \Pi ),
\end{equation}
with $ Q \in \Q $ \citep{Bishop06}. Intuitively, the ELBO trades-off how well the model fits the data (first term) with how close it is to the prior (second term). Picking the parameter that maximizes the ELBO is the variational analogue of empirical Bayes, which selects parameters by maximising the marginal likelihood. Since VI is used exactly when the marginal likelihood is intractable, it is natural to use the ELBO, which measures the quality of the variational approximation \citep{Bishop06}.

Similar to our derivation of the coordinate-ascent update expression for $ \gamma_j $, we exploit the product structure of our variational distribution to evaluate an expression for the ELBO.
{\allowdisplaybreaks
\begin{align}
    \L_Q 
    =&\ - \E_{\mu, \sigma, \gamma} \left[ 
        \log \frac
            {dQ_{\mu, \sigma, \gamma}}
            {d \Pi} (\beta)
        - l_p(\D; \beta)
   \right] \nonumber \\
   % ---
    =&\ - \E_{\mu, \sigma, \gamma} \left[ \sum_{j=1}^p \left( 
            \mathbb{I}_{\{z_j =1 \}} \log \frac
                {\gamma_j d N(\mu_j, \sigma_j^2)}
                {\bar{w}_j d \text{Lap}(\lambda)} (\beta_j)
            + \mathbb{I}_{\{z_j = 0 \}} \log \frac
                {(1-\gamma_j)}
                {(1-\bar{w}_j)}
        \right)
        - l_p(\D; \beta)
   \right] \nonumber \\
   % ---
    = &\ - \sum_{j=1}^p \left( \E_{Q_j} \left[ 
            \mathbb{I}_{\{z_j =1 \}} \log \frac
                {\gamma_j d N(\mu_j, \sigma_j^2)}
                {\bar{w} d \text{Lap}(\lambda)} (\beta_j)
            + \mathbb{I}_{\{z_j = 0 \}} \log \frac
                {(1-\gamma_j)}
                {(1-\bar{w})}
       \right] 
    \right)
    + \E_{\mu, \sigma, \gamma} \left[ l_p(\D; \beta) \right] \nonumber \\
    % ---
    = &\ - \sum_{j=1}^p \Bigg(
        \gamma_j \Bigg\{ 
            \lambda \sigma_j \sqrt{2/\pi} e^{-\mu_j^2/(2\sigma_j^2)}
            + \lambda \mu_j (1 - 2 \Phi( - \mu_j / \sigma_j ))
            + \log \frac{\sqrt{2}}{\sqrt{\pi} \sigma_j \lambda }
            - \frac{1}{2} \nonumber \\
            & + \log \frac{\gamma_j}{1-\gamma_j}
            - \log \frac{a_0}{b_0}
        \Bigg\}
	+ \log(1 - \gamma_j) - \log(1 - \bar{w})
    \Bigg)
    + \E_{\mu, \sigma, \gamma} \left[ l_p(\D; \beta) \right]
\end{align}
} % end of \allowframebreak
As we cannot evaluate a closed-form expression for $ \E_{\mu, \sigma, \gamma} \left[ l_p(\D; \beta) \right] $, we use Monte Carlo integration to estimate this quantity, i.e. we compute
\begin{equation} \label{eq:estimating_elbo}
    \widehat{\L}_Q = 
    \frac{1}{B} \sum_{i=1}^B \log L_p(\D; \beta^{(i)} )
    - \text{KL}(Q \| \Pi )
    , \quad Q \in \Q,
\end{equation}
where $B$ are the number of Monte Carlo samples and $\beta^{(i)} \overset{\text{iid}}{\sim} Q$ for $i = 1,\dots, B$. Notably, the elements of $\beta^{(i)} = (\beta_1^{(i)}, \dots, \beta_p^{(i)})^\top \in \R^p$ are given by sampling $\beta_j^{(i)} \overset{\text{ind}}{\sim} N(\mu_j, \sigma_j^2)$ with probability $\gamma_j$ or taking $\beta_j^{(i)} = 0$ with probability $1-\gamma_j$ for $j = 1,\dots,p$. The remaining term can be evaluated explicitly as
\begin{align}
    & \KL(Q \| \Pi)
    = \ \sum_{j=1}^p \Bigg(
        \gamma_j \Bigg\{ 
	    \lambda \sigma_j \sqrt{\frac{2}{\pi}} e^{-\frac{\mu_j^2}{2\sigma_j^2}}
            + \log \frac{\sqrt{2}}{\sqrt{\pi} \sigma_j \lambda }
	    - \frac{1}{2} \label{eq:elbo_part_2} \\
	    & + \lambda \mu_j (1 - 2 \Phi(-\frac{\mu_j}{\sigma_j} ))
            + \log \frac{\gamma_j}{1-\gamma_j}
            - \log \frac{a_0}{b_0}
        \Bigg\}
        + \log(1 - \gamma_j)
    \Bigg). \nonumber
\end{align}

Although popular, model selection based on the ELBO is not justified theoretically and can be sensitive to the setting; therefore we consider additional goodness of fit measures. One such measure, proposed by \citeauthor{Nott12} (\citeyear{Nott12}), involves using the variational posterior to approximate the log-predictive density score (LPDS) on an held out testing dataset, defined as
\begin{equation*}
    \text{LPDS}
    = \log \int L_p \left( \D_{\text{test}} ; \beta \right) d \tilde{\Pi}(\beta | \D_{\text{train}}) 
    = \log \E_{\tilde{\Pi}} \left[ L_p \left( \D_{\text{test}} ; \beta \right) \right],
\end{equation*}
where
$\D_{\text{test}}$ is the held out test set and $\D_{\text{train}}$ is the training set used to compute $\tilde{\Pi}$ \citep{Nott12}. By Jensen's inequality,
\begin{equation*}
    \log \E_{\tilde{\Pi}} \left[ L_p \left( \D_{\text{test}} | \beta \right) \right]
    \leq
    \E_{\tilde{\Pi}} \left[ \log L_p \left( \D_{\text{test}} | \beta \right) \right]
\end{equation*}
Hence, given that computing the ELBO for a validation set involves computing $\E_{\tilde{\Pi}} \left[ \log L_p \left( \D_{\text{test}} ; \beta \right) \right]$, we can cheaply obtain an upper bound for the LPDS to use as a goodness of fit measure.

A further goodness of fit measure specific to survival models is the concordance index (c-index), defined as 
$$
    k = \mathbb{P}(\boldsymbol{T}_i > \boldsymbol{T}_j | \eta_j > \eta_i),\quad i \neq j, 
$$
where $ \eta_k = \beta_0^\top x_k $ is referred to as the \textit{prognostic index}. Intuitively, for two observations $(t_i, \delta_i=1, x_i)$ and $(t_j, \delta_j=1, x_j)$, when $ \eta_j > \eta_i $ we would expect $ t_i > t_j $. This remark follows from the form of the hazard function under the PHM, where we assume $h(t) = h_0(t) \exp(\beta^\top x) = h_0(t) \exp( \eta) $. Therefore, when $ \eta_j > \eta_i $ the hazard rate $h(t_j) > h(t_i) $ and thus we would expect $t_j$ to have failed before $t_i$. Often the c-index is estimated by,
\begin{equation*} \label{eq:cindex_estimator} 
    \widehat{k} = \frac
	{
	\sum_{i=1}^n \sum_{j > i} \
	    \I(t_i < t_j) \I(\widehat{\eta}_i > \widehat{\eta}_j) \delta_i +
	    \I(t_j < t_i) \I(\widehat{\eta}_j > \widehat{\eta}_i) \delta_j}
	{
	\sum_{i=1}^n \sum_{j > i} \
	    \I(t_i < t_j) \delta_i + \I(t_j < t_i) \delta_j
	},
\end{equation*}
where the prognostic index is estimated using $ \widehat{\eta}_j = \widehat{\beta}^\top x_j $ for a given point estimate $\widehat{\beta}$ of $\beta$, and $ \mathbb{I}(\cdot) $ is the indicator function \citep{Harrell82}. Notably, the c-index is not robust to censoring and tends to overestimate $k$ when there is a high degree of censoring \citep{Gonen05}.

\section{Simulation study}

\subsection{Markov chain Monte Carlo sampler} \label{appendix:mcmc}

To construct our sampler we note that the model given by \eqref{eq:posterior_phm} can be reformulated such that the prior is given by:
\begin{equation}
\begin{aligned}
    \beta_j \overset{\text{iid}}{\sim} & \ \text{Laplace}(\lambda) \\
    z_j | w_j \overset{\text{ind}}{\sim} & \ \text{Bernoulli}(w_j) \\
    w_j \overset{\text{iid}}{\sim} & \ \text{Beta}(a_0, b_0)
\end{aligned}
\end{equation}
and likelihood as $p(\D | \beta, z) = L_p(\D; \beta \circ z)$ where $\circ$ denotes the element wise product, i.e. $ (\beta \circ z)_j = \beta_j z_j$.

\Cref{alg:mcmc_sampler} details our Gibbs sampler for the posterior in \eqref{eq:posterior_phm} based on the above formulation. Notably, we introduce the notation $ x_{k:l} := (x_i)_{i=k}^l $, for $ 1 \leq l < k \leq p,\ x \in \R^p$, and denote $\beta^{(i)} \in \R^p$, $z^{(i)} \in \{0, 1\}^p$ and $w^{(i)} \in [0,1]^p$ as the MCMC samples.

\begin{algorithm}[htp]
    \caption{Spike and Slab MCMC sampler}
    \label{alg:mcmc_sampler}
    \begin{algorithmic}[1]
    \State \textbf{Require:} $ K $, the Metropolis-Hastings proposal kernel, N: number of samples.
	\State Initialise $ z^{(0)}, \beta^{(0)}, w^{(0)} $
	\For {$ i = 1, \dots N $}
	    \NoDo \For {$ j = 1, \dots p $}
    	    \State $ w^{(i)}_j \sim w_j | \D, \beta_{1:p}^{(i-1)}, z_{1:p}^{(i-1)}, w^{(i-1)}_{1:j-1}, w^{(i-1)}_{j+1:p} $
	    \EndFor
	    
	    \NoDo \For {$ j = 1, \dots p $}
    	    \State $ z_j^{(i)} \sim z_j | \D, \beta_{1:p}^{(i-1)}, z_{1:(j-1)}^{(i)},                                  z_{(j+1):p}^{(i-1)}, w_{1:p}^{(i)} $
	    \EndFor
	    \NoDo \For {$ j = 1, \dots p $}
    	    \State $ \beta_j^{(i)} \sim \beta_j | \D, \beta_{1:(j-1)}^{(i-1)},                                     \beta_{(j+1):p}^{(i-1)}, z_{1:p}^{(i)}, w_{1:p}^{(i)} $
	    \EndFor
	\EndFor
    \end{algorithmic}
\end{algorithm}

Ignoring the superscript for clarity, the distribution $ w_j | \D, \beta, z, w_{-j} $ is conditionally independent of $ \D, \beta, z$ and $w_{-j}$. Therefore, $w^{(i)}_j$ is sampled iid. from the prior of $w_j$, i.e. $w_j^{(i)} \overset{\text{iid.}}{\sim} \text{Beta}(a_0, b_0)$. Regarding $z_j^{(i)}$, the conditional density
\begin{align} 
    p(z_j | \D, \beta, z_{-j}, w) \propto &\ p(\D | \beta, z_{-j}, z_j, w) \pi(z_j | \beta, z_{-j}, w). \nonumber \\
    = &\ p(\D; \beta, z) \pi(z_j | w_j). \label{eq:mcmc_zj} 
\end{align}
As $z_j$ is discrete, evaluating the RHS of \eqref{eq:mcmc_zj} for $ z_j = 0 $ and $ z_j = 1 $, gives the unnormalised conditional probabilities. Summing gives the normalisation constant and thus we can sample $ z_j $ from a Bernoulli distribution with parameter 
\begin{equation}
    p = \frac
	{p(z_j = 1 | \D, \beta, z_{-j}, w)}
	{p(z_j = 0 | \D, \beta, z_{-j}, w) + p(z_j = 1 | \D, \beta, z_{-j}, w)}.
\end{equation}
Finally, to sample from $ \beta_j^{(i)} $ we use a Metropolis-Hastings within Gibbs step, wherein a proposal $ \beta_j^{(i)} $ is sampled from a random-walk proposition kernel $ K $. The proposal is then accepted with probability $A$ or rejected with probability $1-A$, in which case $ \beta_j^{(i)} \leftarrow \beta_j^{(i-1)} $. Noting $A$ is given by,
\begin{equation}
    A = \min \left(1, \frac
	{p(\D; \beta_{-j}, \beta_j^{(i)}, z^{(i)}) \pi(\beta_j^{(i)}) } 
	{p(\D; \beta_{-j}, \beta_j^{(i-1)}, z^{(i)}) \pi(\beta_j^{(i-1)})} 
	\frac
	    {K(\beta_j^{(i-1)} |\beta_j^{(i)})}
	    {K(\beta_j^{(i)} |\beta_j^{(i-1)})}
    \right)
\end{equation}
% and 
% \begin{equation}
%     p(\beta_j | \D, \beta_{-j}, z, w) \propto L_p(\D; \beta, z, w) \pi(\beta_j).
% \end{equation}
Within our implementation we let $ K = N\left( \beta_j^{(i-1)}, \sigma_k^2 \sigma_s ^{2 (1 - z_j^{(i-1)})} \right) $ with $ \sigma_k = 0.2 $ and $ \sigma_s = 10$. The implementation of the sampler is available as an \texttt{R} package that can be installed from \url{https://github.com/mkomod/survival-ss}.

\newpage
%-------------------------------------------------------------------------------

\section{Sensitivity analysis} \label{appendix:sensitivity_analysis}

\subsection{Sensitivity to starting values}

To examine the sensitivity with respect to the initialization of $\mu$, $\sigma$ and $\gamma$ we generate data as describe in \Cref{sec:simulation_design} taking $(n,p,s,c)=(200,1000,10,0.25)$.

\subsubsection{Sensitivity to $\mu$}

To examine the sensitivity with respect to initialization of $\mu$ we compared four different methods: 
\begin{itemize}
    \itemsep0em
    \item Random: where $\mu_j \overset{\text{iid.}}{\sim} N(0,1)$ for $j=1,\dots,p$.
    \item Ridge: where $\mu$ is the MLE under the ridge penalty.
    \item Elastic Net: where $\mu$ is the MLE under the elastic net penalty (equal mixture of ridge and LASSO penalties).
    \item LASSO: where $\mu$ is the MLE under the LASSO penalty.
\end{itemize}
In the last three cases a regularization hyperparameter of $100 \lambda_{\min}$ is used, where $\lambda_{\min}$ is the hyperparameter wherein all estimates coefficients are equal to zero. In order to compute the MLE under the different penalties we use \texttt{glmnet}. To evaluate the different initialization methods we compute the $\ell_2$-error, $\ell_1$-error, TPR, FDR, AUC and runtime, presenting the median, lower ($5\%$) and upper ($95\%$) quantiles across 100 runs in \Cref{tab:sen_mu}. 

Examining \Cref{tab:sen_mu}, we notice that within setting 1 the initialization method does not impact the performance of the method with all methods performing equally. It is worth noting the runtime for random initialization is largest, meaning it takes the algorithm longer to converge given a poor initialization. Within settings 2-4, the initialization methods can have a substantial effect on performance. For instance, within settings 3 initialization with ridge yields a median $\ell_2$-error of $1.019$, whereas initialization with the LASSO penalty gives a median $\ell_2$-error of $0.394$. Furthermore, poor initialization has an effect on the variable selection of the method (\Cref{tab:sen_mu}). Overall, initialization of $\mu$ using the LASSO gave the best models, obtaining best metrics across the different settings. In some cases however, initialization with the LASSO can be slower than other methods e.g. the elastic net, which produced comparable models.

\sidewaystable{
\begin{table}[H]
    \centering
    \resizebox{1.4\textwidth}{!}{ %
{\setlength{\tabcolsep}{2.7em} 
\begin{tabular}{| l c| c c c c c c |}
    \hline
    Setting & Init. Meth. & $\ell_2$-error & $\ell_1$-error & TPR & FDR & AUC & Runtime\\
    \hline
    % --- Indep
    \rule{0pt}{1\normalbaselineskip}
    \multirow{4}{*}{\textit{Setting 1}}
    & Random & 
0.368 (0.21, 0.70)& 0.999 (0.52, 1.86)& 1.000 (0.90, 1.00)& 0.000 (0.00, 0.00)& 1.000 (1.00, 1.00)&  19.3s  (15.6s,31.3s)\\
    & Ridge & 
 0.368 (0.21, 0.70)& 0.999 (0.52, 1.86)& 1.000 (0.90, 1.00)& 0.000 (0.00, 0.00)& 1.000 (1.00, 1.00)&  7.1s  (4.8s,13.0s)\\
    & Elas. Net. & 
 0.368 (0.21, 0.70)& 0.999 (0.52, 1.86)& 1.000 (0.90, 1.00)& 0.000 (0.00, 0.00)& 1.000 (1.00, 1.00)&  7.3s  (4.5s,12.3s)\\
    & LASSO & 
 0.368 (0.21, 0.70)& 1.000 (0.52, 1.86)& 1.000 (0.90, 1.00)& 0.000 (0.00, 0.00)& 1.000 (1.00, 1.00)&  13.8s  (9.4s,19.8s)\\
    [.4em]
    \hline 
    % --- Block
    \rule{0pt}{1\normalbaselineskip}
    \multirow{4}{*}{\textit{Setting 2}}
    & Random & 
 0.848 (0.26, 3.06)& 1.988 (0.68, 10.34)& 0.900 (0.20, 1.00)& 0.000 (0.00, 0.50)& 1.000 (0.75, 1.00)&  29.3s  (19.0s,44.9s)\\
    & Ridge & 
 0.838 (0.26, 3.43)& 1.970 (0.69, 11.70)& 0.900 (0.30, 1.00)& 0.000 (0.00, 0.50)& 1.000 (0.74, 1.00)&  6.9s  (3.0s,12.3s)\\
    & Elas. Net. & 
 0.565 (0.24, 1.72)& 1.566 (0.67, 4.58)& 1.000 (0.70, 1.00)& 0.000 (0.00, 0.15)& 1.000 (0.90, 1.00)&  7.4s  (4.9s,12.2s)\\
    & LASSO & 
 0.445 (0.23, 1.13)& 1.204 (0.64, 3.07)& 1.000 (0.80, 1.00)& 0.000 (0.00, 0.10)& 1.000 (0.95, 1.00)&  12.3s  (8.4s,20.5s)\\
    [.4em]
    \hline 
    % --- N(m, S)
    \rule{0pt}{1\normalbaselineskip}
    \multirow{4}{*}{\textit{Setting 3}}
    & Random & 
 0.669 (0.21, 3.66)& 1.856 (0.58, 12.96)& 1.000 (0.30, 1.00)& 0.000 (0.00, 0.55)& 1.000 (0.75, 1.00)&  15.7s  (10.4s,26.6s)\\
    & Ridge & 
 1.019 (0.21, 3.58)& 2.680 (0.62, 11.69)& 0.900 (0.40, 1.00)& 0.000 (0.00, 0.33)& 0.999 (0.75, 1.00)&  8.5s  (3.0s,19.1s)\\
    & Elas. Net. & 
 0.423 (0.19, 1.68)& 1.170 (0.53, 4.19)& 1.000 (0.80, 1.00)& 0.000 (0.00, 0.10)& 1.000 (0.95, 1.00)&  10.6s  (6.0s,17.9s)\\
    & LASSO & 
 0.394 (0.18, 1.44)& 1.118 (0.53, 3.28)& 1.000 (0.90, 1.00)& 0.000 (0.00, 0.09)& 1.000 (0.95, 1.00)&  11.8s  (7.0s,19.8s)\\
    [.4em]
    \hline 
    % --- Real
    \rule{0pt}{1\normalbaselineskip}
    \multirow{4}{*}{\textit{Setting 4}}
    & Random & 
 0.464 (0.21, 2.88)& 1.289 (0.57, 9.30)& 1.000 (0.60, 1.00)& 0.000 (0.00, 0.33)& 1.000 (0.84, 1.00)&  17.1s  (11.2s,27.4s)\\
    & Ridge & 
 0.562 (0.22, 3.08)& 1.562 (0.63, 9.86)& 1.000 (0.50, 1.00)& 0.000 (0.00, 0.33)& 1.000 (0.80, 1.00)&  10.9s  (4.8s,17.5s)\\
    & Elas. Net. & 
 0.398 (0.19, 1.41)& 1.071 (0.50, 3.40)& 1.000 (0.90, 1.00)& 0.000 (0.00, 0.10)& 1.000 (0.95, 1.00)&  10.6s  (6.8s,17.0s)\\
    & LASSO & 
 0.390 (0.18, 1.15)& 1.058 (0.50, 3.05)& 1.000 (0.90, 1.00)& 0.000 (0.00, 0.10)& 1.000 (0.95, 1.00)&  12.0s  (6.5s,21.3s)\\
    [.4em]
    \hline 
\end{tabular} %
}}

    \caption{Sensitivity to different initialization methods for $\mu$. Presented are the median and $(5\%, 95\%)$ quantiles.\\ Data is generated taking $(n,p,s,c)=(200,1000,10,0.25)$.}
    \label{tab:sen_mu}
\end{table}
}
\break

\subsubsection{Sensitivity to $\sigma$ and $\gamma$}

To examine the sensitivity with respect to $\sigma$ and $\gamma$ we performed a grid search examining starting values of $S \times \Gamma$ where $S = \{ 0.01, 0.05, 0.10, 0.25, 0.5, 0.75, 1.0 \}$ and $\Gamma = \{0.01, 0.05, 0.10, 0.25, 0.5, 0.75 \}$. To examine the sensitivity we compute and present the mean $\ell_2$-error, $\ell_1$-error, TPR, FDR, AUC and runtime (in seconds) across 100 runs.

\begin{figure}[htp]
    \centering
    \includegraphics[width=.49\textwidth]{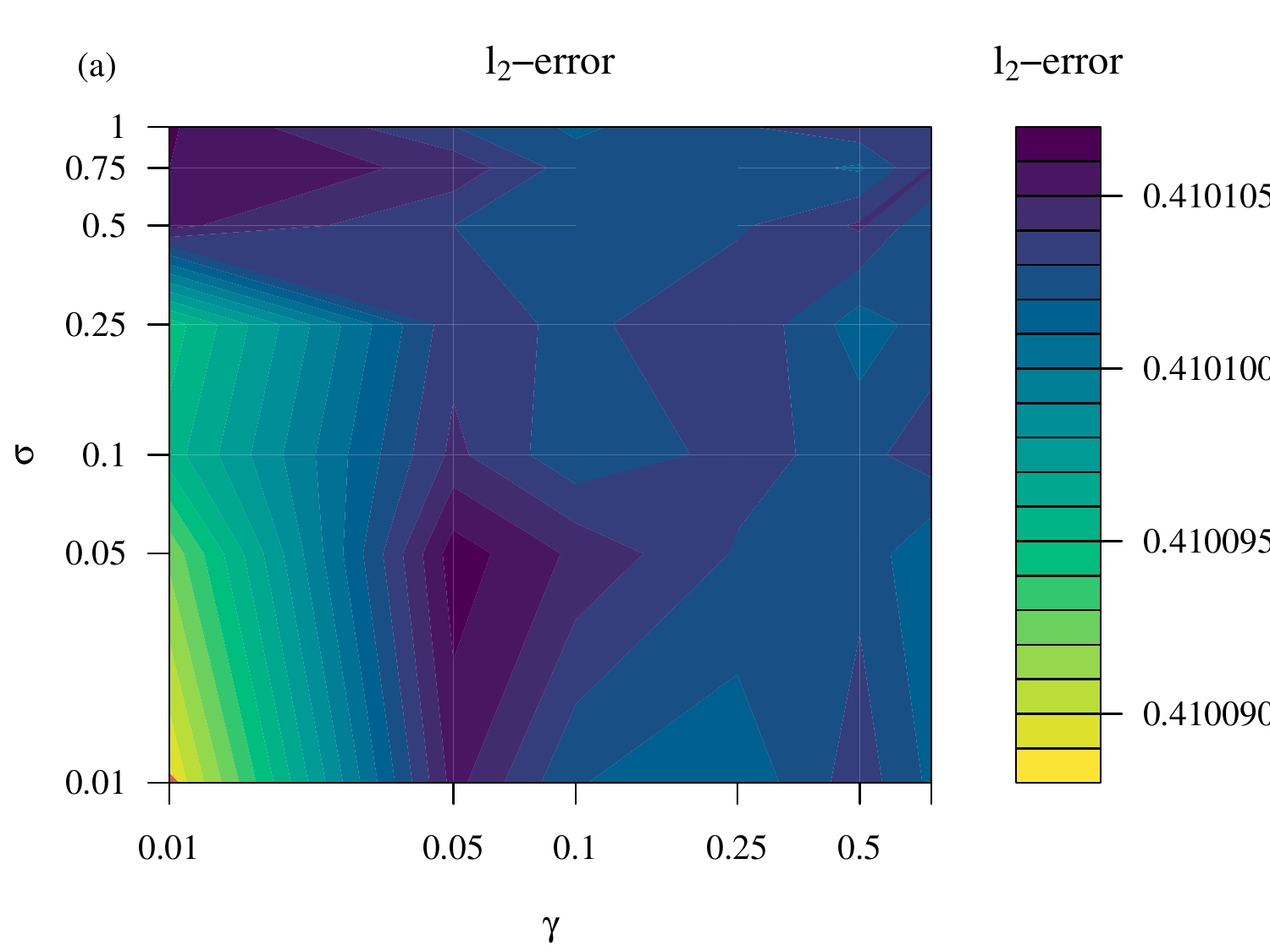}
    \includegraphics[width=.49\textwidth]{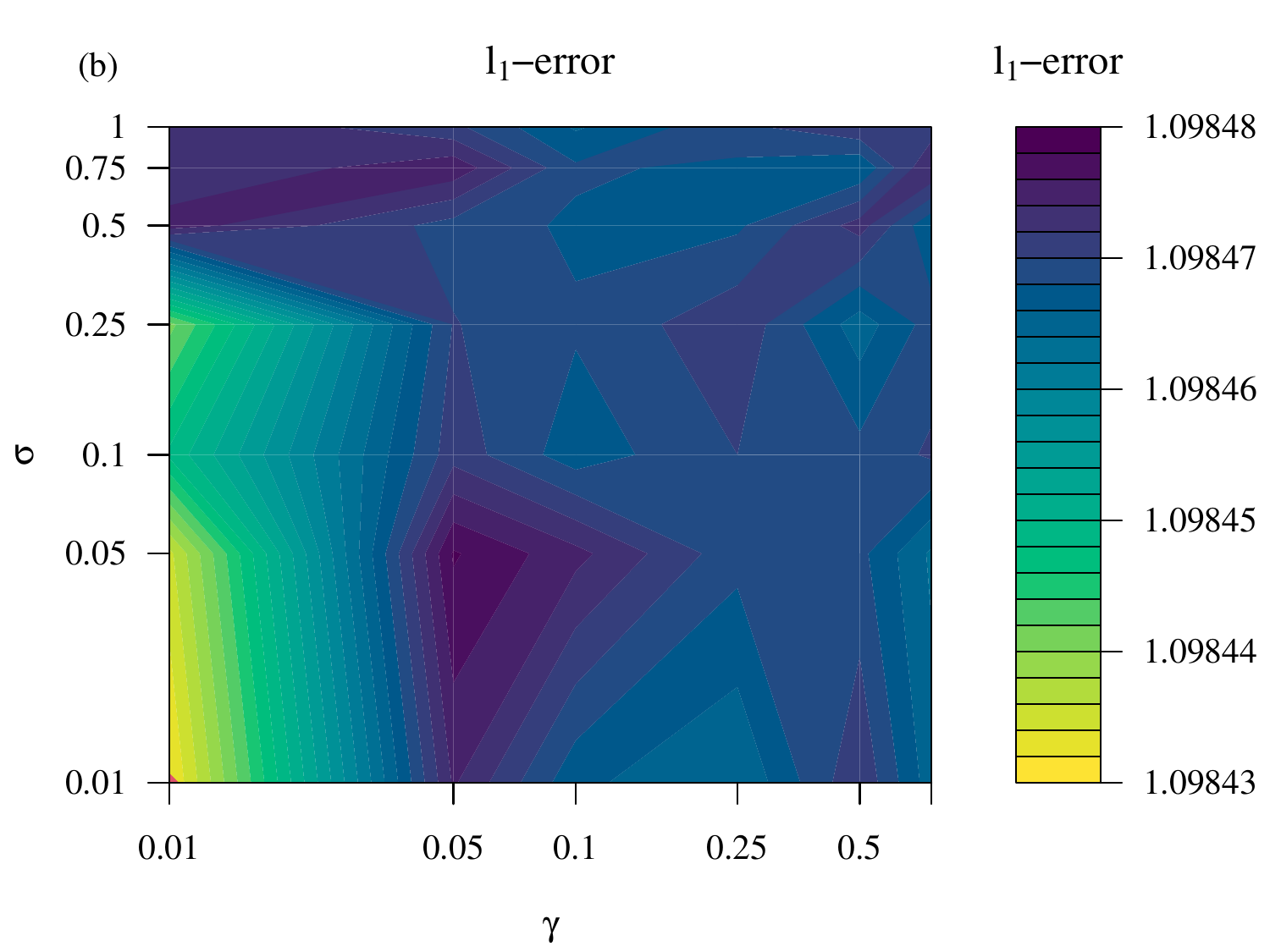}
    \includegraphics[width=.49\textwidth]{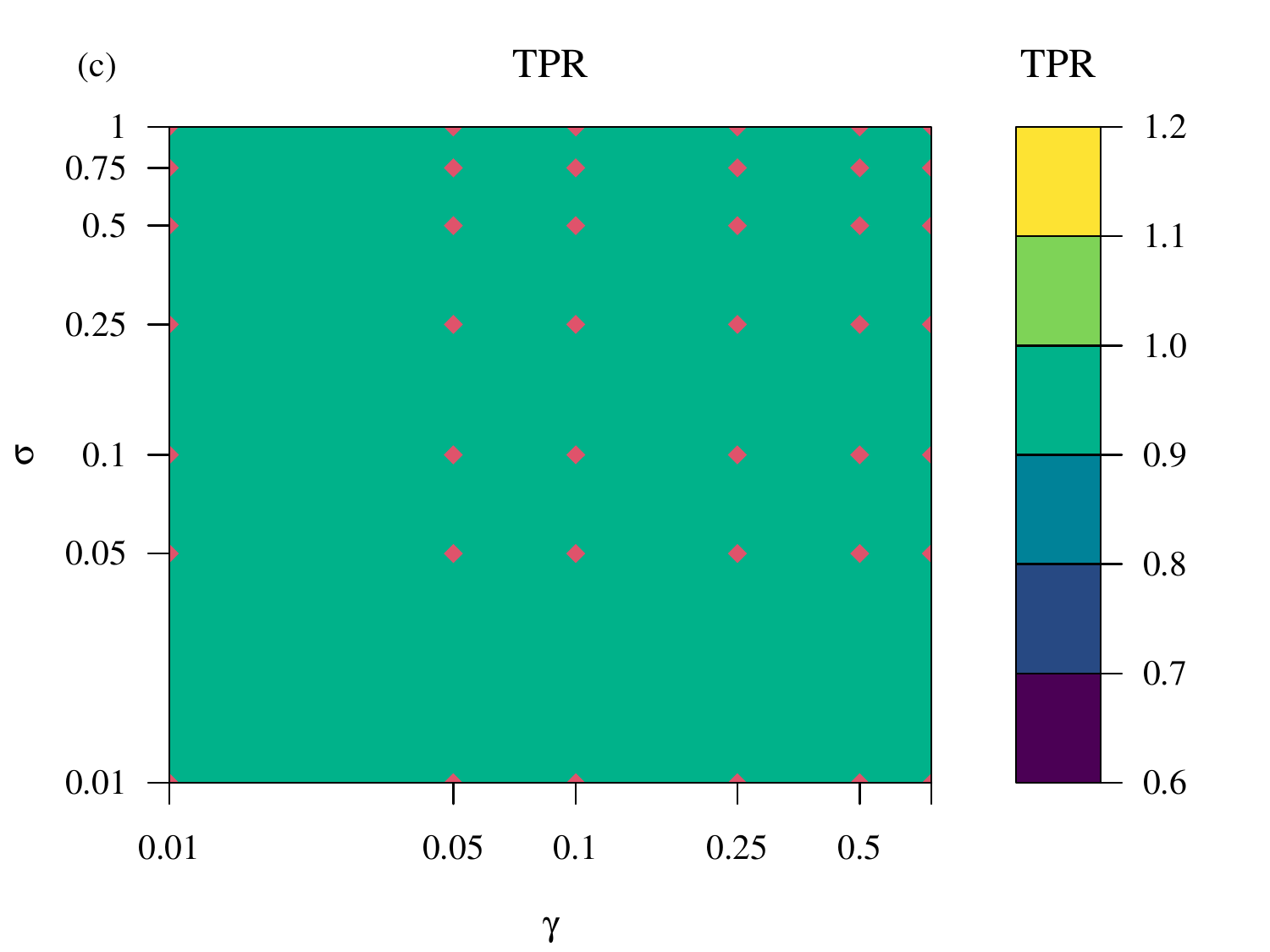}
    \includegraphics[width=.49\textwidth]{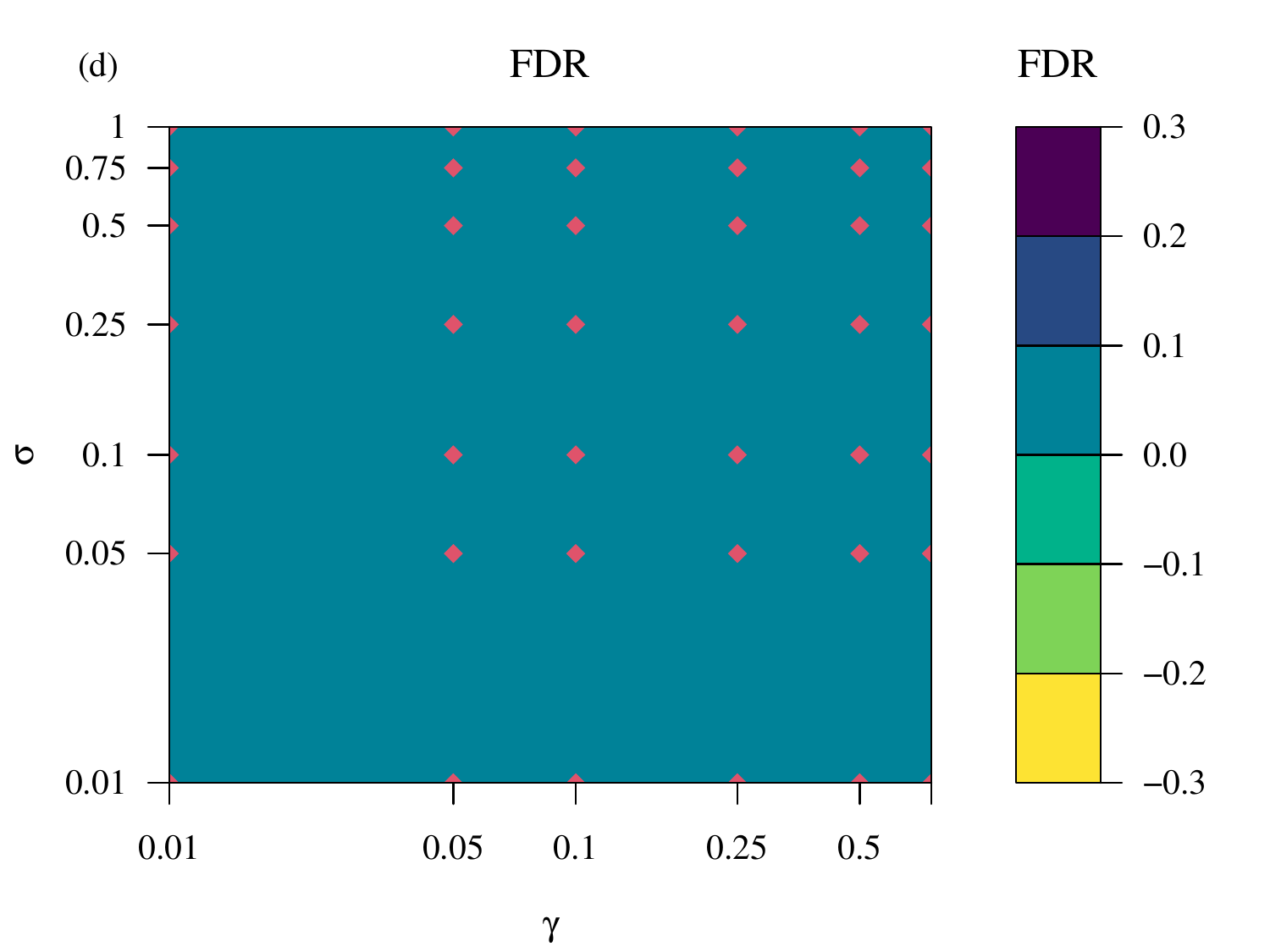}
    \includegraphics[width=.49\textwidth]{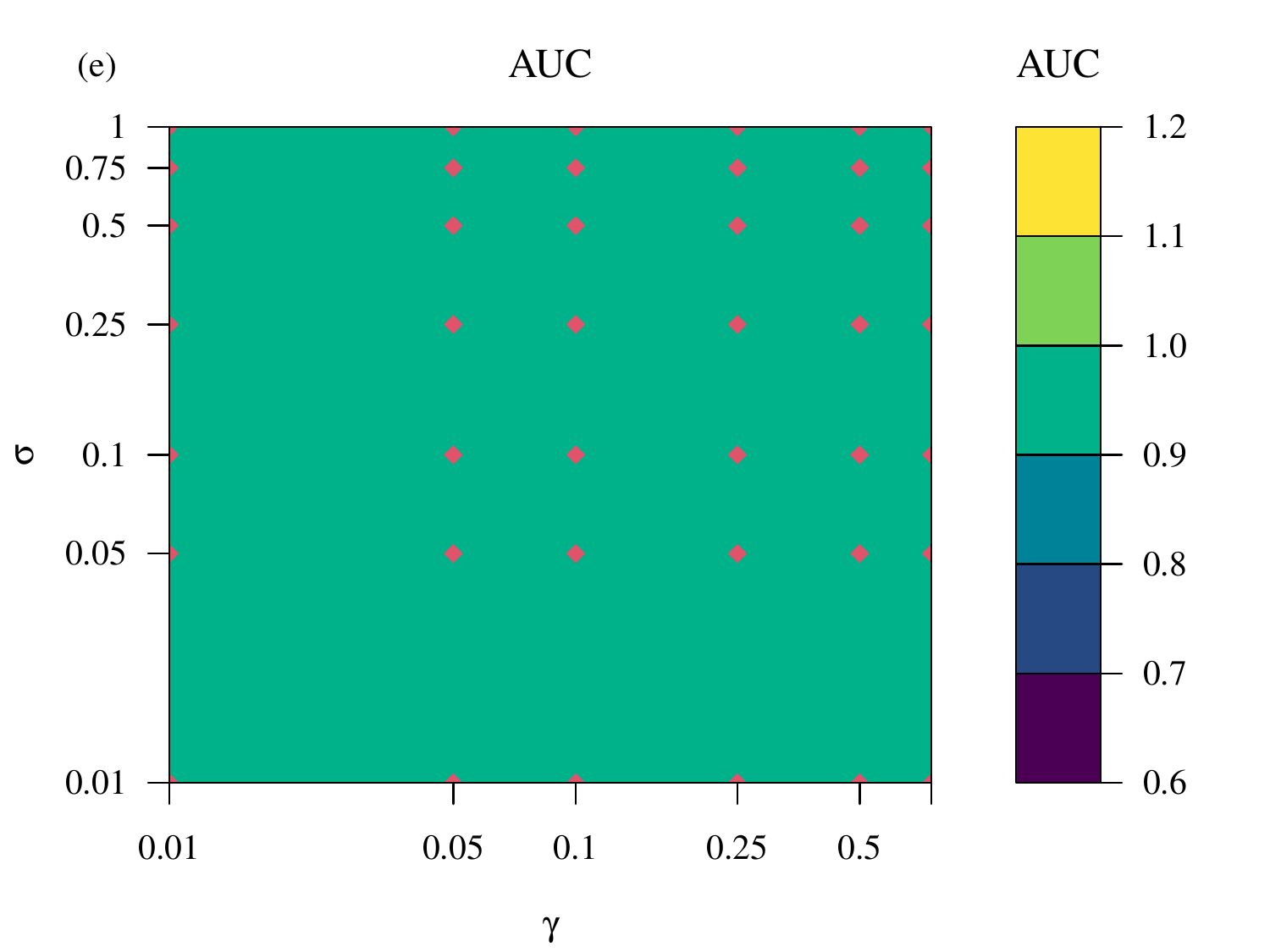}
    \includegraphics[width=.49\textwidth]{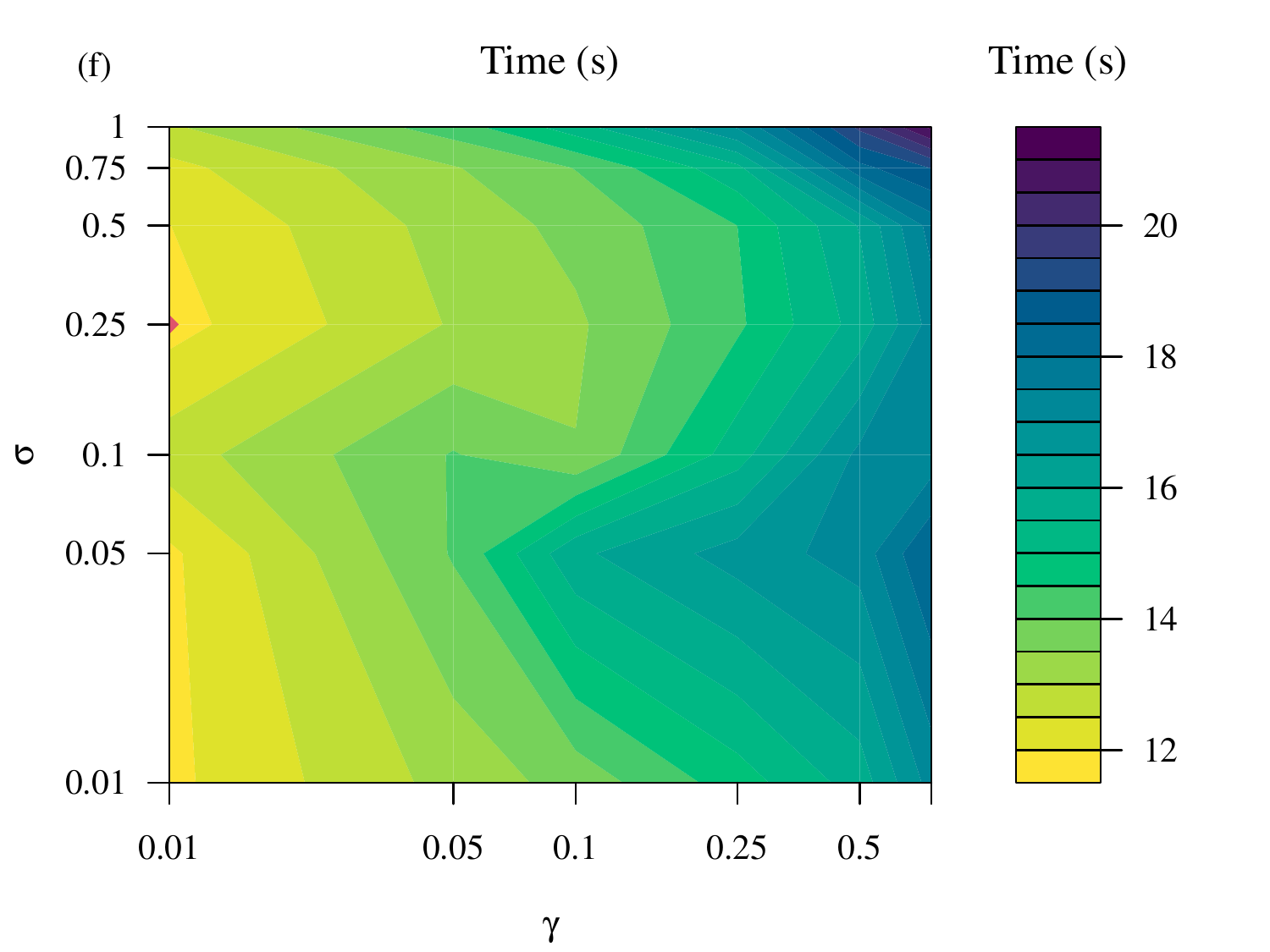}
    \caption{\textbf{Setting 1}: sensitivity with respect to $\sigma$ and $\gamma$}
    \label{fig:sensitivity_s_g_1}
\end{figure}

\begin{figure}[htp]
    \centering
    \includegraphics[width=.49\textwidth]{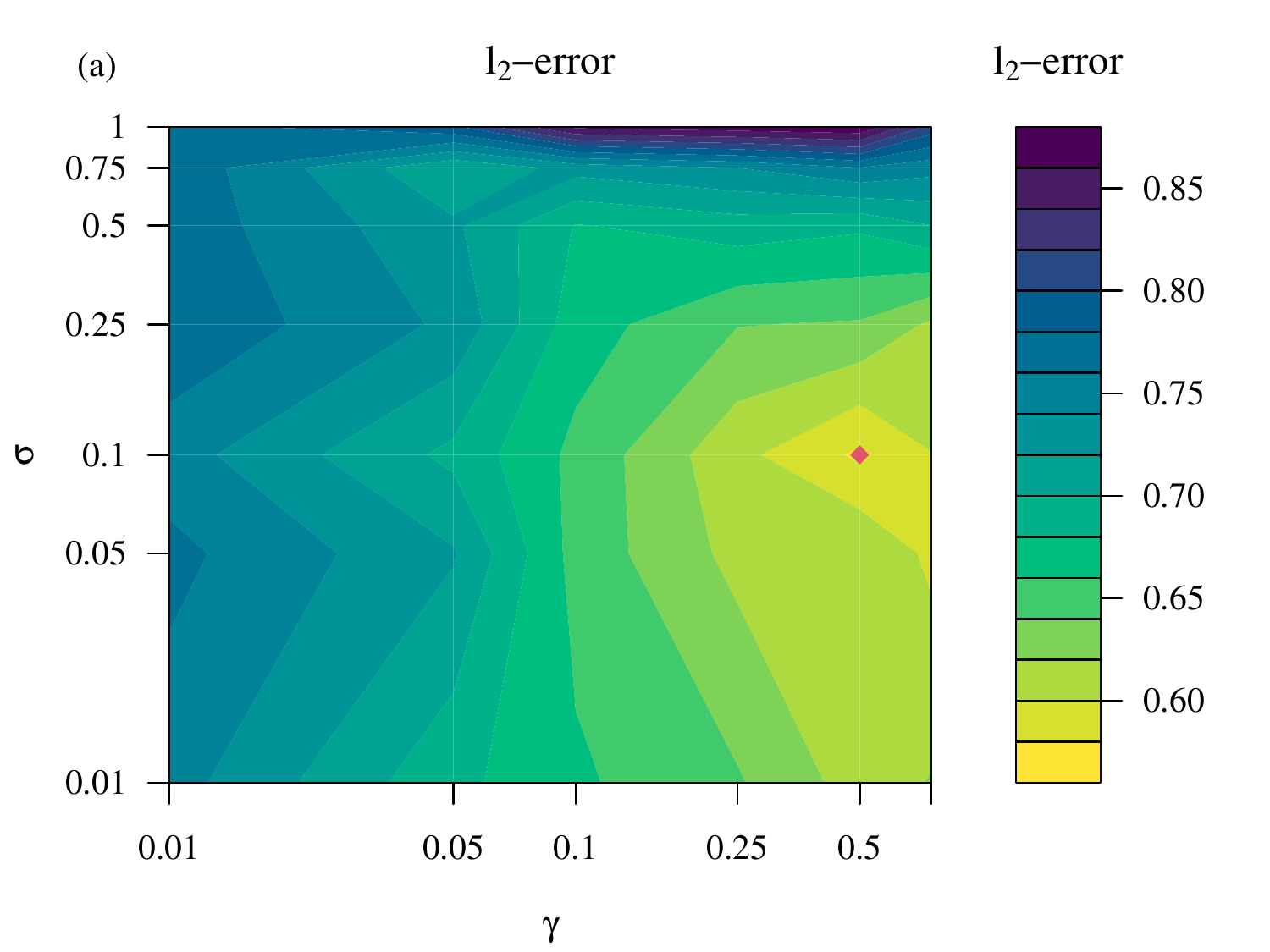}
    \includegraphics[width=.49\textwidth]{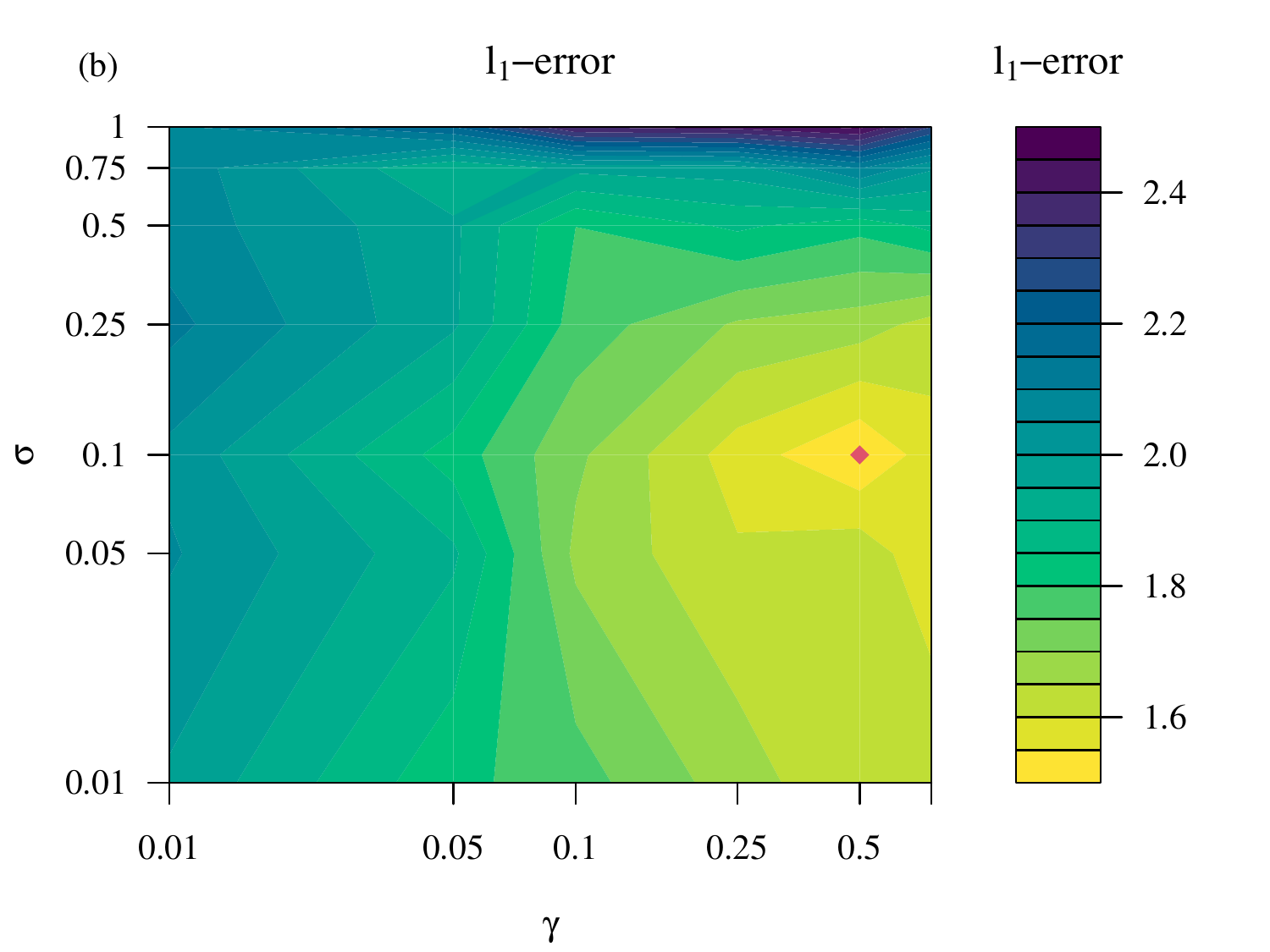}
    \includegraphics[width=.49\textwidth]{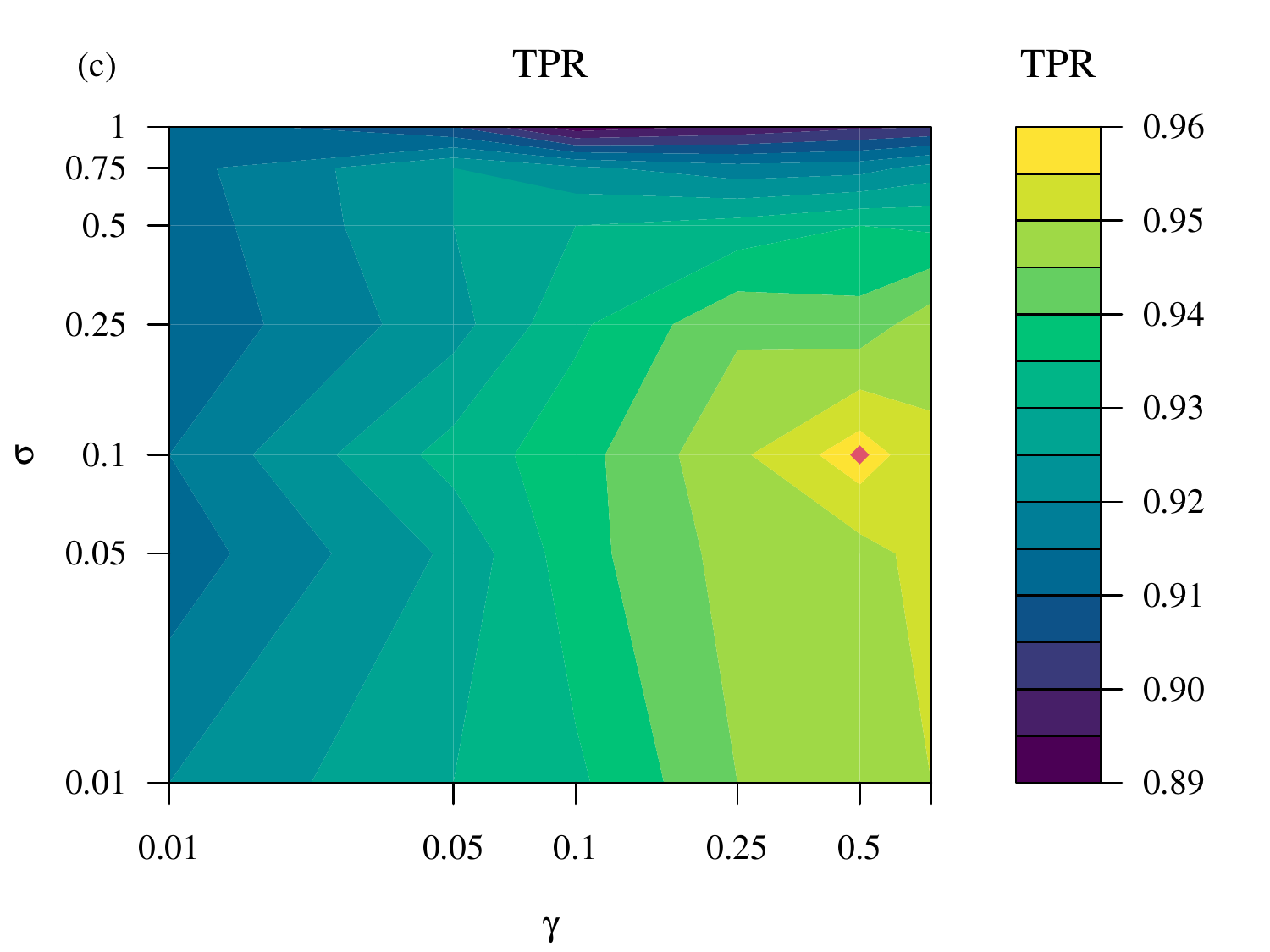}
    \includegraphics[width=.49\textwidth]{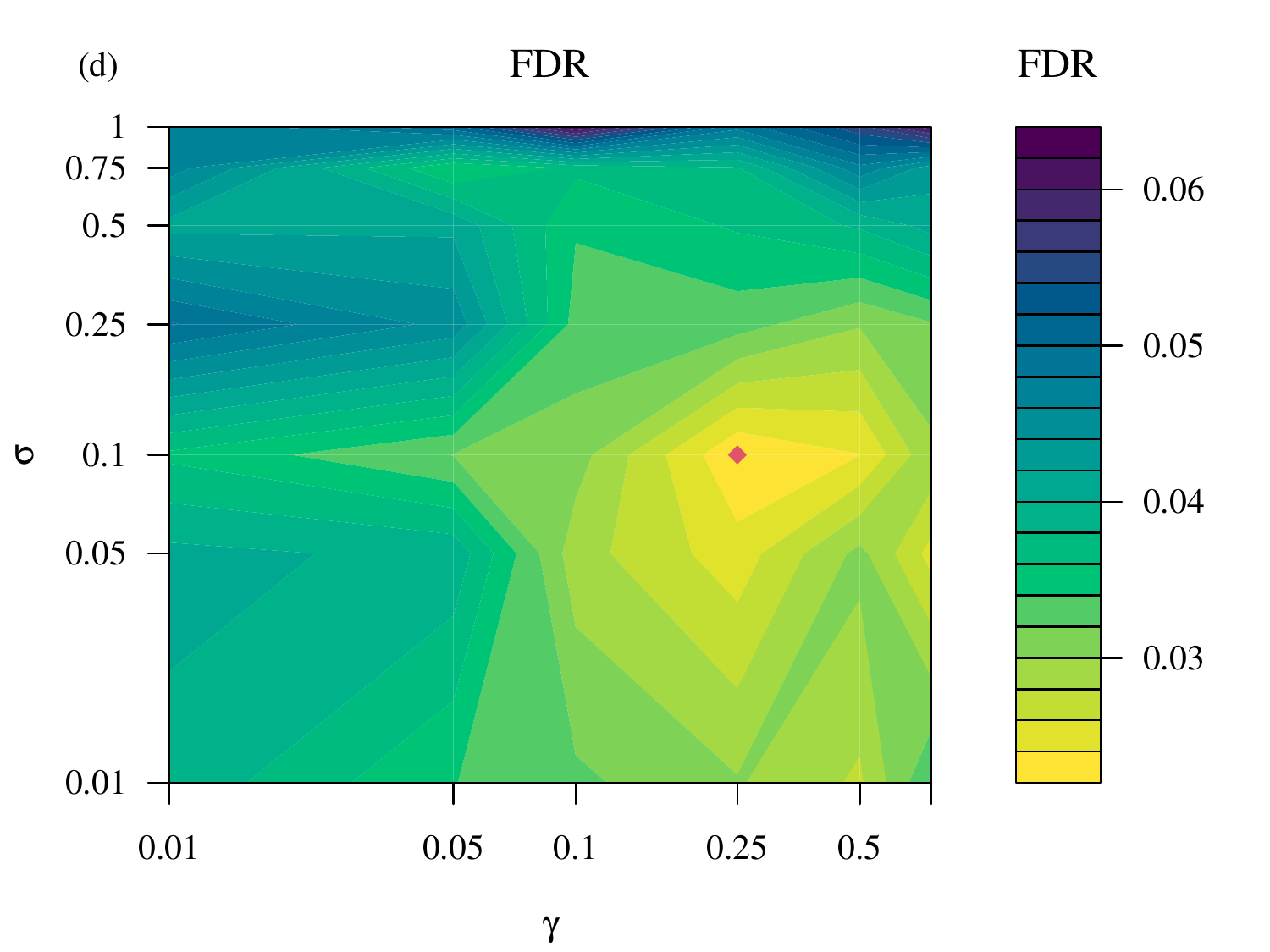}
    \includegraphics[width=.49\textwidth]{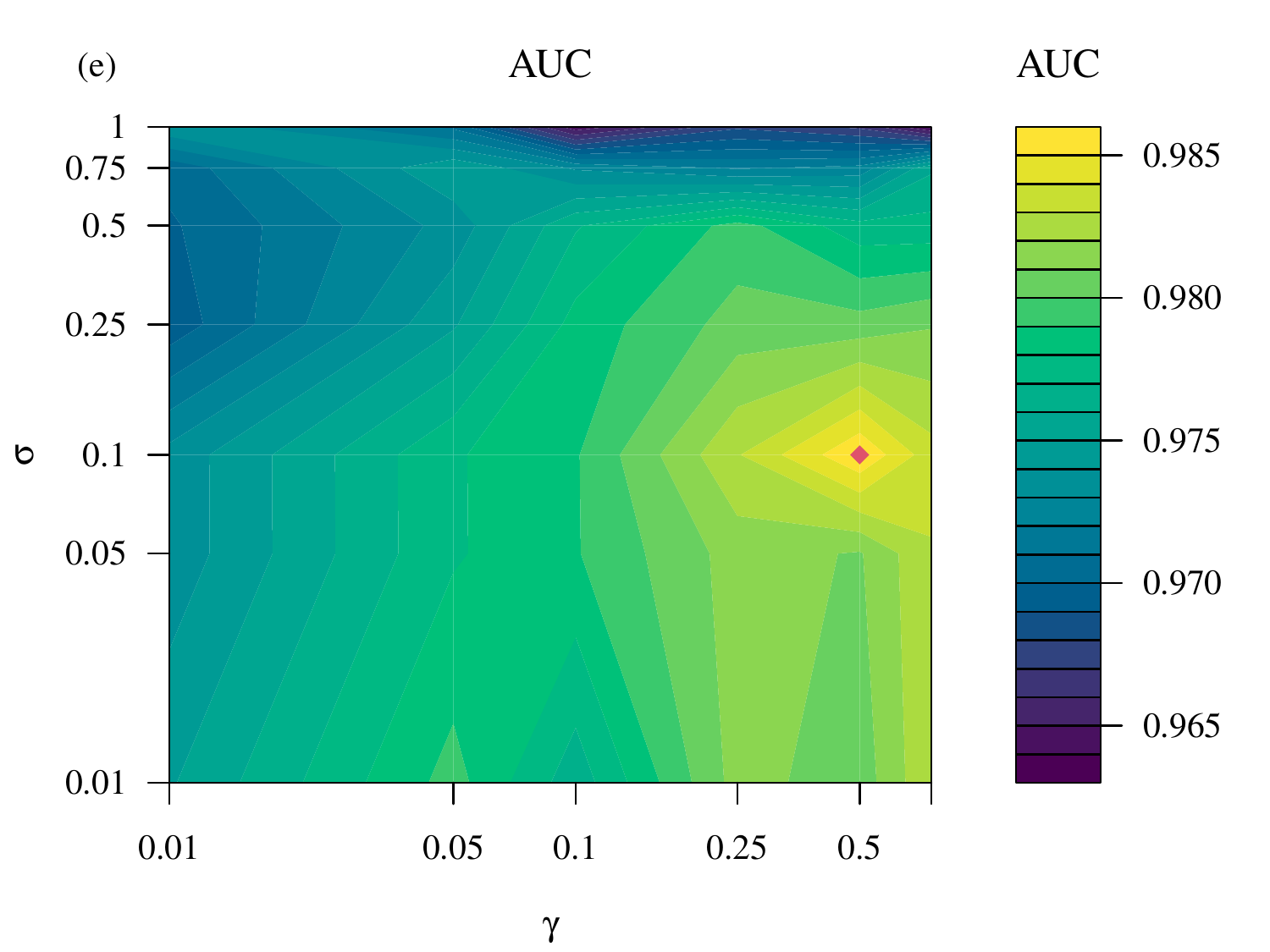}
    \includegraphics[width=.49\textwidth]{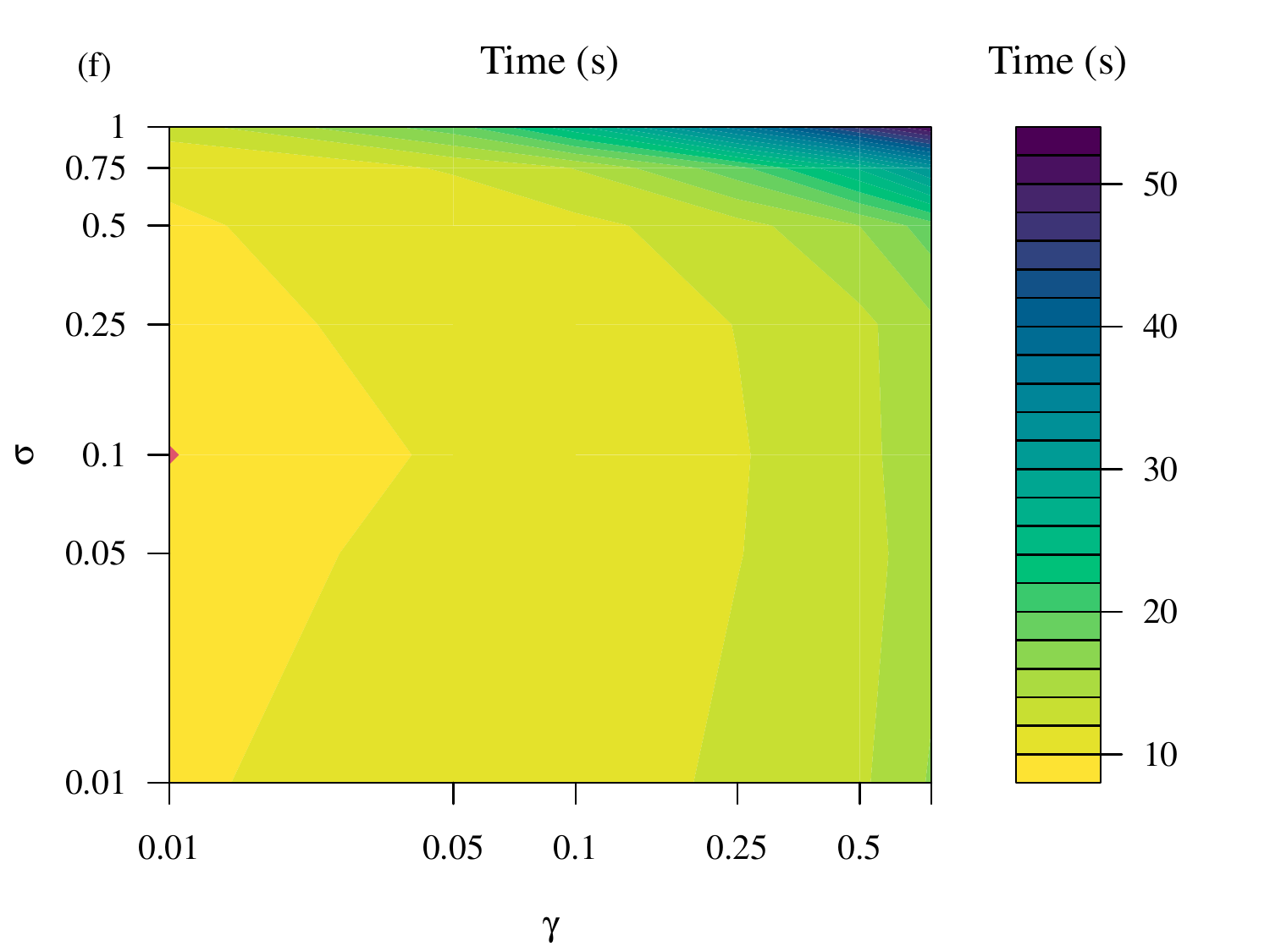}
    \caption{\textbf{Setting 2}: sensitivity with respect to $\sigma$ and $\gamma$}
    \label{fig:sensitivity_s_g_2}
\end{figure}

\begin{figure}[htp]
    \centering
    \includegraphics[width=.49\textwidth]{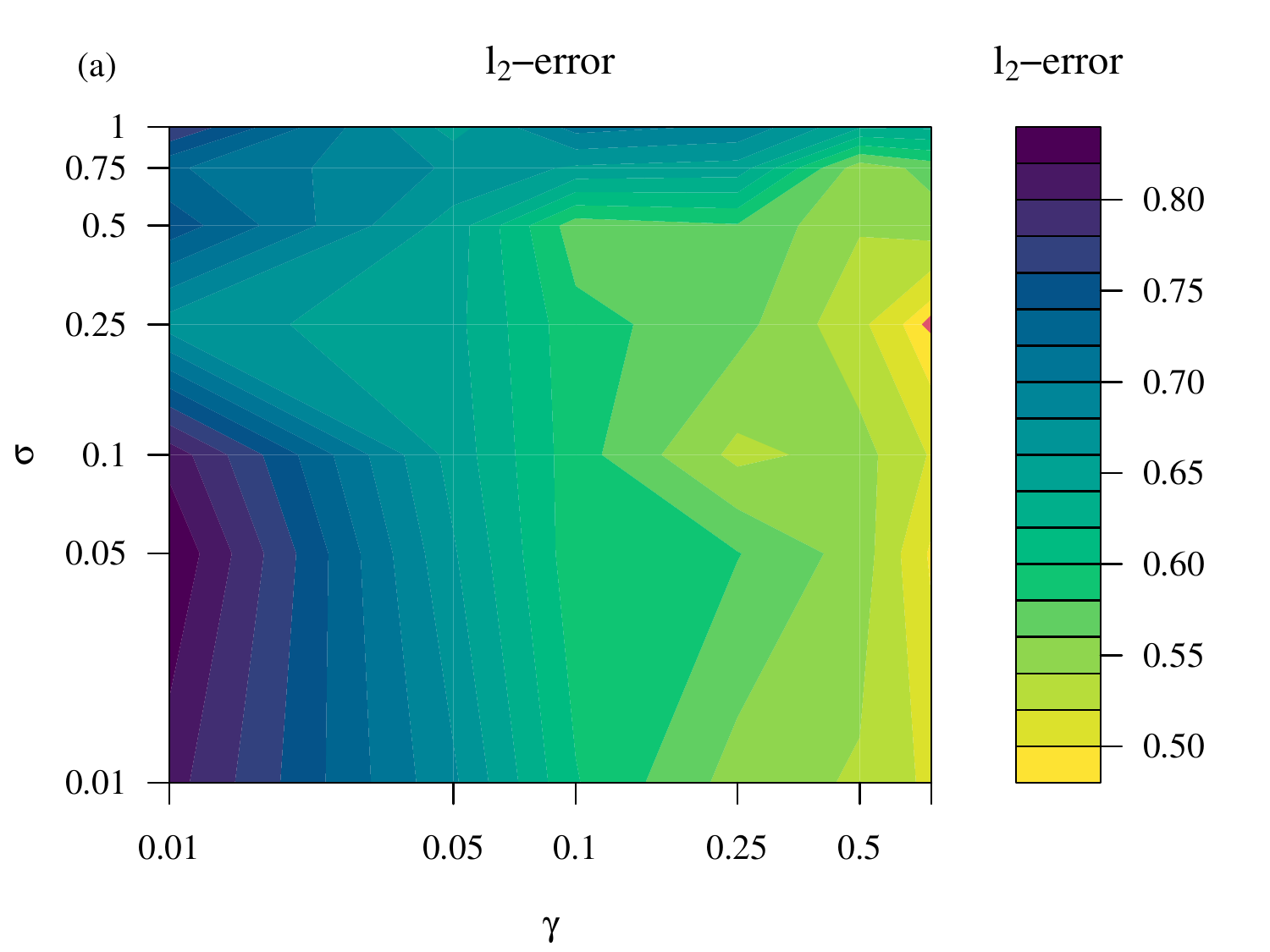}
    \includegraphics[width=.49\textwidth]{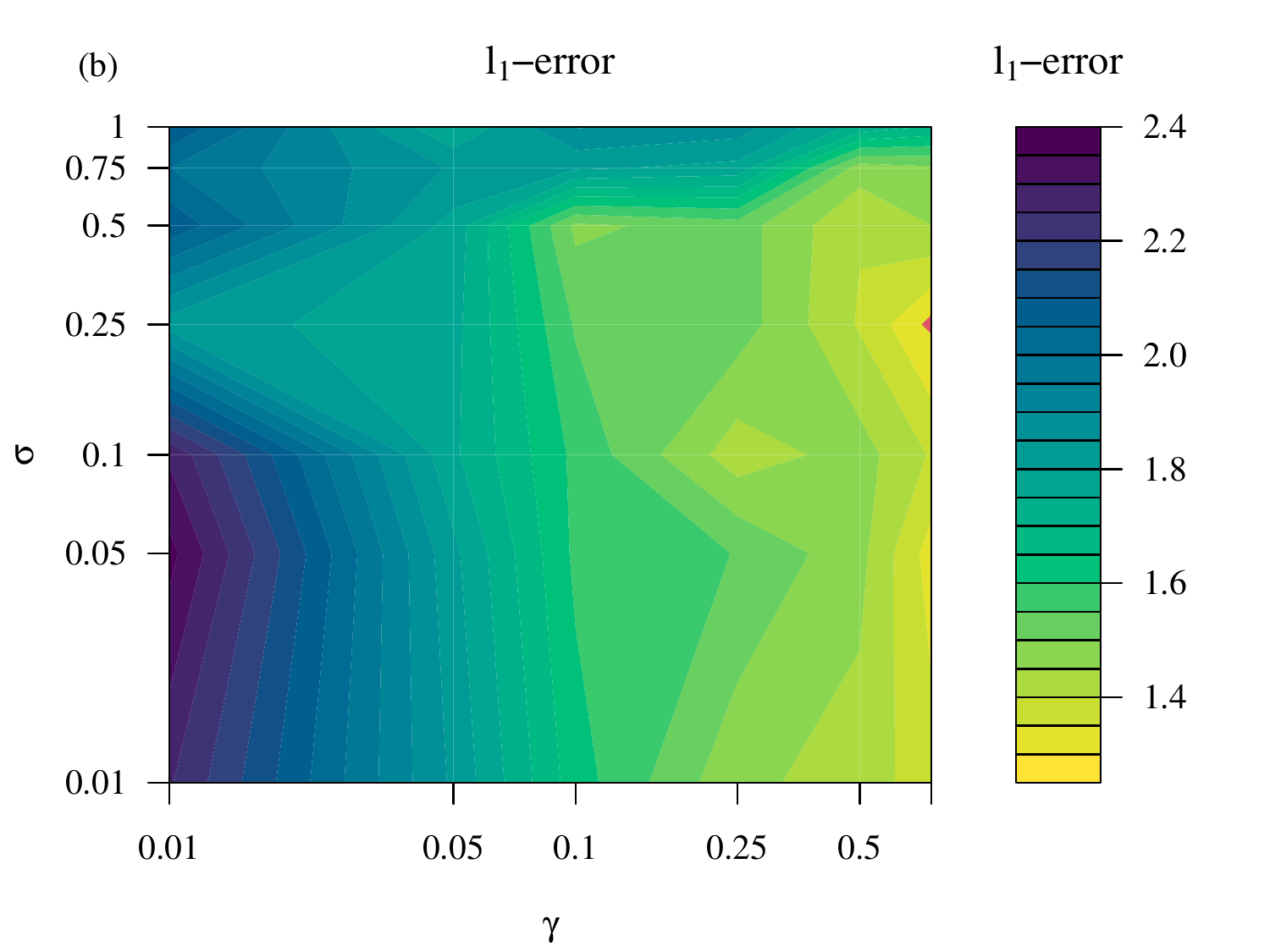}
    \includegraphics[width=.49\textwidth]{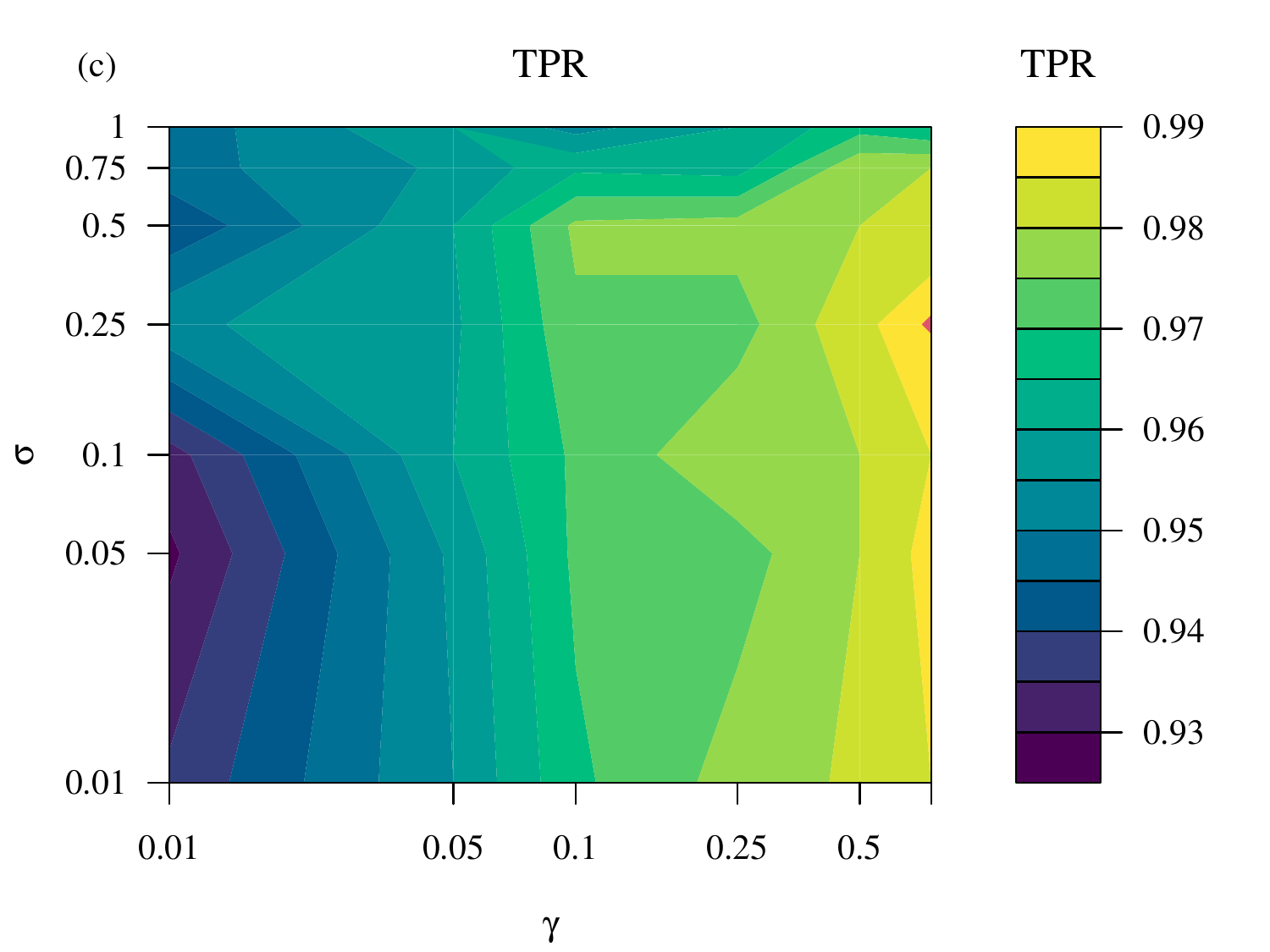}
    \includegraphics[width=.49\textwidth]{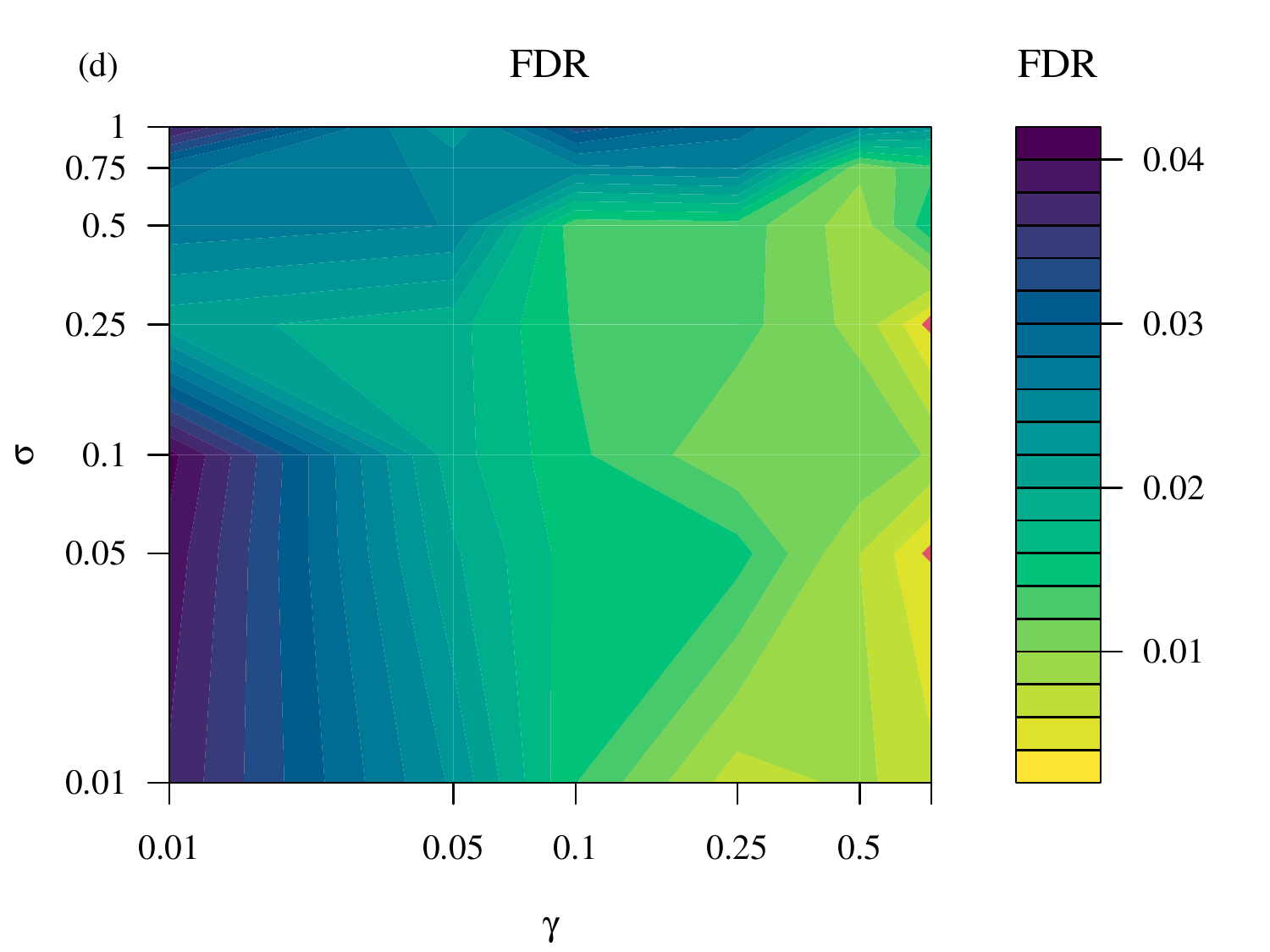}
    \includegraphics[width=.49\textwidth]{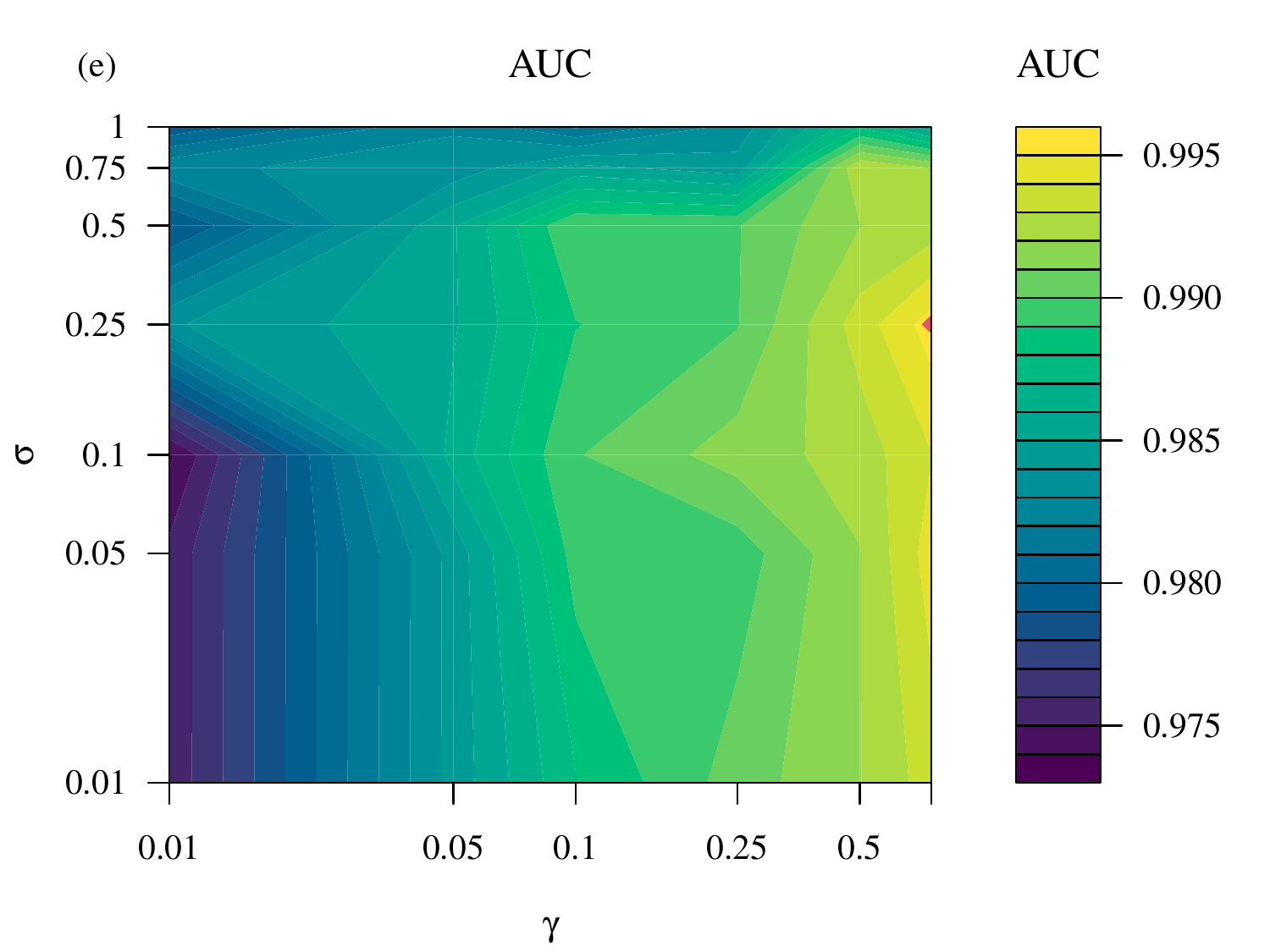}
    \includegraphics[width=.49\textwidth]{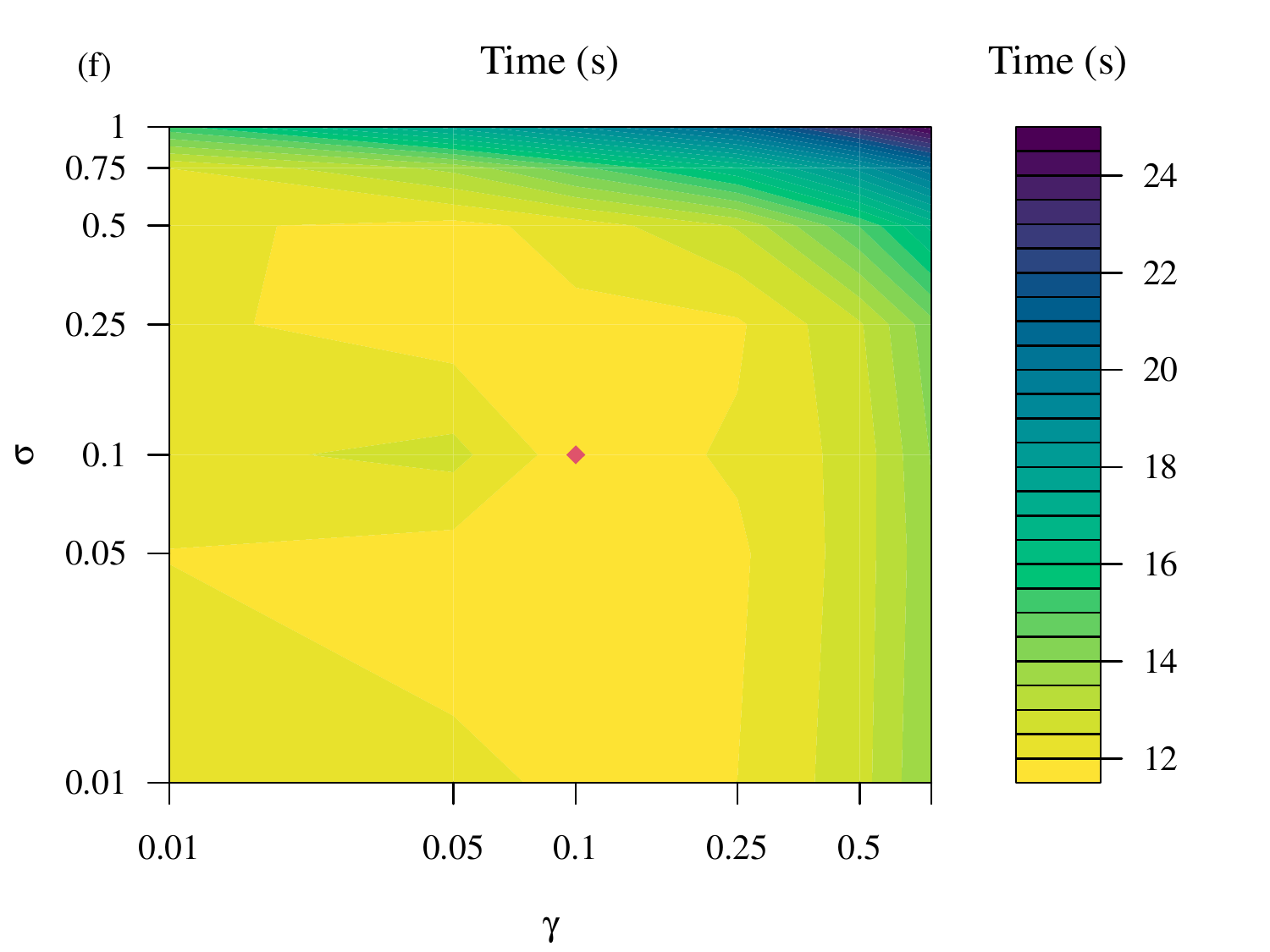}
    \caption{\textbf{Setting 3}: sensitivity with respect to $\sigma$ and $\gamma$}
    \label{fig:sensitivity_s_g_3}
\end{figure}

\begin{figure}[htp]
    \centering
    \includegraphics[width=.49\textwidth]{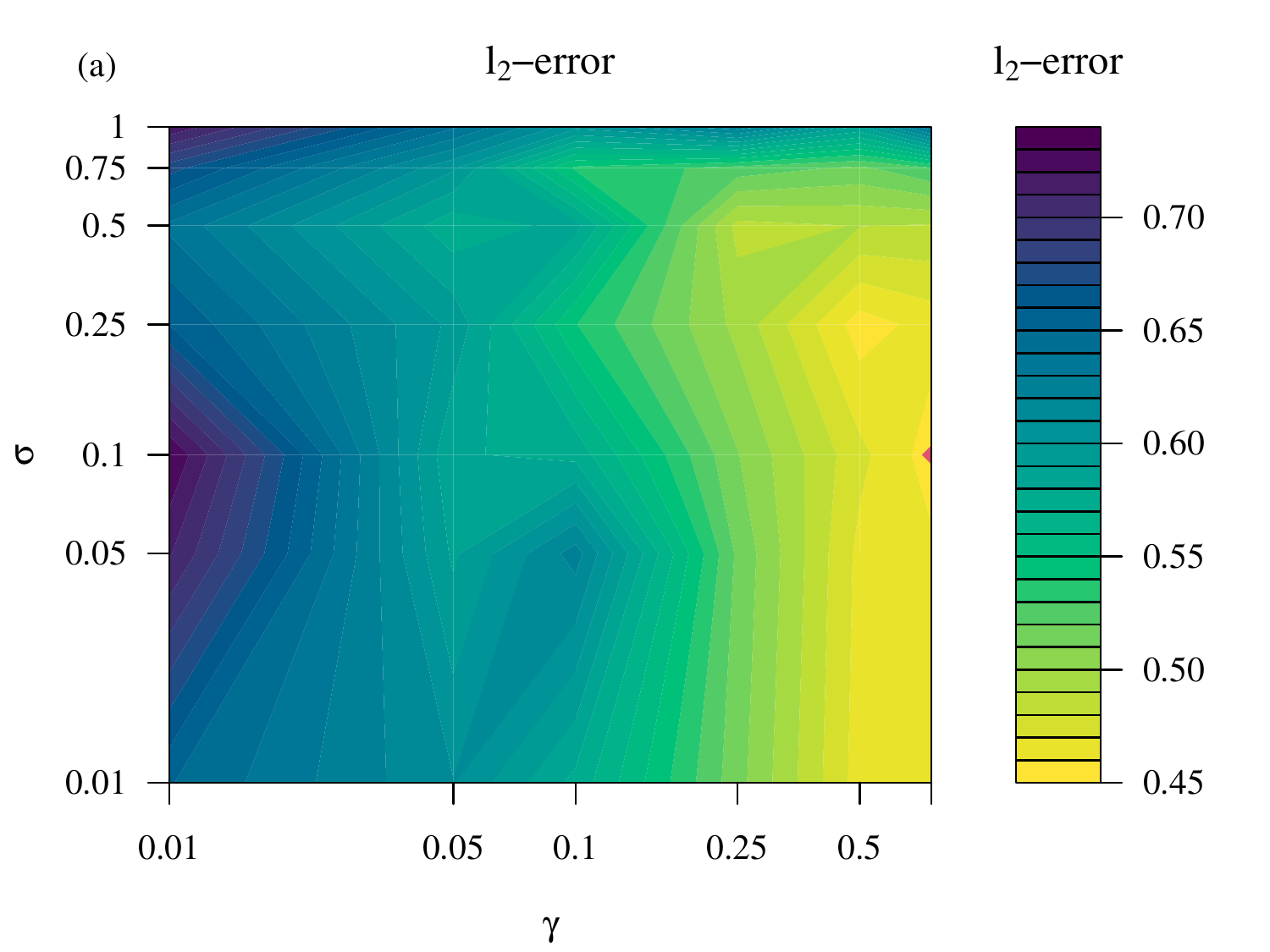}
    \includegraphics[width=.49\textwidth]{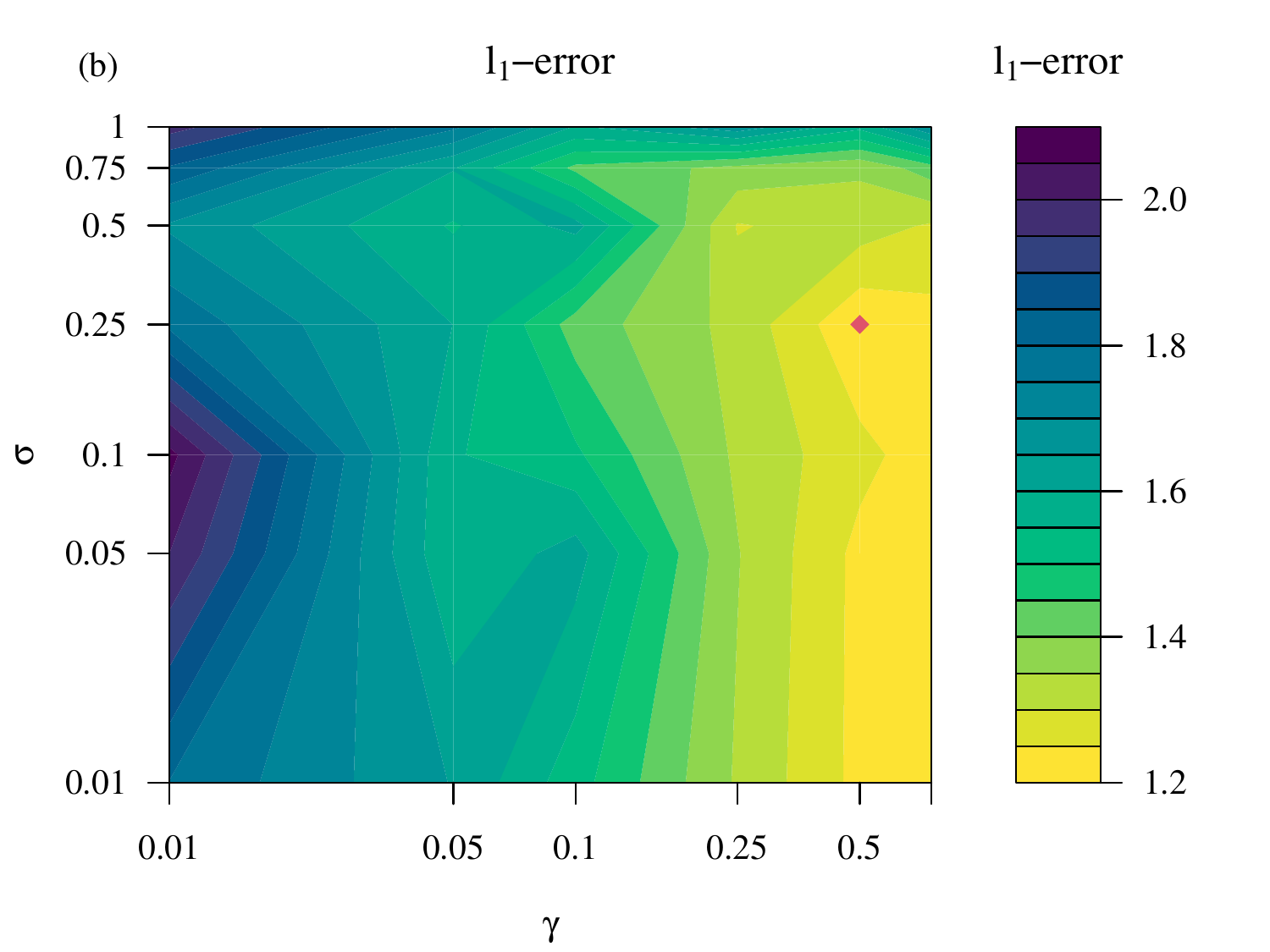}
    \includegraphics[width=.49\textwidth]{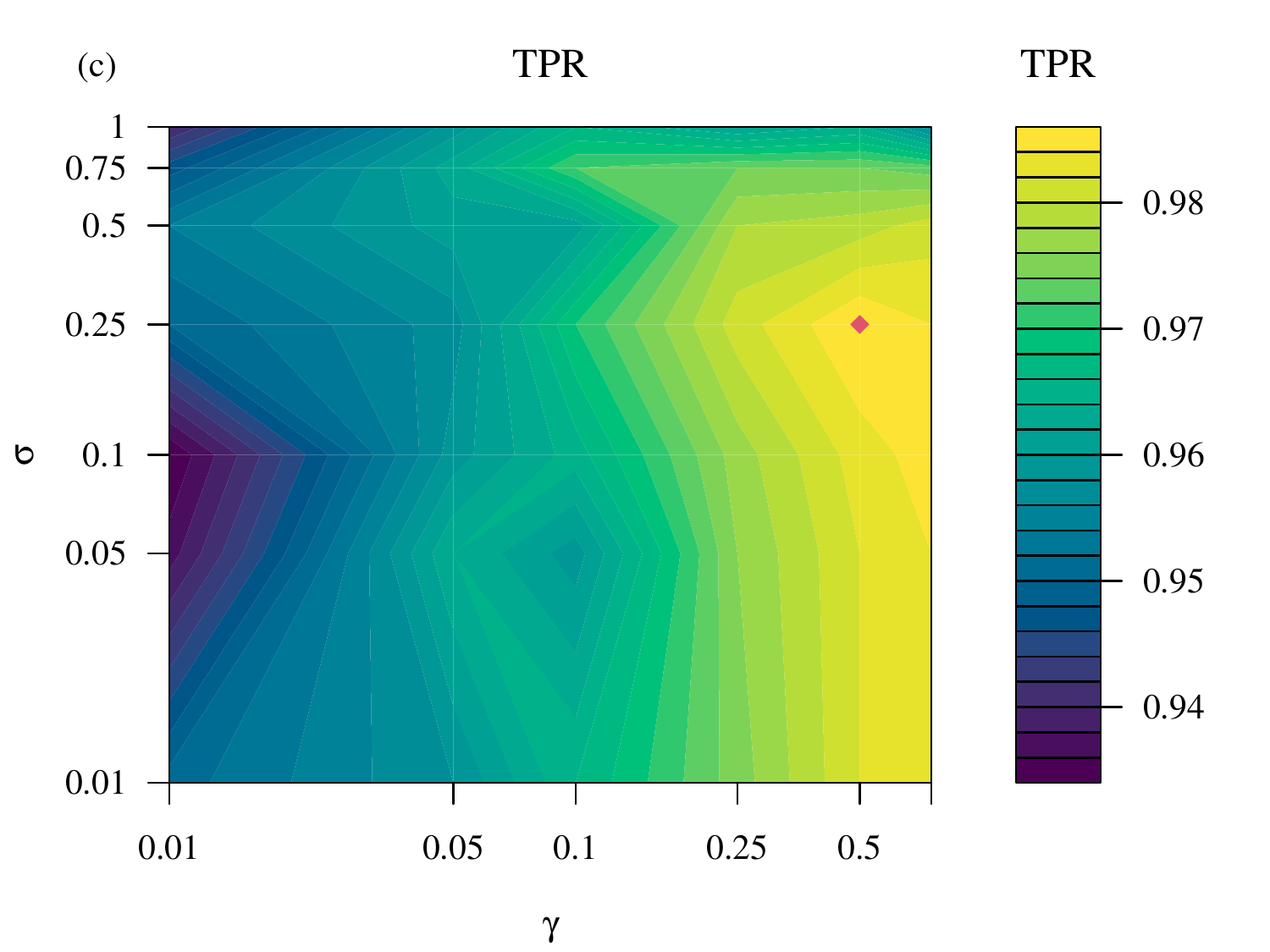}
    \includegraphics[width=.49\textwidth]{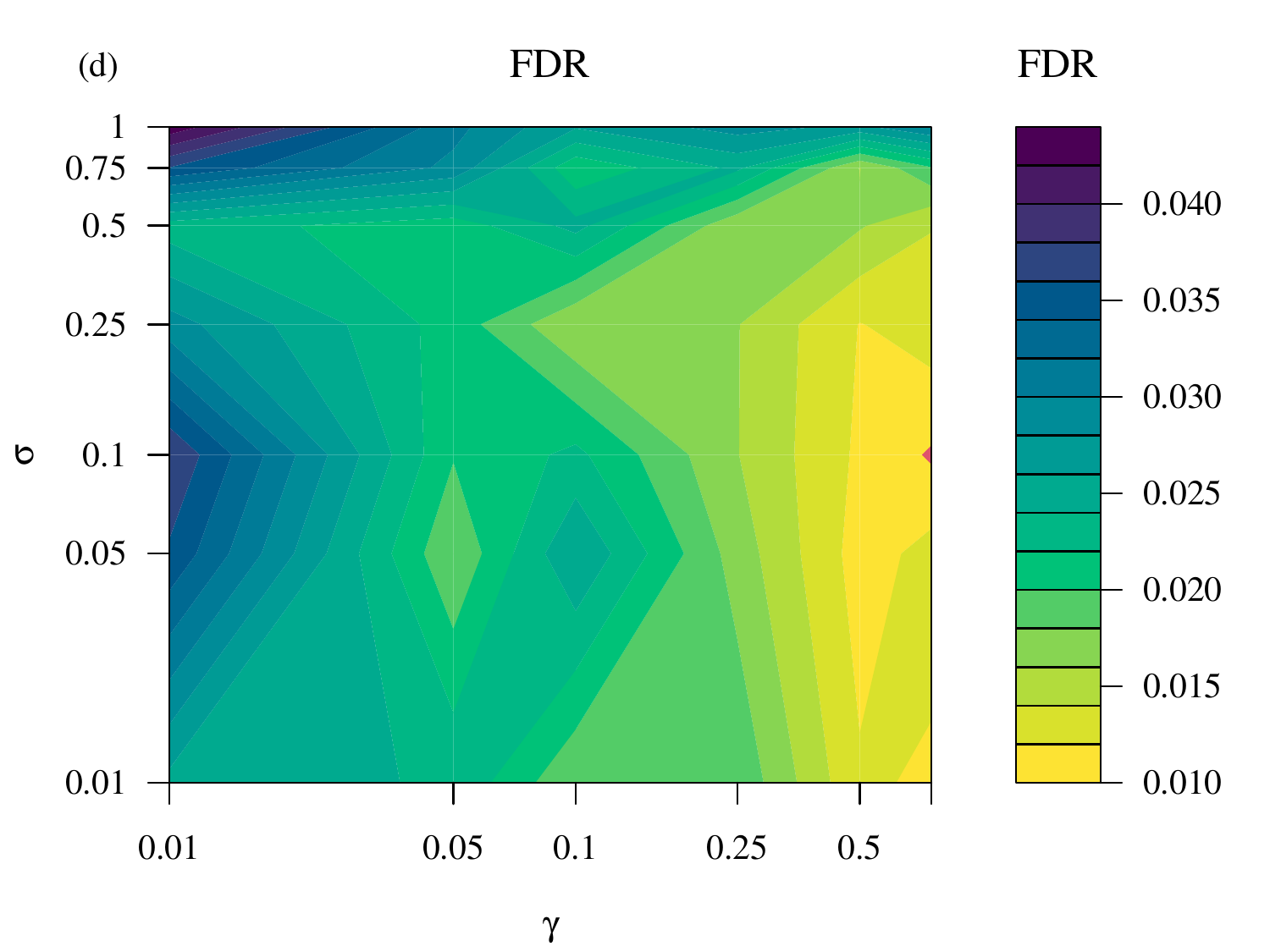}
    \includegraphics[width=.49\textwidth]{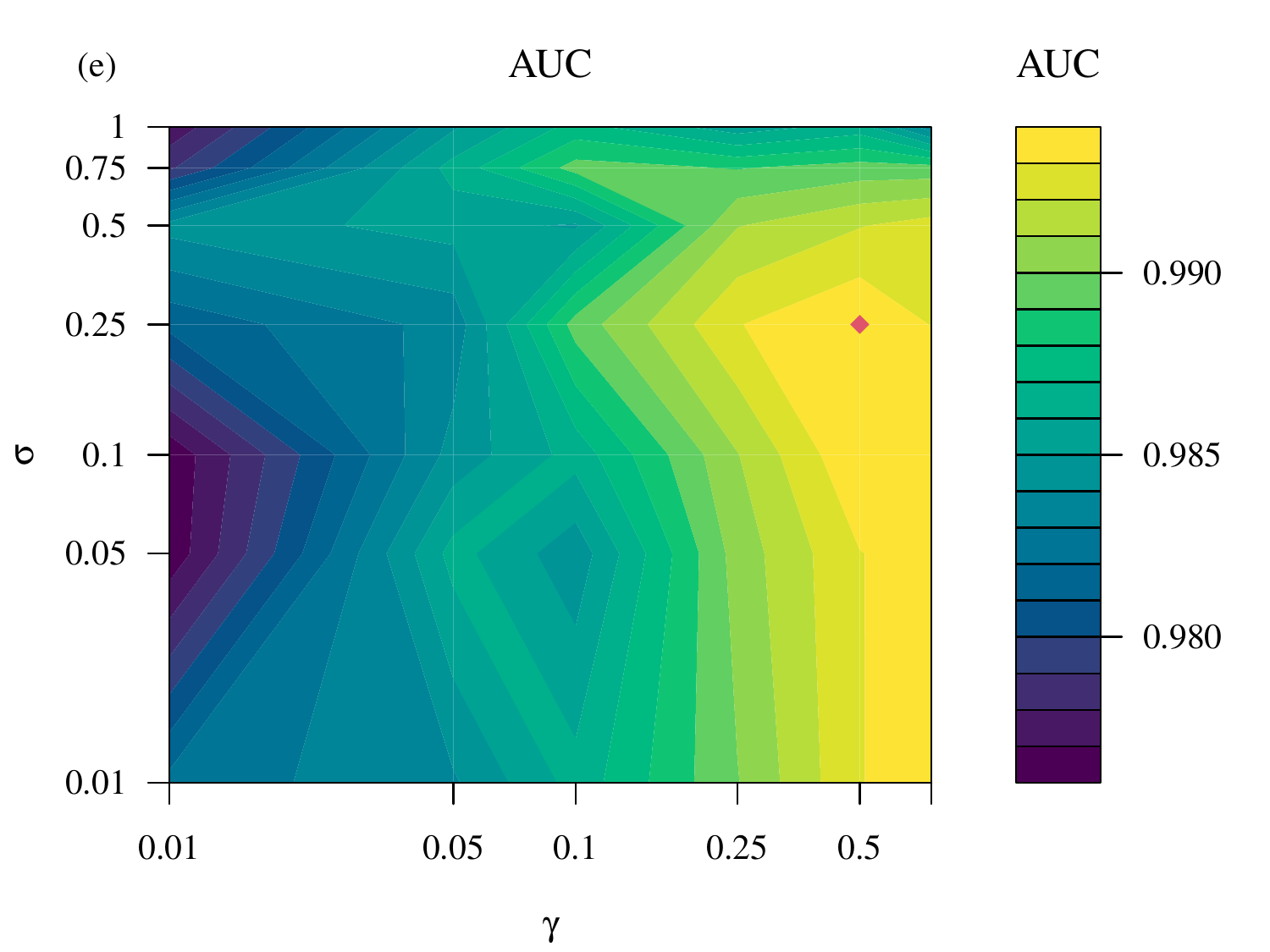}
    \includegraphics[width=.49\textwidth]{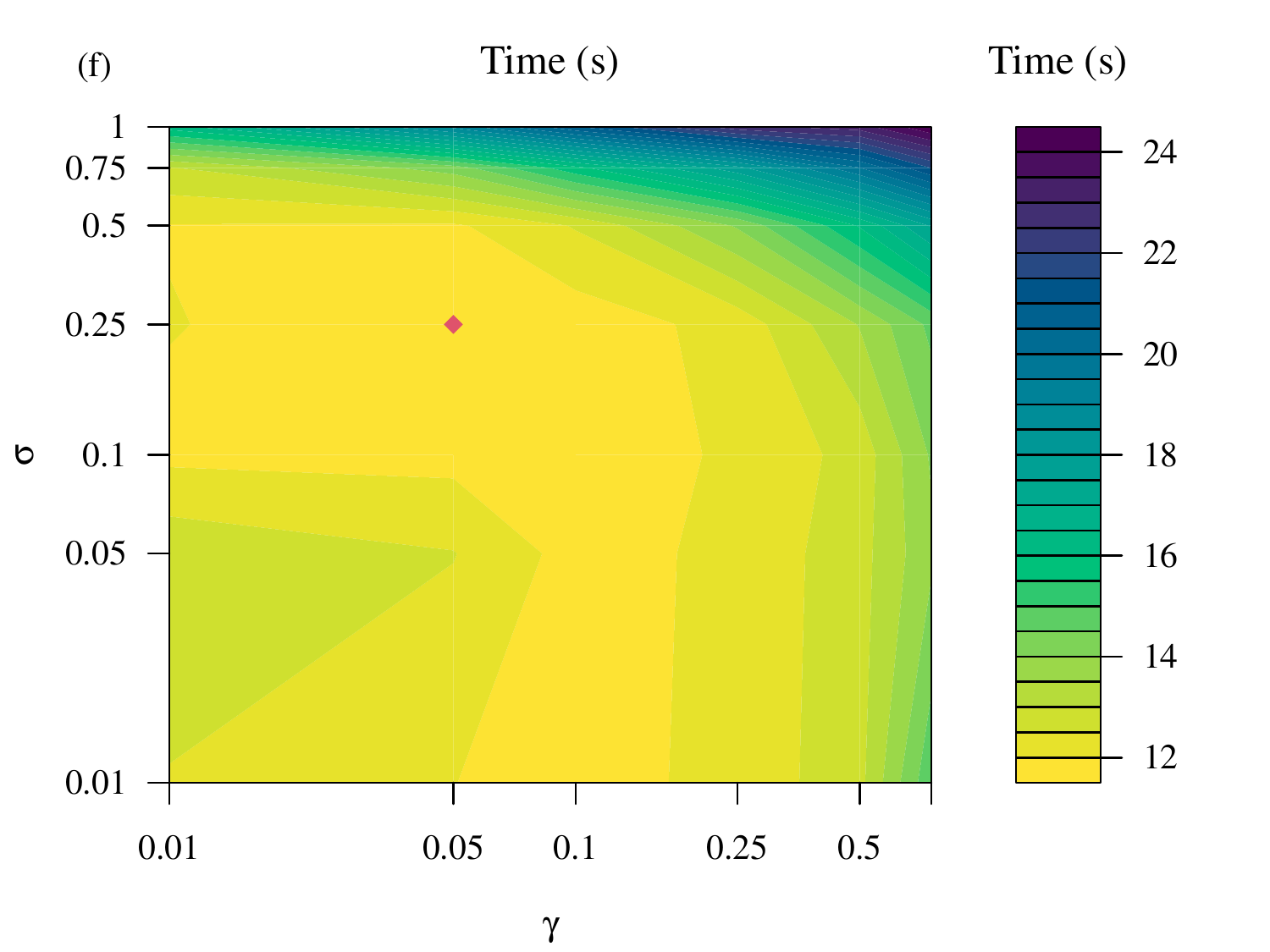}
    \caption{\textbf{Setting 4}: sensitivity with respect to $\sigma$ and $\gamma$}
    \label{fig:sensitivity_s_g_4}
\end{figure}

Within the simplest setting we notice the method is not sensitive to the starting values, obtaining an $\ell_2$ error of $0.41$ and $\ell_1$-error of $1.09$ and the ideal values for the TPR, FDR and AUC across the different values of $\sigma$ and $\gamma$ (\Cref{fig:sensitivity_s_g_1}). More interestingly, the method can be sensitive to the initial values of $\sigma$ and $\gamma$ in more complicated settings (settings 2-4). Specifically, when the $\gamma_j$s are small the performance is worse than when they are larger (Figures \ref{fig:sensitivity_s_g_2} - \ref{fig:sensitivity_s_g_4}), for instance using a value of $\gamma_j=0.01, j=1,\dots,p$, would give a worse performance across all metrics in comparison to a value of $\gamma_j=0.5$. Furthermore, within setting 2, we notice the performance is sensitive to both the values of $\sigma$ and $\gamma$, with the optimal across all metrics except the FDR given by $\sigma_j=0.1$ and $\gamma_j=0.5$ (\Cref{fig:sensitivity_s_g_2}). Finally, for settings 3 and 4 the method is not particularly sensitive to the value of $\sigma$, however can be sensitive to the value of $\gamma$. Therefore choosing a starting value of $\gamma$ of at least $0.5$ is appropriate within these settings. It is worth pointing out, that in some cases overflow issues were encountered when the values of $\sigma_j$ and $\gamma_j$ were too large, for instance when $\gamma_j = 0.75$ and $\sigma_j = 1$.

\newpage

\subsection{Sensitivity to prior parameters}

To evaluate the sensitivity to the prior parameters we ran simulations for Settings 1-4 taking $(n, p, s, c)=(500,5000,30,0.25)$ and fit the VB posterior. To evaluate the performance we compute: (i) $\ell_2$-error, (ii) $\ell_1$-error, (iii) TPR, (iv) FDR, (v) AUC and (vi) runtime (ran on Intel$^\circledR$ Xeon$^\circledR$ E5-2680 v4 2.40GHz CPUs), reporting the mean across 100 replications. To assess the model's goodness of fit we compute the: (i) ELBO, (ii) ELL under the VB posterior, (iii) c-index, and report the mean across the 100 replications in the following figures.

\subsubsection{Sensitivity to $\lambda$}

\begin{figure}[htp]
    \centering
    \includegraphics[width=\textwidth]{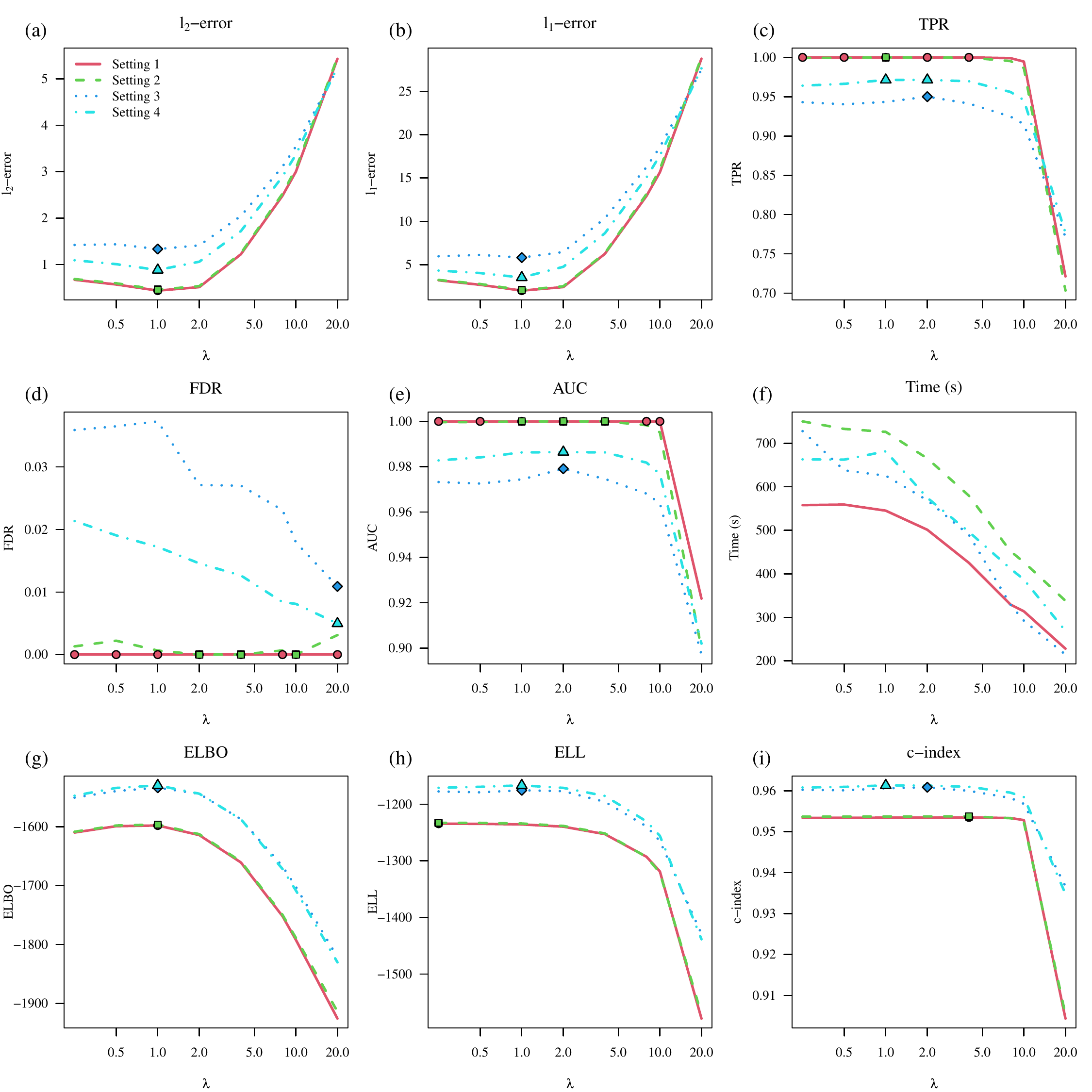}
    \caption{Sensitivity to $\lambda$ for $a_0=1$.}
    \label{fig:sensitivity_lambda}
\end{figure}

We consider a grid of values $\Lambda =\{0.25, 0.5, 1, 2, 4, 8, 10, 20\}$ and fix $a_0=1$. For each $\lambda \in \Lambda$ we fit the VB posterior and compute the above performance metrics and goodness of fit measures, reporting the means across 100 replications in \Cref{fig:sensitivity_lambda}.

Generally, the proposed method is not particularly sensitive to the value of $\lambda$ and performs comparably for values between $0.25$ and $2$, \Cref{fig:sensitivity_lambda} (a)-(e). However, as $\lambda$ increases there is an increase in the $\ell_2$-error, $\ell_1$-error, meaning the resulting point estimate for $\beta_0$ is being overly shrunk, i.e. high prior probability is placed about zero. Further, as $\lambda$ increases above $10$ there is a decrease in TPR and AUC, meaning that fewer features are correctly identified. The decrease in TPR simultaneously leads to an increase in FDR as there are proportionally more incorrectly identified features than correctly identified features. Finally, we note, as $\lambda$ increases there is a decrease in runtime, \Cref{fig:sensitivity_lambda} (f).

Overall, the ELBO seems to be the most appropriate goodness of fit measure, as it is maximized for model parameter $\lambda$ that yields the lowest $\ell_2$-error $\ell_1$-error and largest TPR. Within these settings the ELL and c-index do not identify the model with the best performance metric and in turn should be used cautiously.

\subsubsection{Sensitivity to $a_0$}

\begin{figure}[htp]
    \centering
    \includegraphics[width=\textwidth]{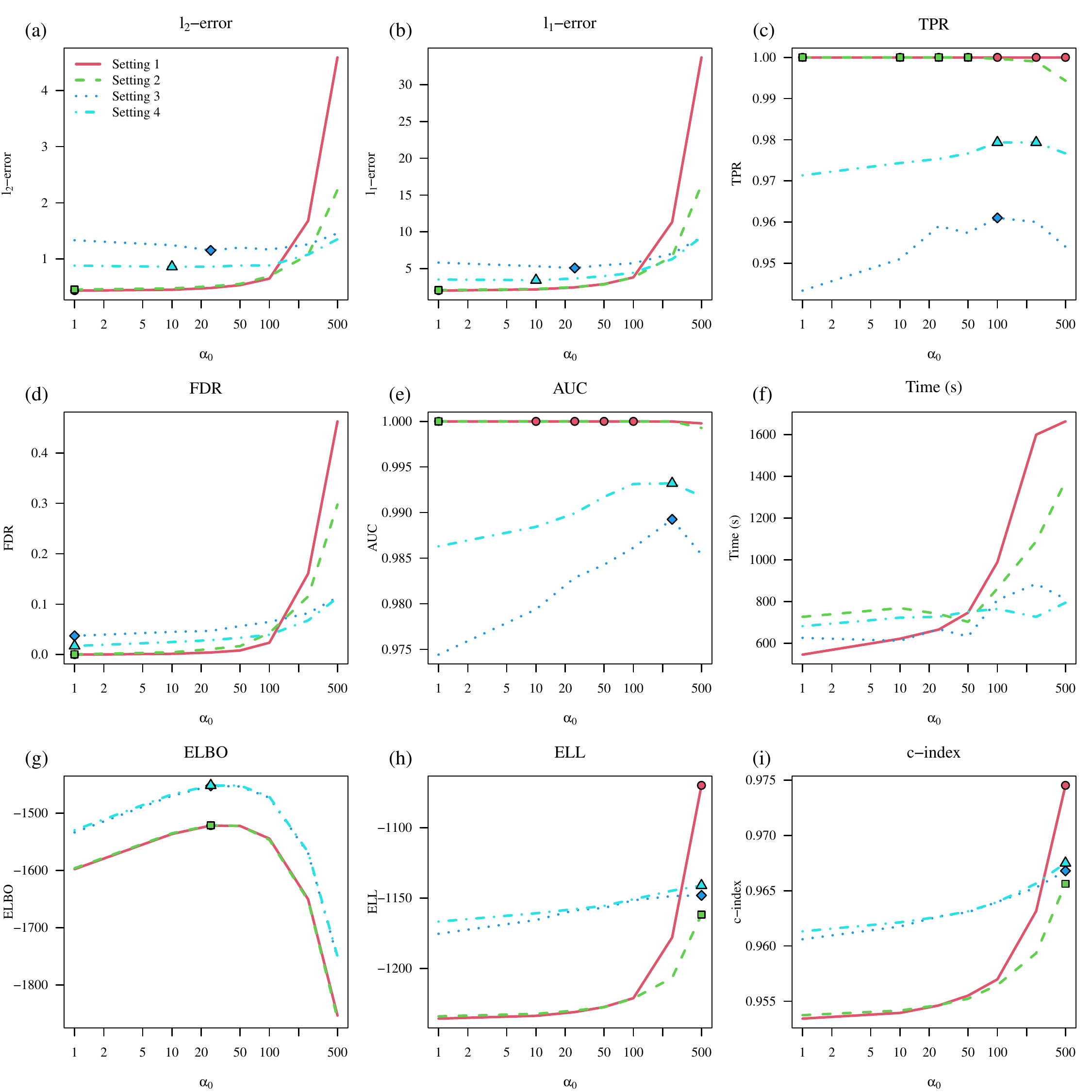}
    \caption{Sensitivity to $a_0$ for $\lambda=1$}
    \label{fig:sensitivity_alpha}
\end{figure}

To examine the sensitivity with respect to $a_0$, we fix $\lambda=1$ and examine the grid of values of $A_0 = \{1, 2, 5, 10, 20, 50, 100, 500 \}$. For each $a_0 \in A_0$ we compute the VB posterior and report the respective performance metrics and goodness of fit measures (averaged across 100 replications) in \Cref{fig:sensitivity_alpha}.

Generally, the proposed method is not particularly sensitive to the value of $a_0$, performing equivalently for values between $1$ and $100$, Figures \ref{fig:sensitivity_alpha} (a)-(e). However, for large values of $a_0$ the performance decreases, arising as a large value of $a_0$ corresponds to us believing \textit{a priori} that many coefficients are non-zero, meaning in the resulting models many coefficients will be non-zero. This behaviour is observed in our results through the increased $\ell_1$-error, $\ell_2$-error and FDR. Finally, as $a_0$ increases the runtime increases \Cref{fig:sensitivity_alpha} (f).

Regarding goodness of fit measures, the ELBO is maximized for models with correspondingly good performance metrics. Further, the ELL and c-index increase as $a_0$ increases, meaning models with many non-zero parameters are favoured. These performance metrics should therefore be used cautiously when tuning $a_0$.

\subsubsection{Sensitivity to $\lambda$ and $a_0$}

Finally, we consider the sensitivity for both parameters. To do so, we fit the VB posterior for $(\lambda, a_0) \in \Lambda \times A_0$, where $\Lambda$ and $A_0$ are the same sets defined earlier. The results for the respective performance metrics and goodness of fit measures are presented in Figures \ref{fig:sensitivity_alpha_lambda_1} - \ref{fig:sensitivity_alpha_lambda_4} for settings 1 to 4 respectively.

Generally our method is not sensitive to the hyperparameter values in the simpler settings (1 and 2). For instance, examining Figures \ref{fig:sensitivity_alpha_lambda_1} and \ref{fig:sensitivity_alpha_lambda_2} (a) - (e), we notice that the optimal value of the TPR, FDR and AUC is attained for multiple combination of $a_0$ and $\lambda$. Regarding the more complicated settings (3 and 4), our method can be sensitive to $\lambda$ and $a_0$, for instance, in setting 3, poor selection of $a_0$ and $\lambda$ can have an impact on the FDR (\Cref{fig:sensitivity_alpha_lambda_3} (d)), in addition, metrics are not consistently optimal for a single combination of $\lambda$ and $a_0$. For example, in setting 3 the optimal value of $\ell_2$-error is obtained when $(\lambda, a_0)=(2, 100)$ whereas the optimal AUC is obtained when $(\lambda, a_0)=(4, 100)$, meaning in practice a trade-off may need to be made between better point estimates and variable selection.

As before, the ELBO seems to the most appropriate goodness of fit measure for most settings. However, in highly correlated settings, where there is no single ``best'' parameter values, the optimal ELBO corresponds the model with smallest $\ell_1$-error [Figures \ref{fig:sensitivity_alpha_lambda_1} (g) and \ref{fig:sensitivity_alpha_lambda_4} (g)]. Meaning, other measures are more appropriate if the practitioner wishes to optimise other performance metrics, e.g. the TPR. Finally, the ELL and c-index should be used cautiously as these favour models with many non-zero coefficients and should therefore be used for tuning $\lambda$ when $a_0$ is fixed.

\sidewaysfigure{\begin{figure}[htp]
    \centering
    \includegraphics[width=.41\textwidth]{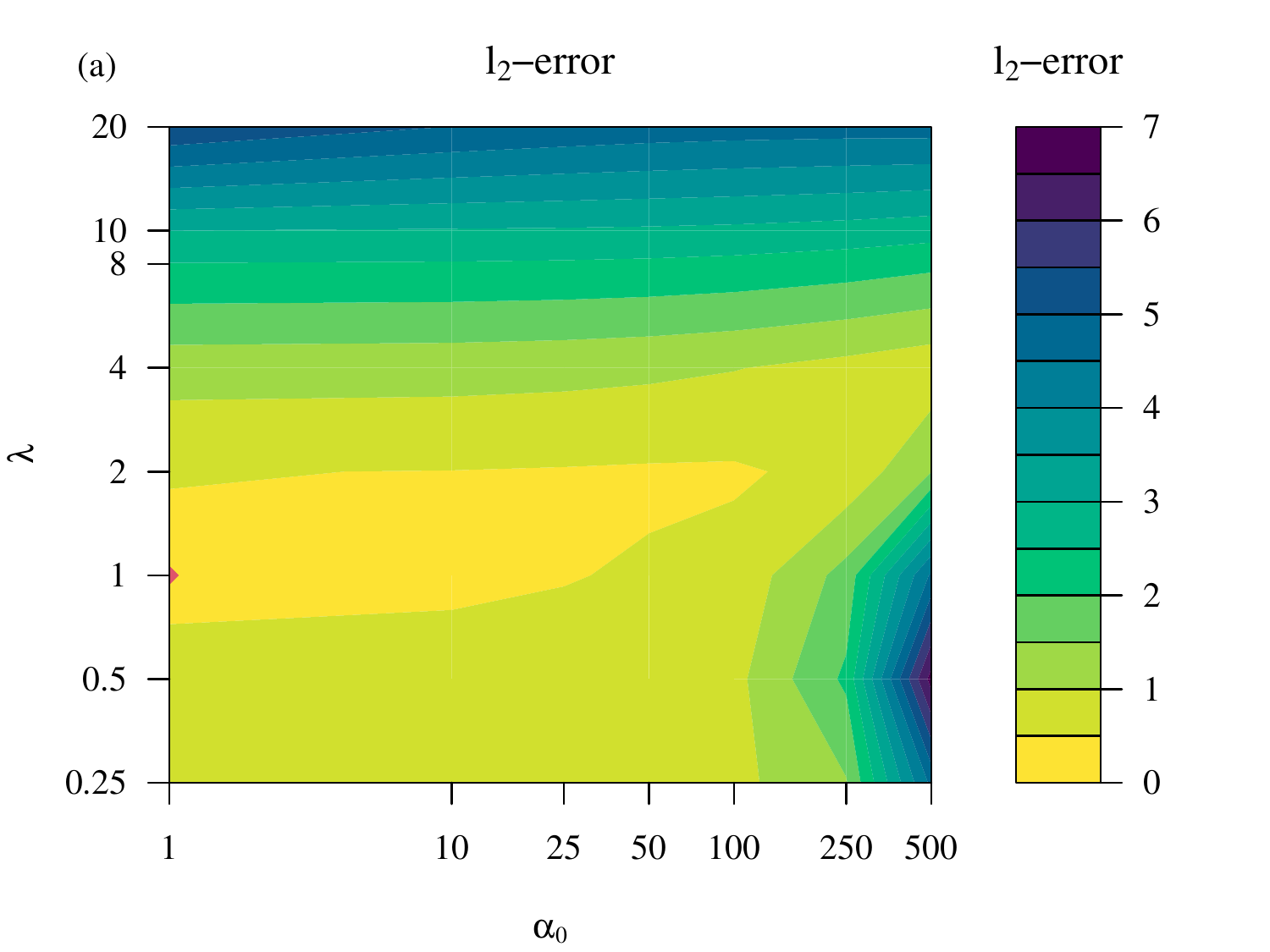}
    \includegraphics[width=.41\textwidth]{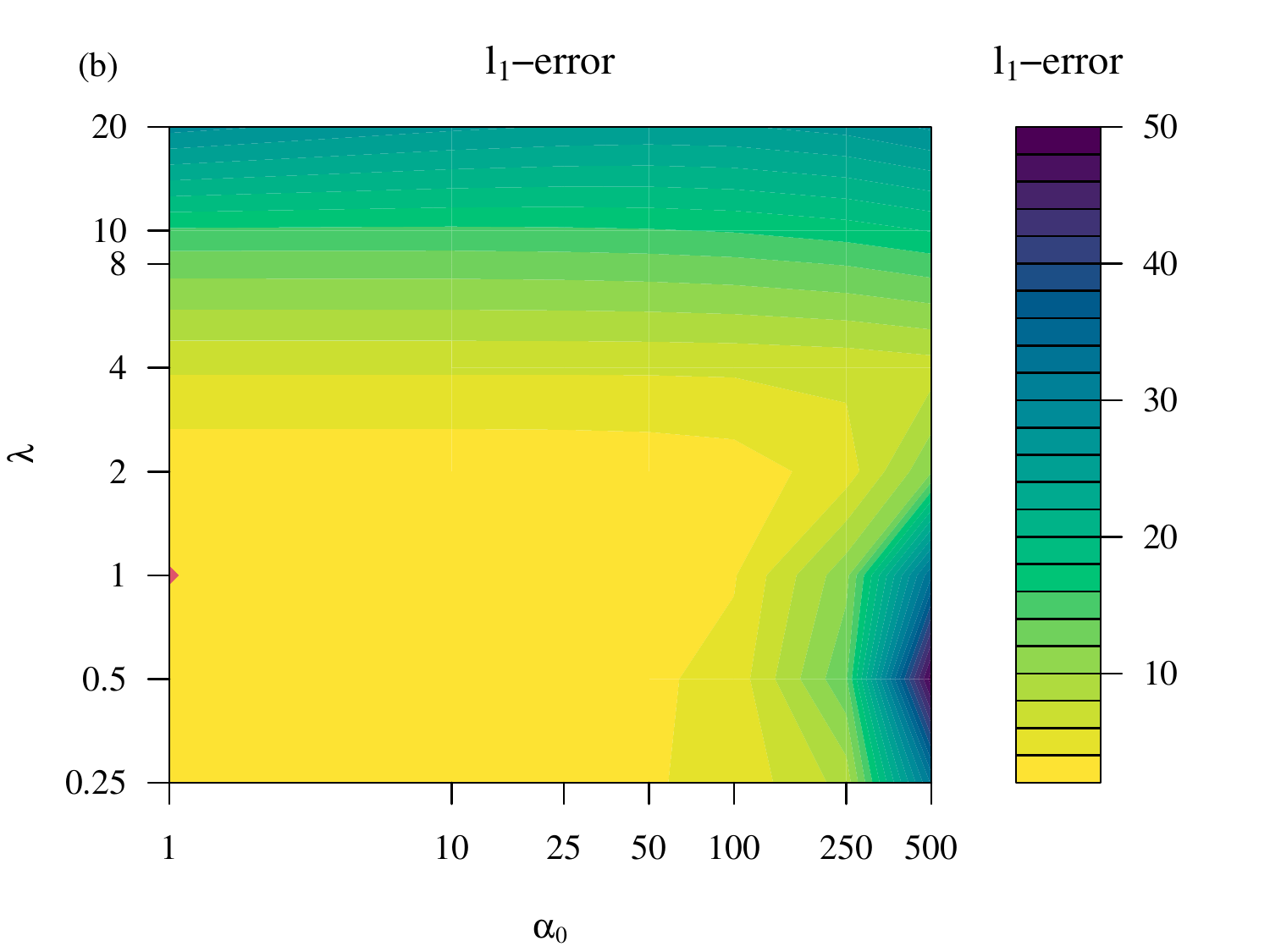}
    \includegraphics[width=.41\textwidth]{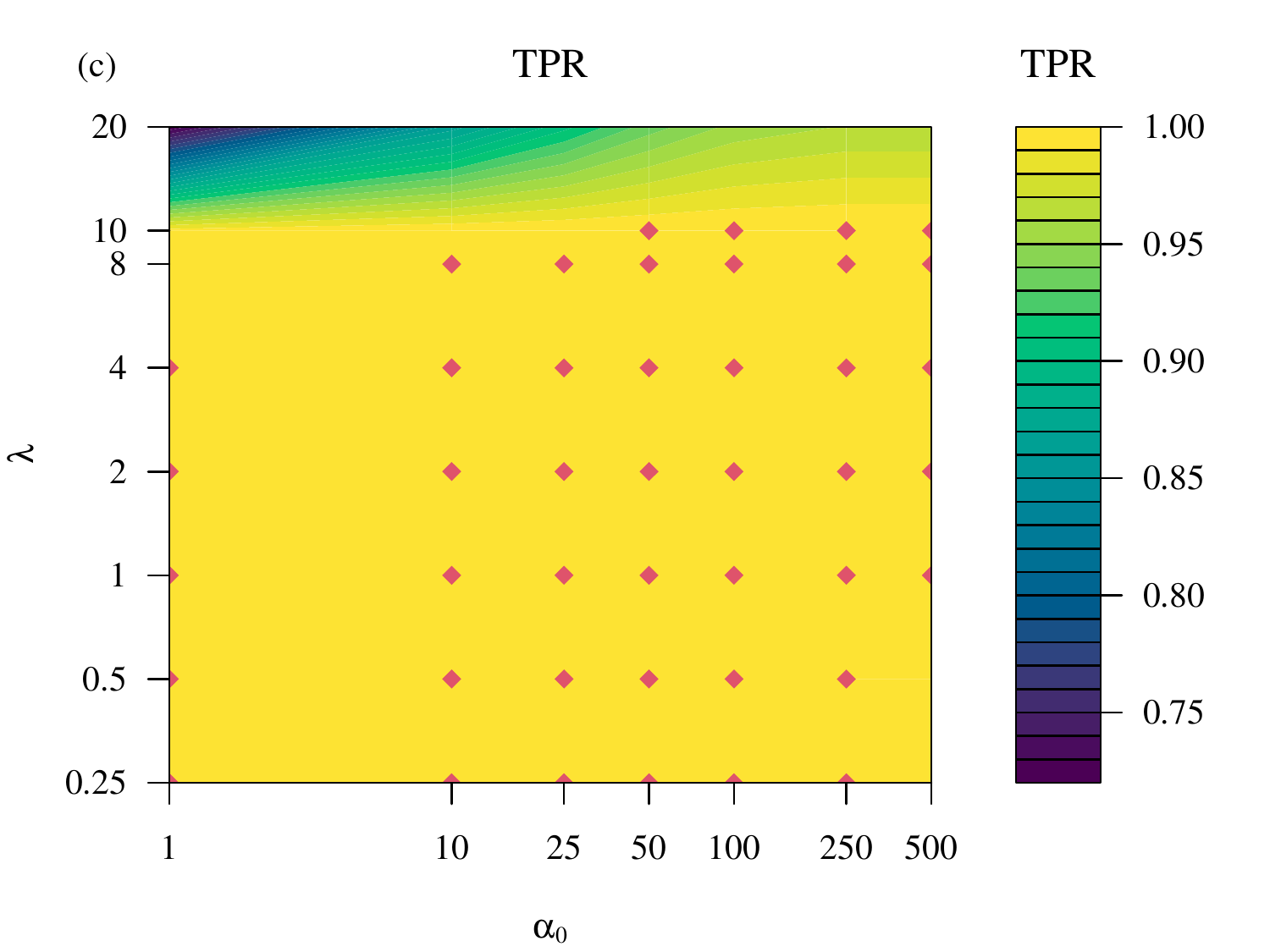}
    \includegraphics[width=.41\textwidth]{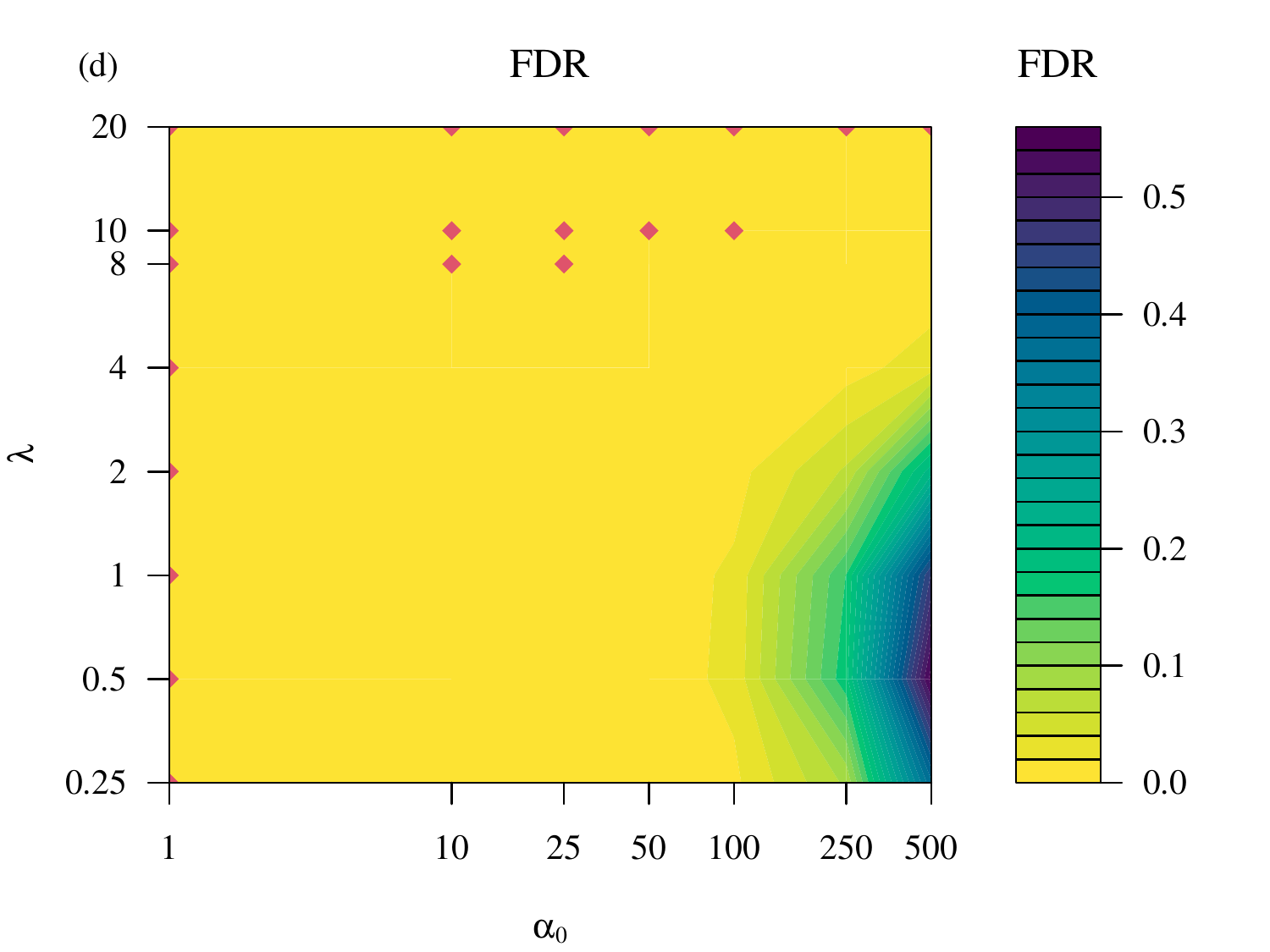}
    \includegraphics[width=.41\textwidth]{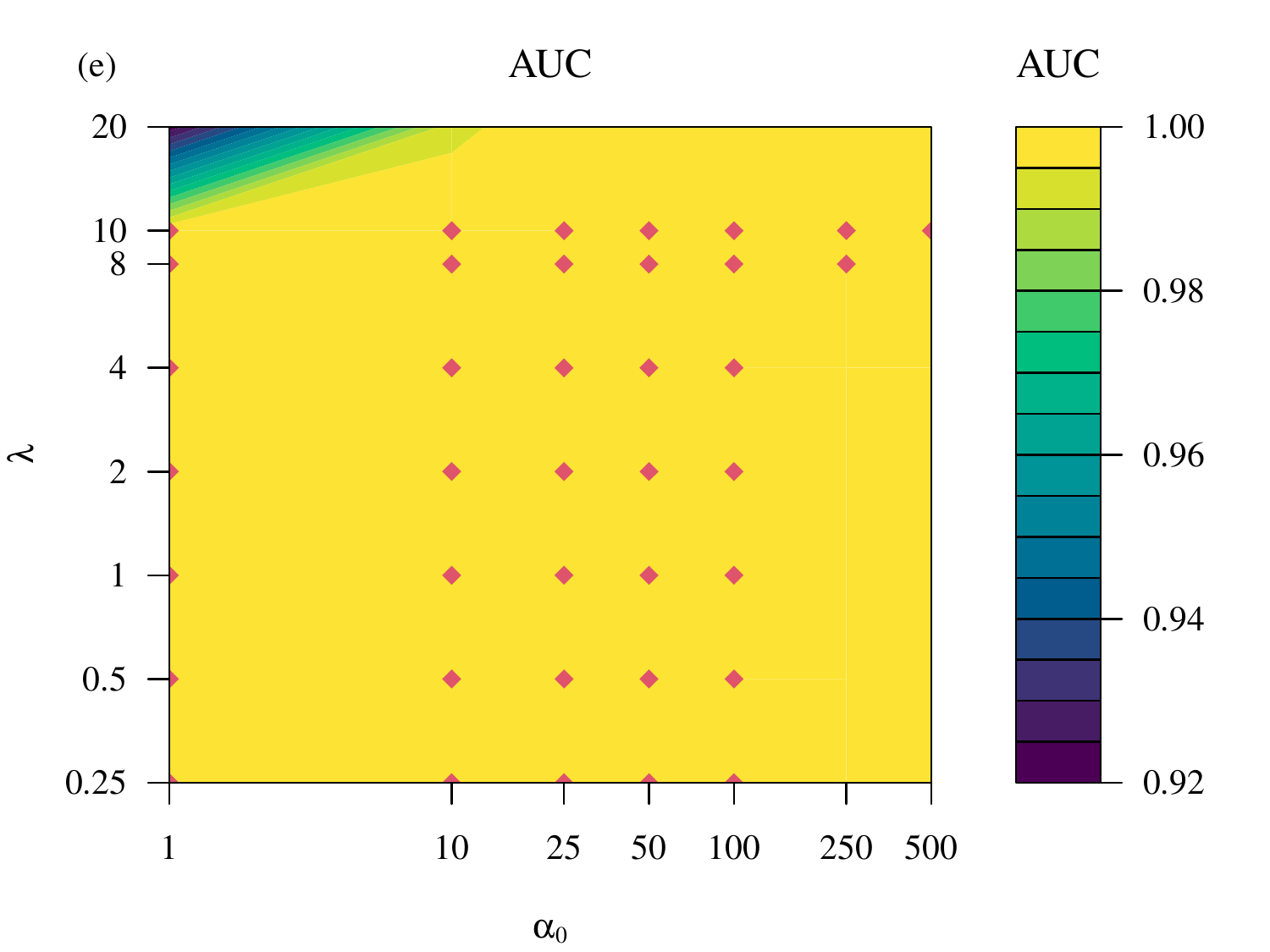}
    \includegraphics[width=.41\textwidth]{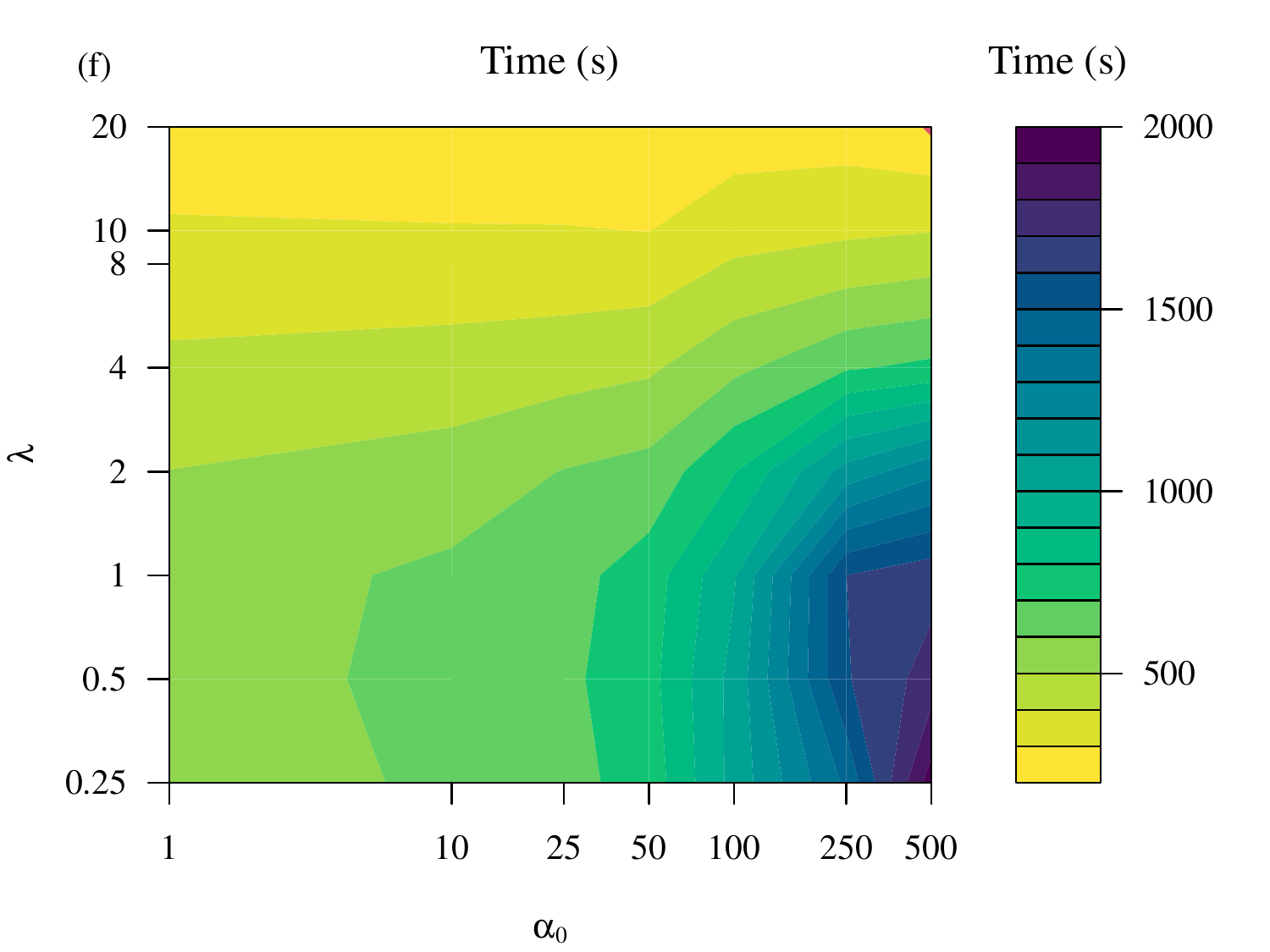}
    \includegraphics[width=.41\textwidth]{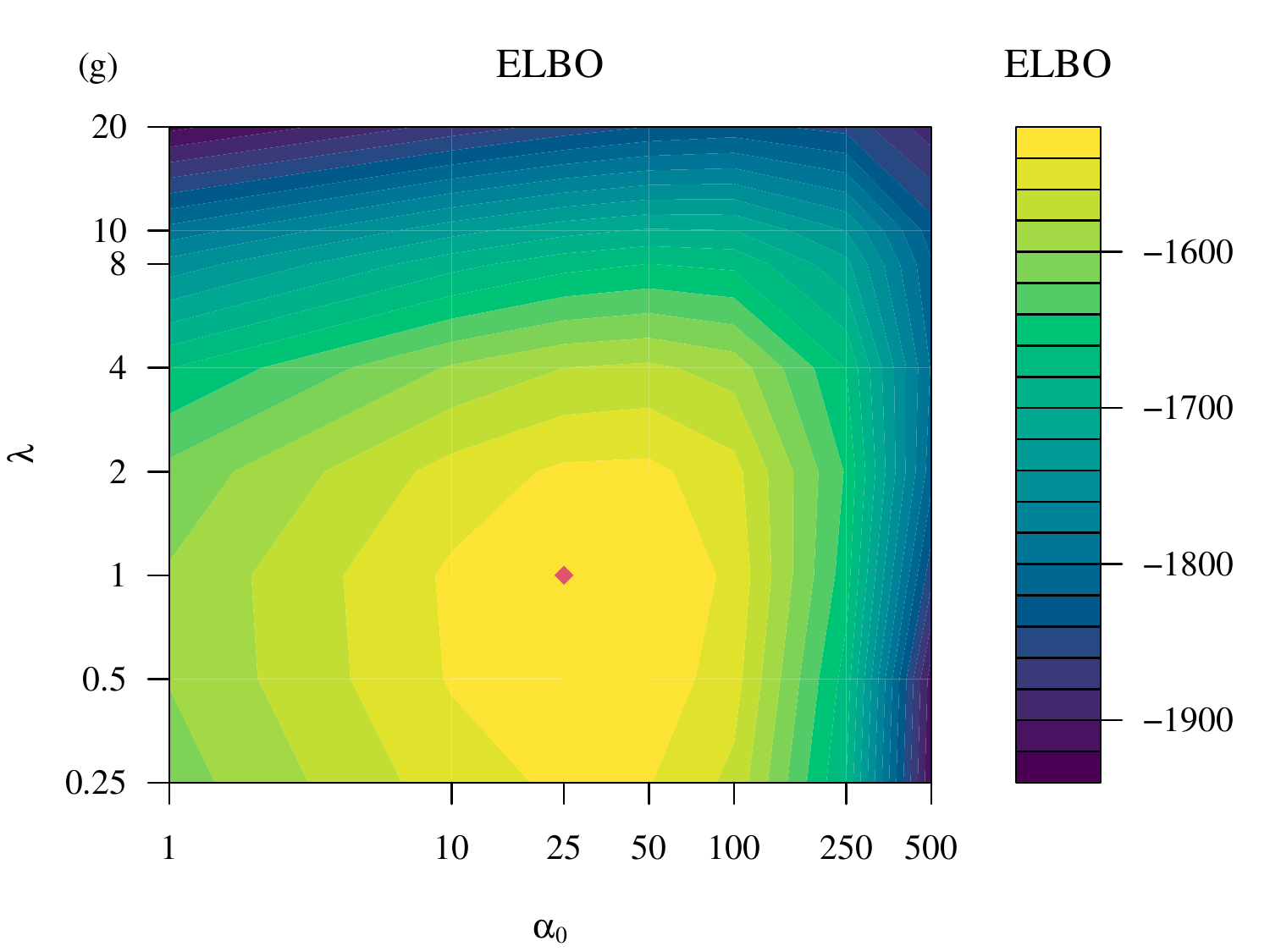}
    \includegraphics[width=.41\textwidth]{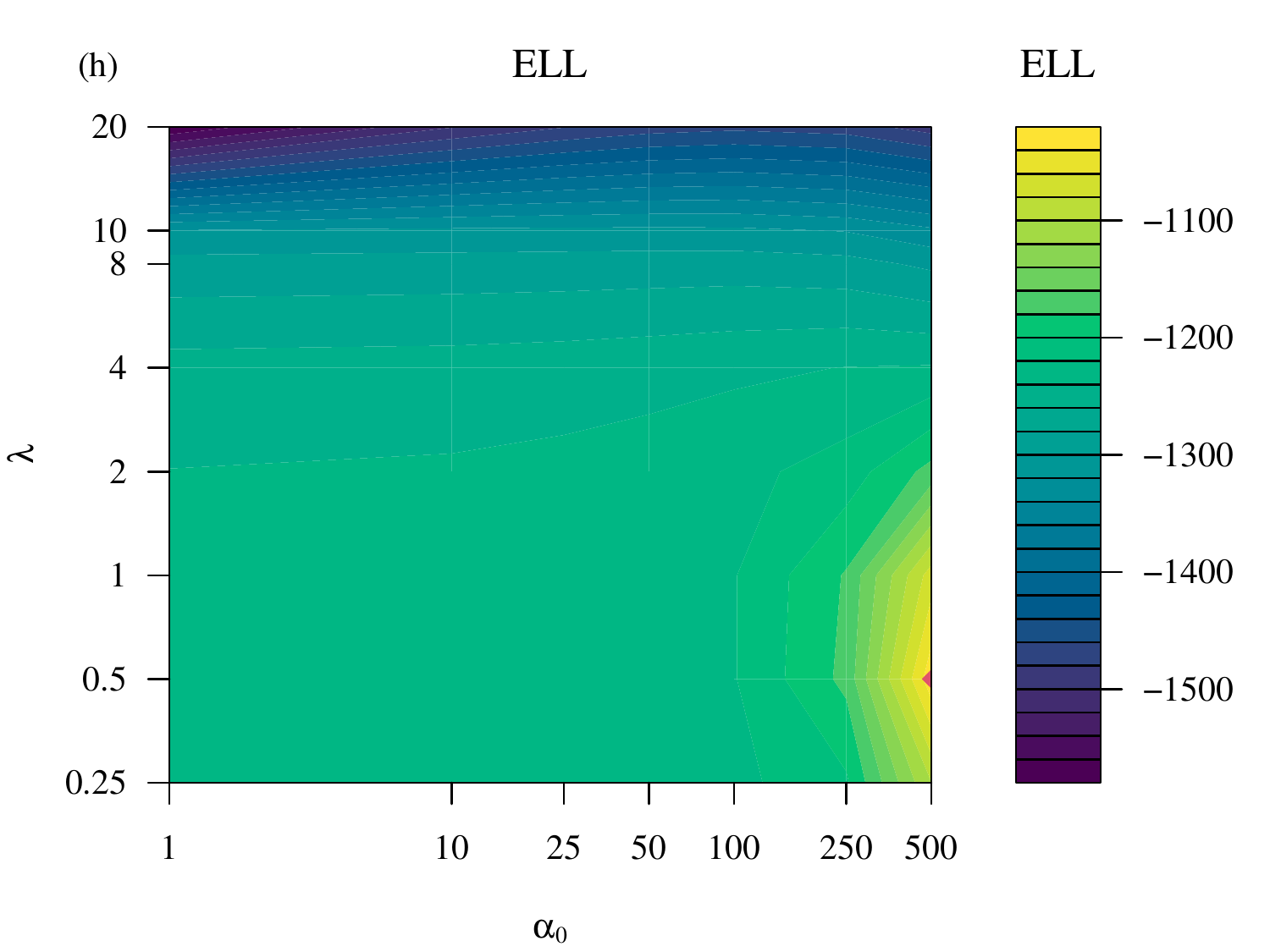}
    \includegraphics[width=.41\textwidth]{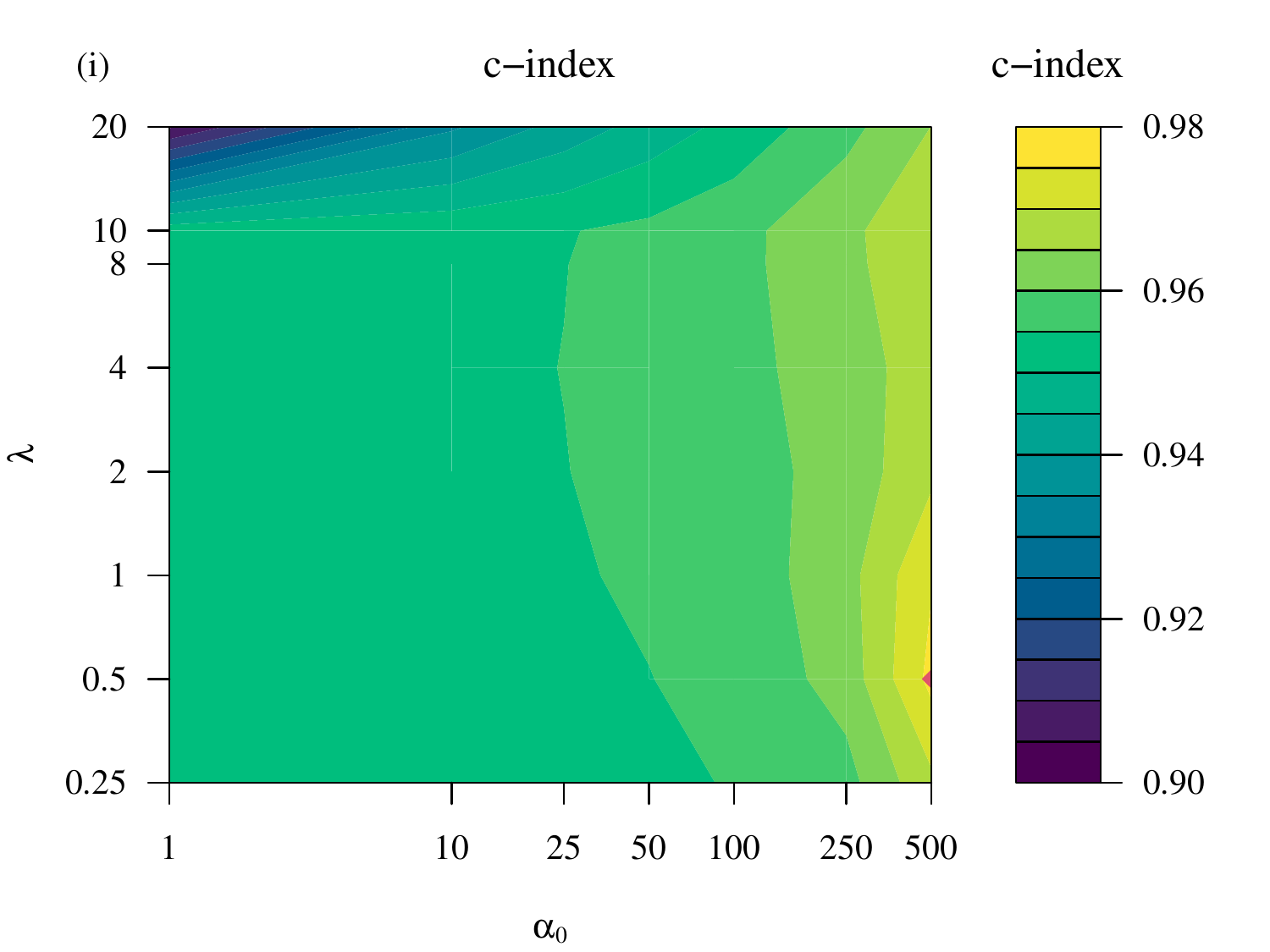}
    \caption{\textbf{Setting 1}: sensitivity with respect to $\lambda$ and $a_0$.}
    \label{fig:sensitivity_alpha_lambda_1}
\end{figure}}

\sidewaysfigure{\begin{figure}[htp]
    \centering
    \includegraphics[width=.41\textwidth]{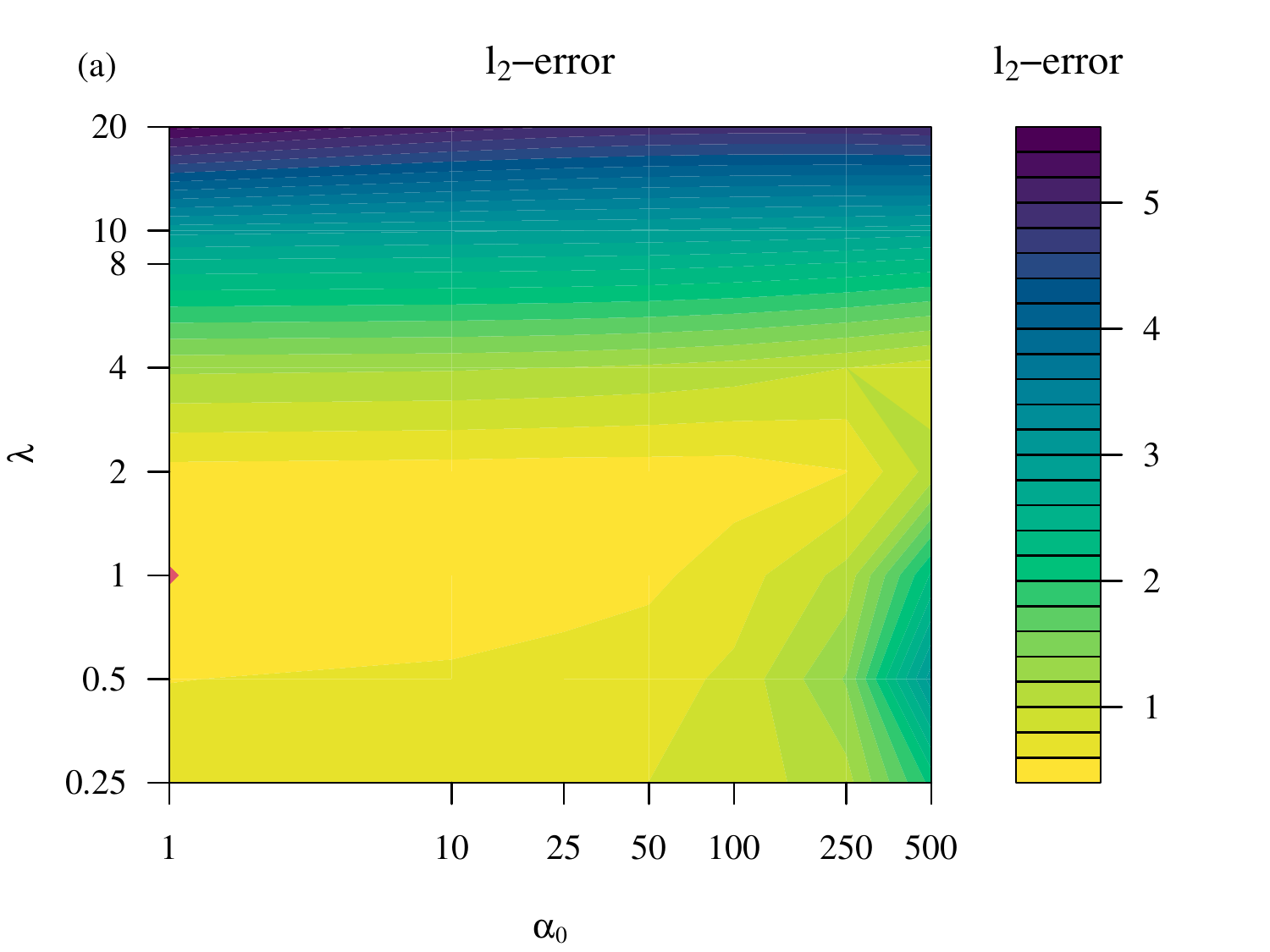}
    \includegraphics[width=.41\textwidth]{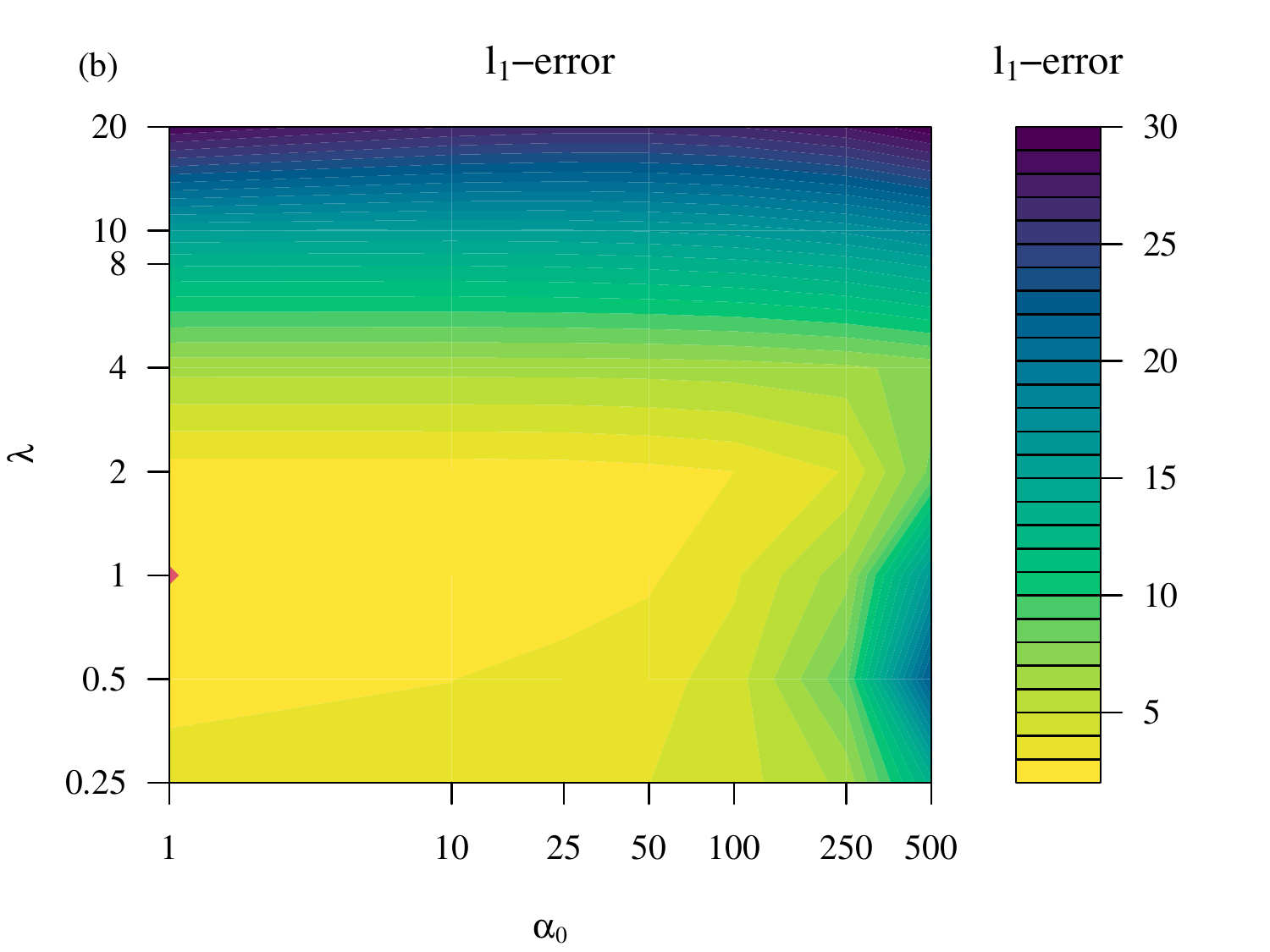}
    \includegraphics[width=.41\textwidth]{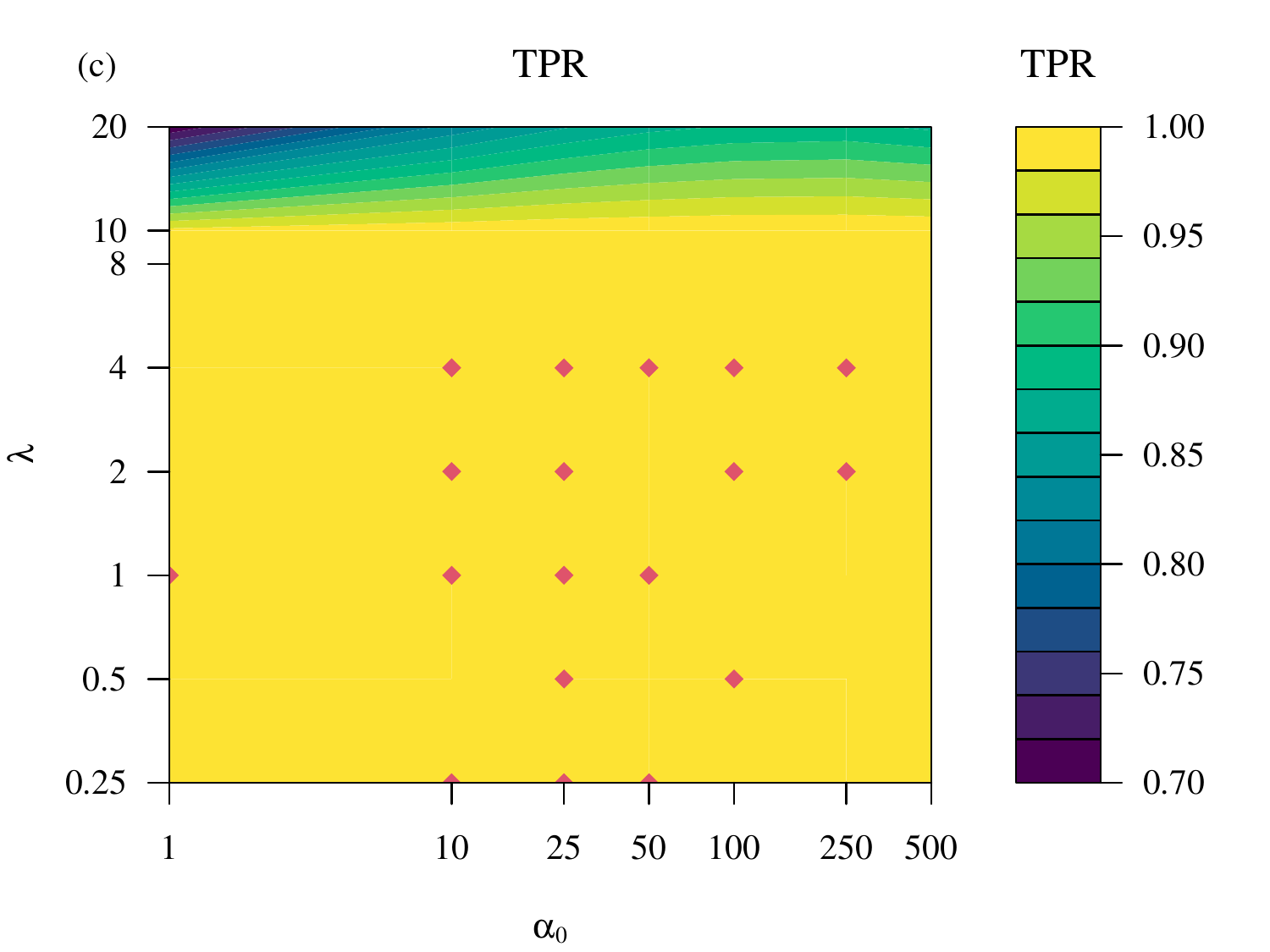}
    \includegraphics[width=.41\textwidth]{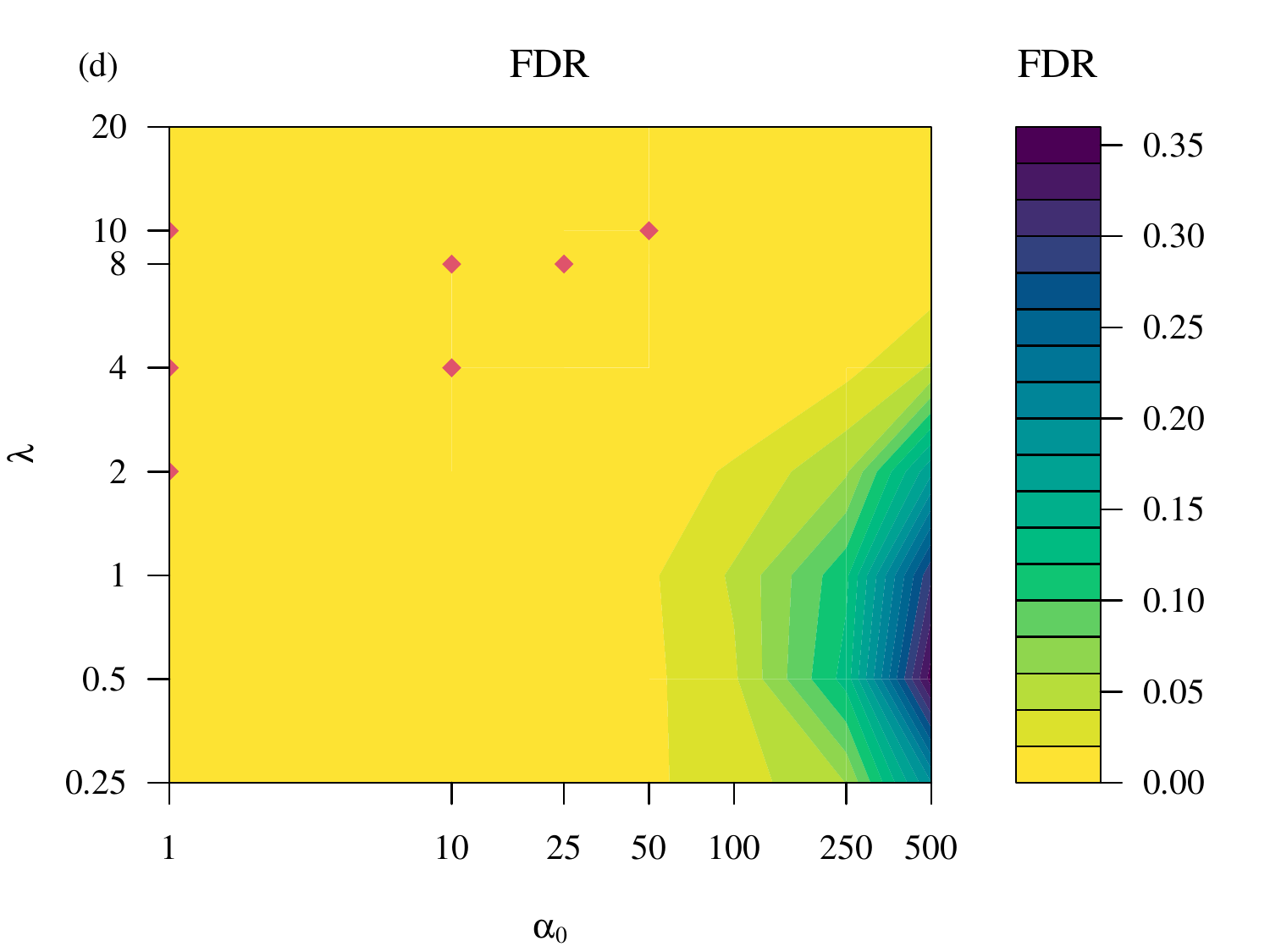}
    \includegraphics[width=.41\textwidth]{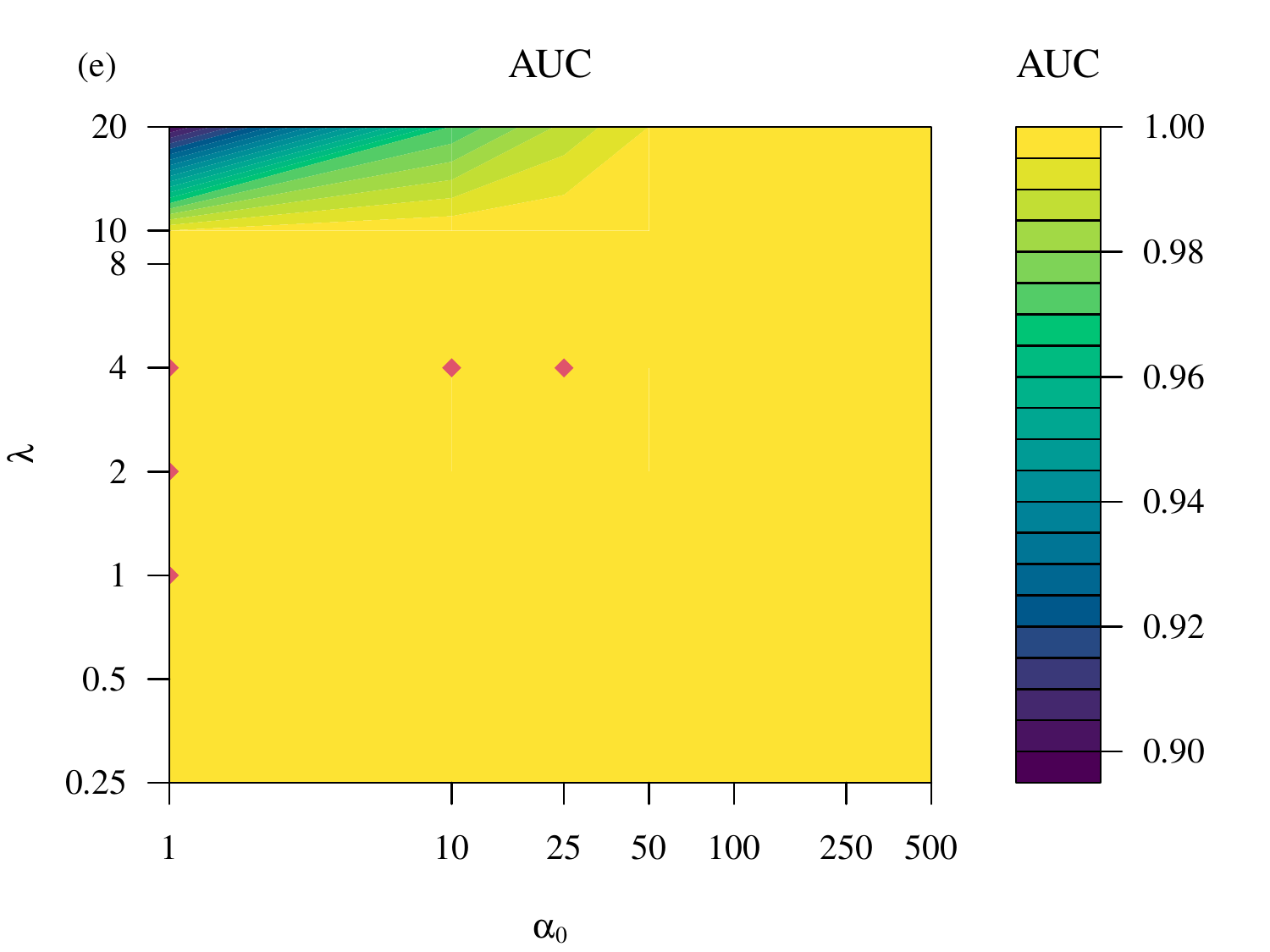}
    \includegraphics[width=.41\textwidth]{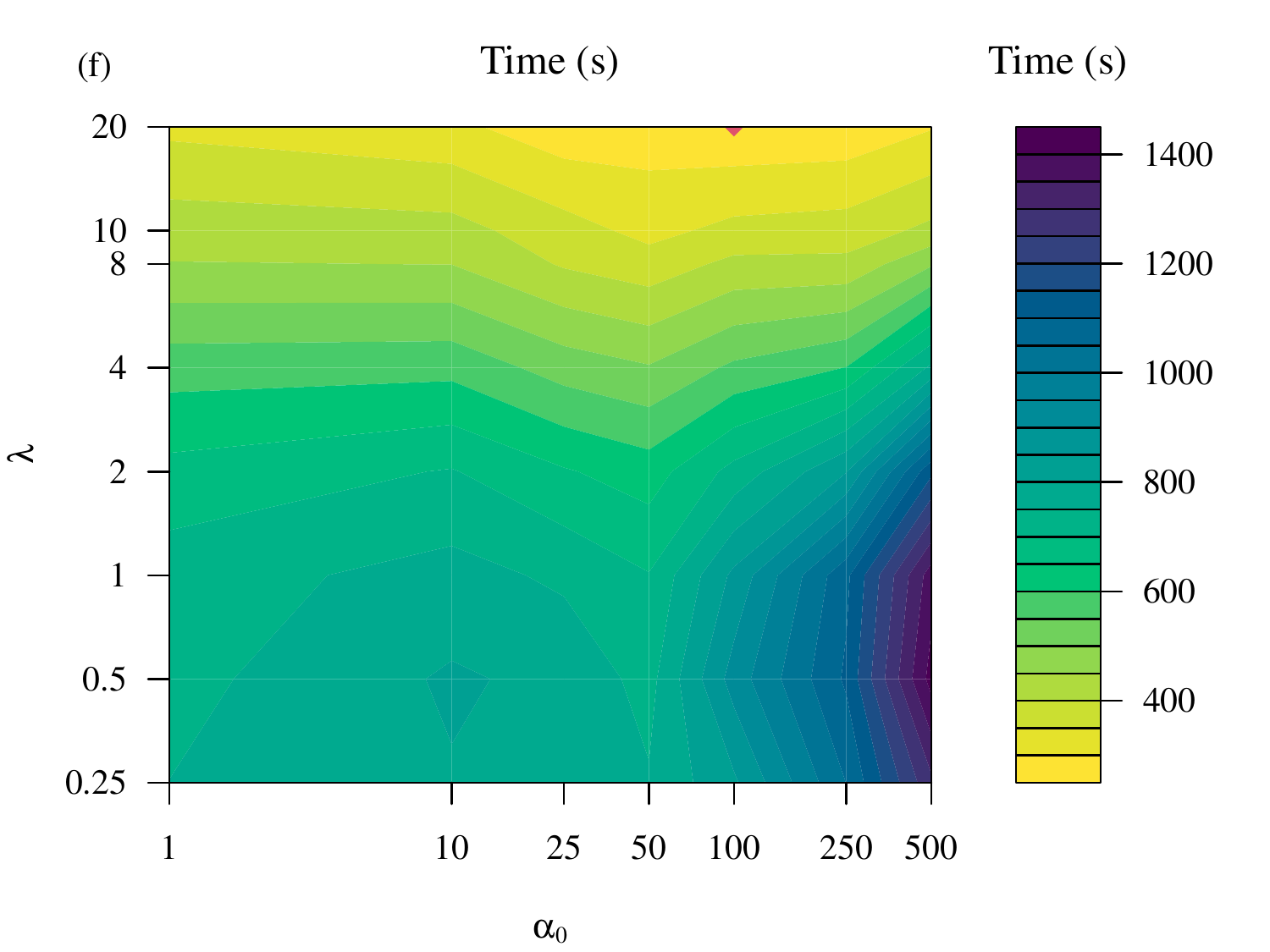}
    \includegraphics[width=.41\textwidth]{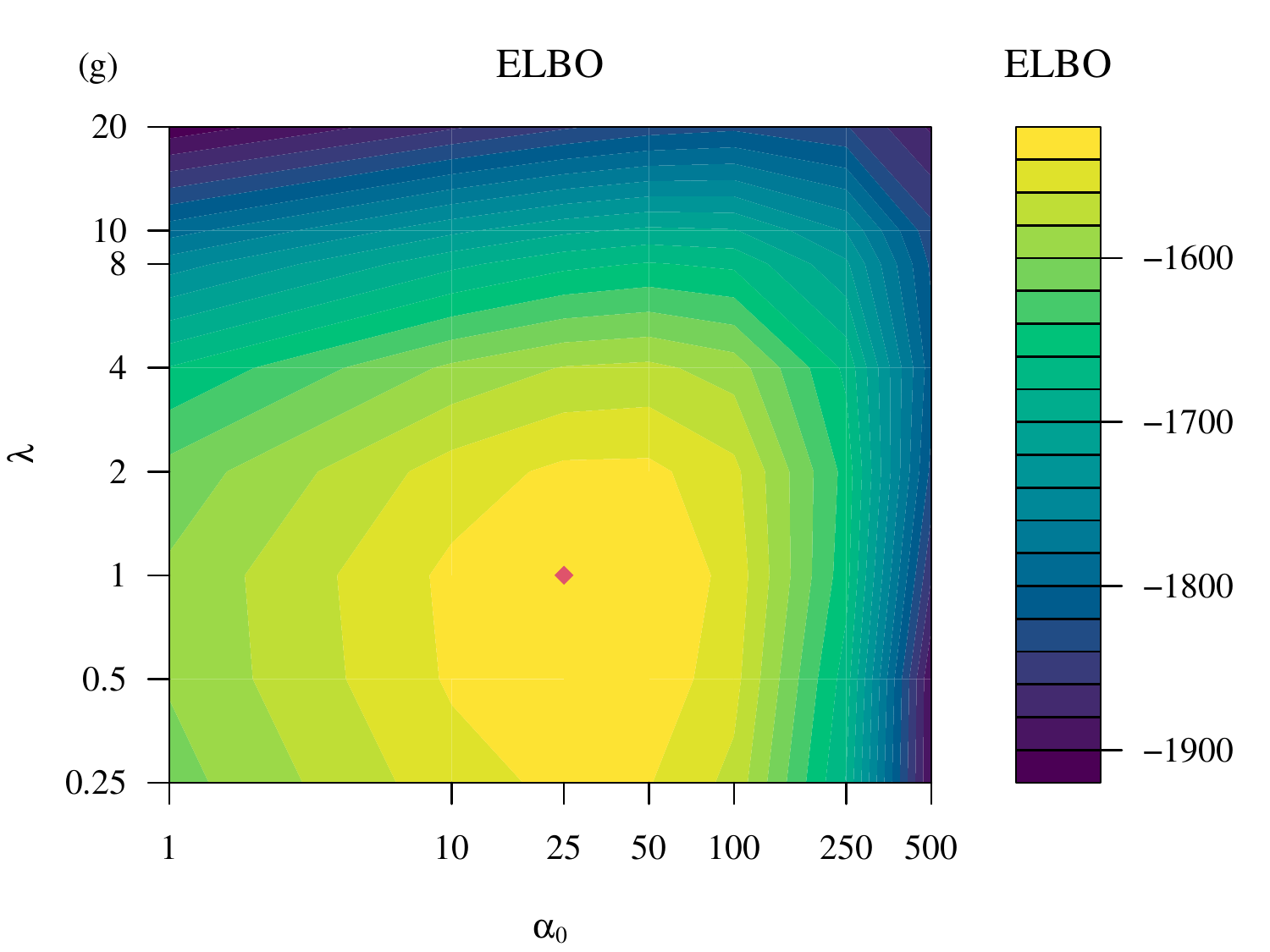}
    \includegraphics[width=.41\textwidth]{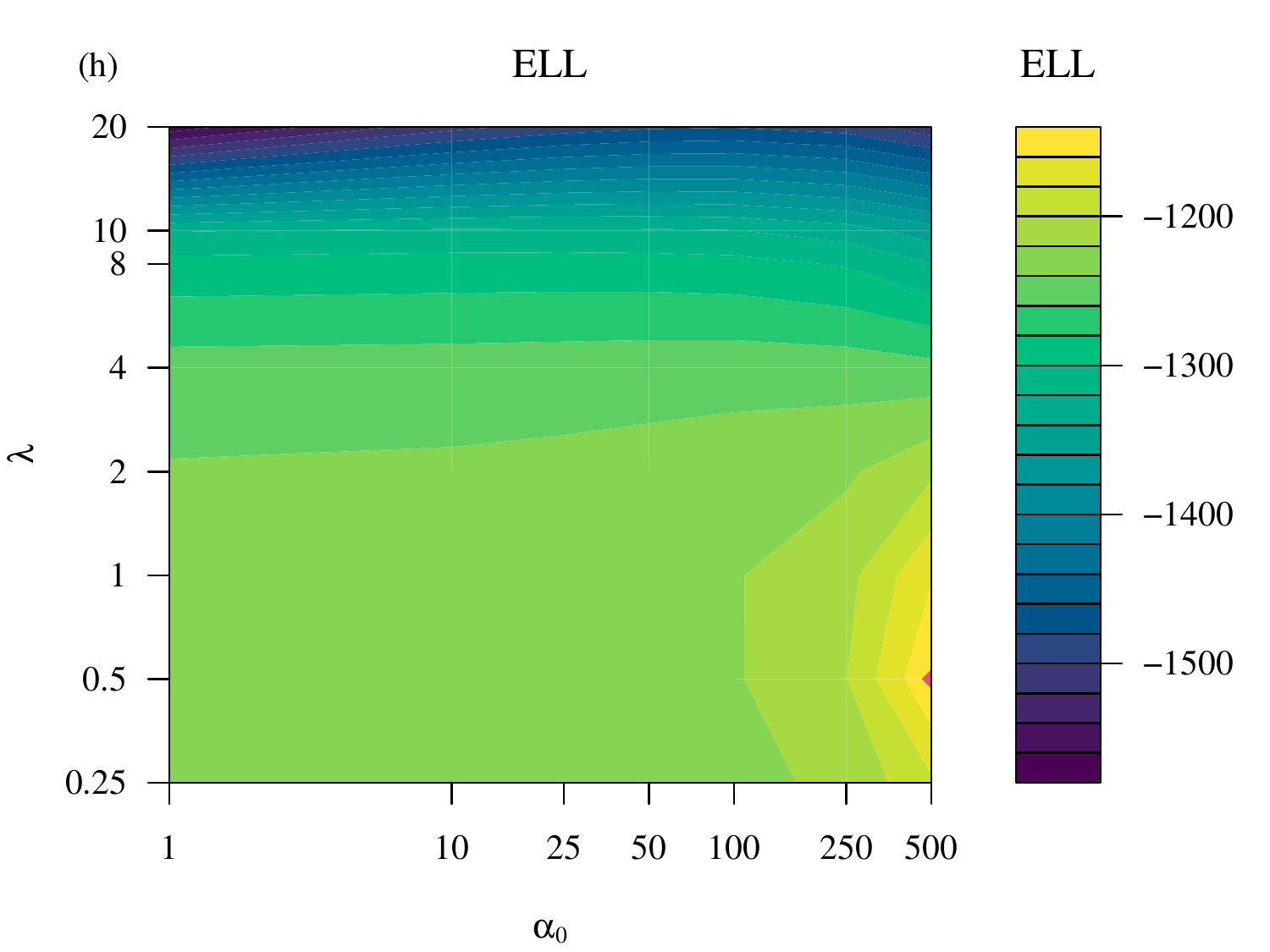}
    \includegraphics[width=.41\textwidth]{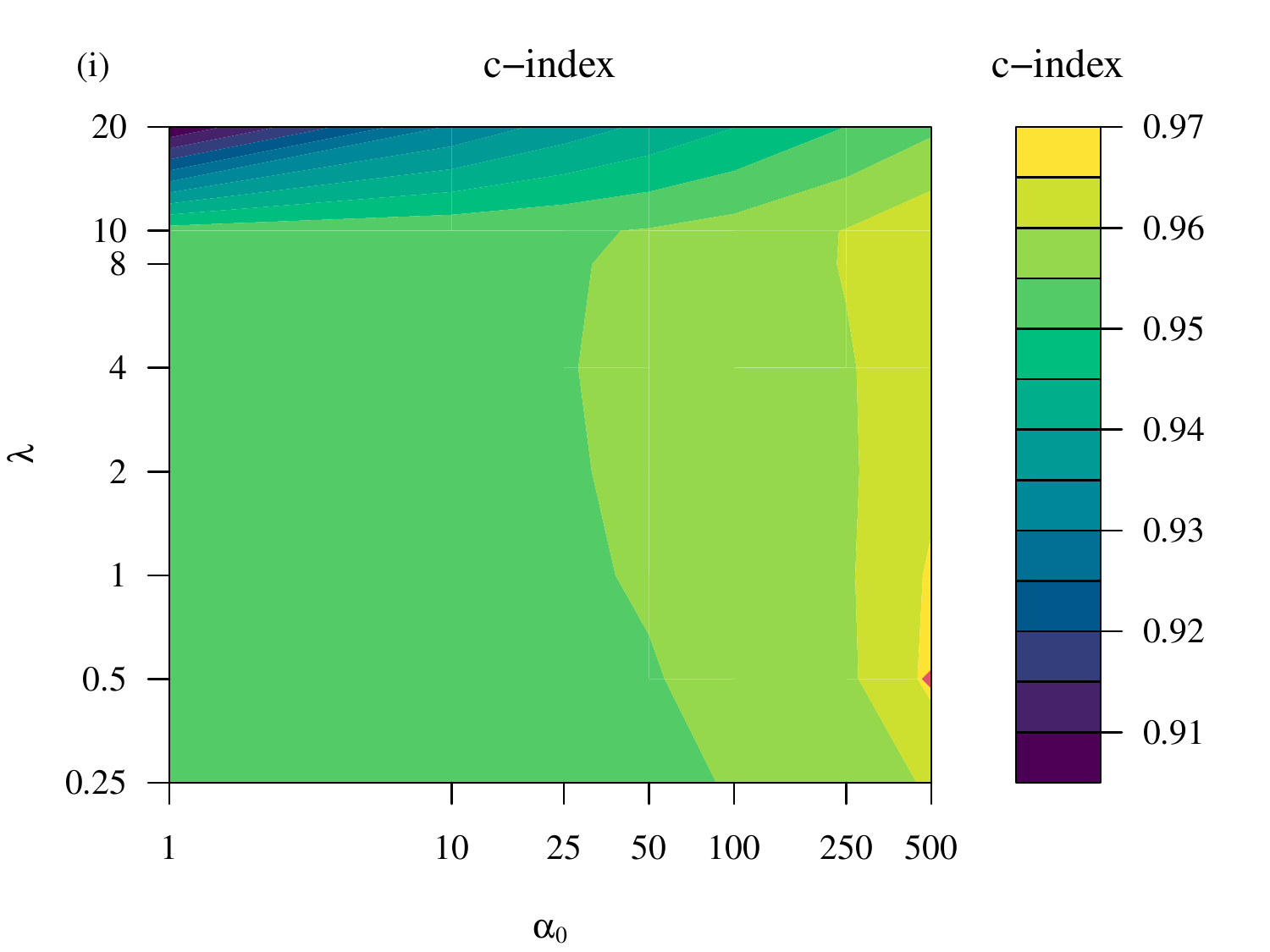}
    \caption{\textbf{Setting 2}: sensitivity with respect to $\lambda$ and $a_0$.}
    \label{fig:sensitivity_alpha_lambda_2}
\end{figure}}

\sidewaysfigure{\begin{figure}[htp]
    \centering
    \includegraphics[width=.41\textwidth]{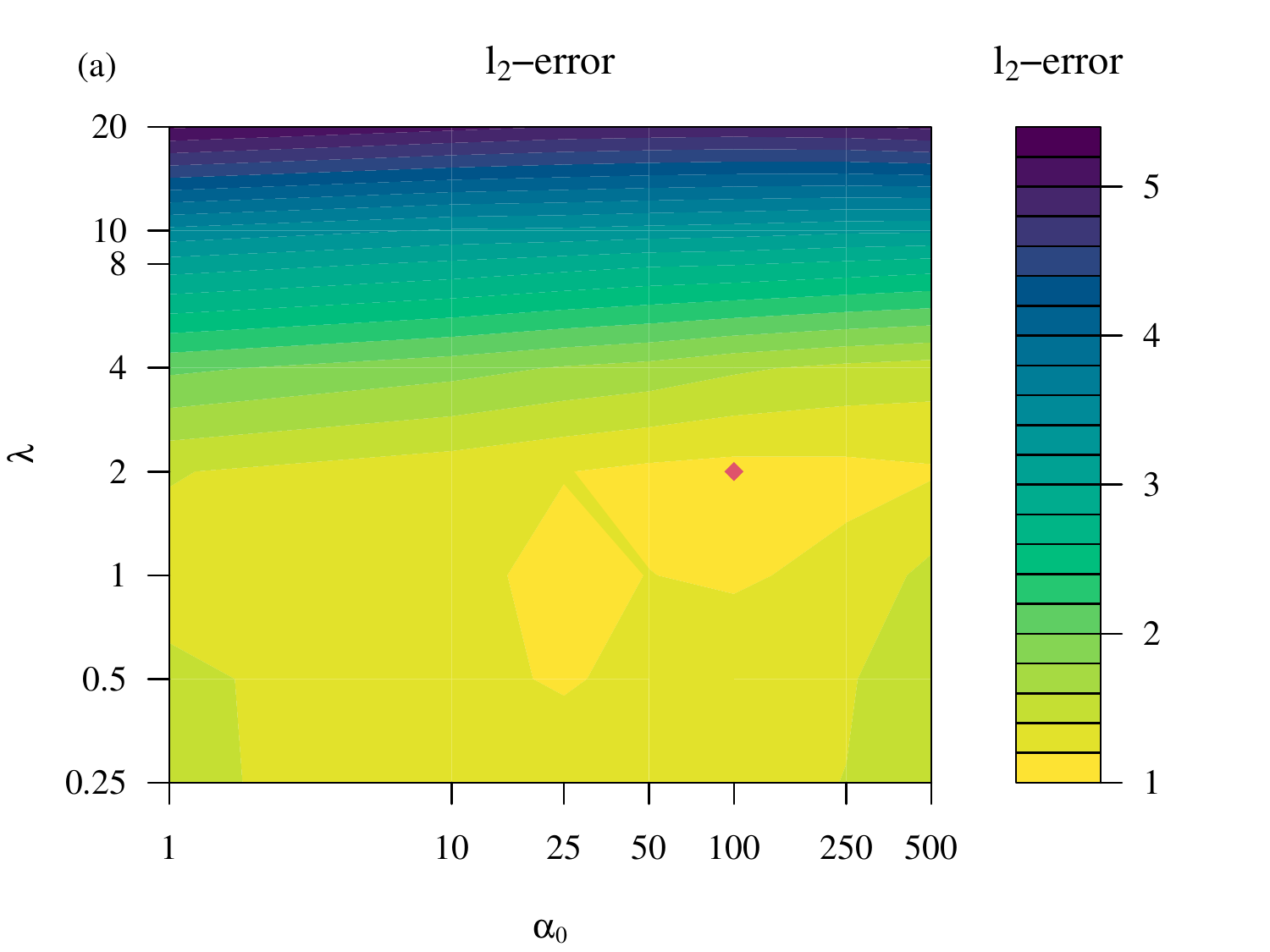}
    \includegraphics[width=.41\textwidth]{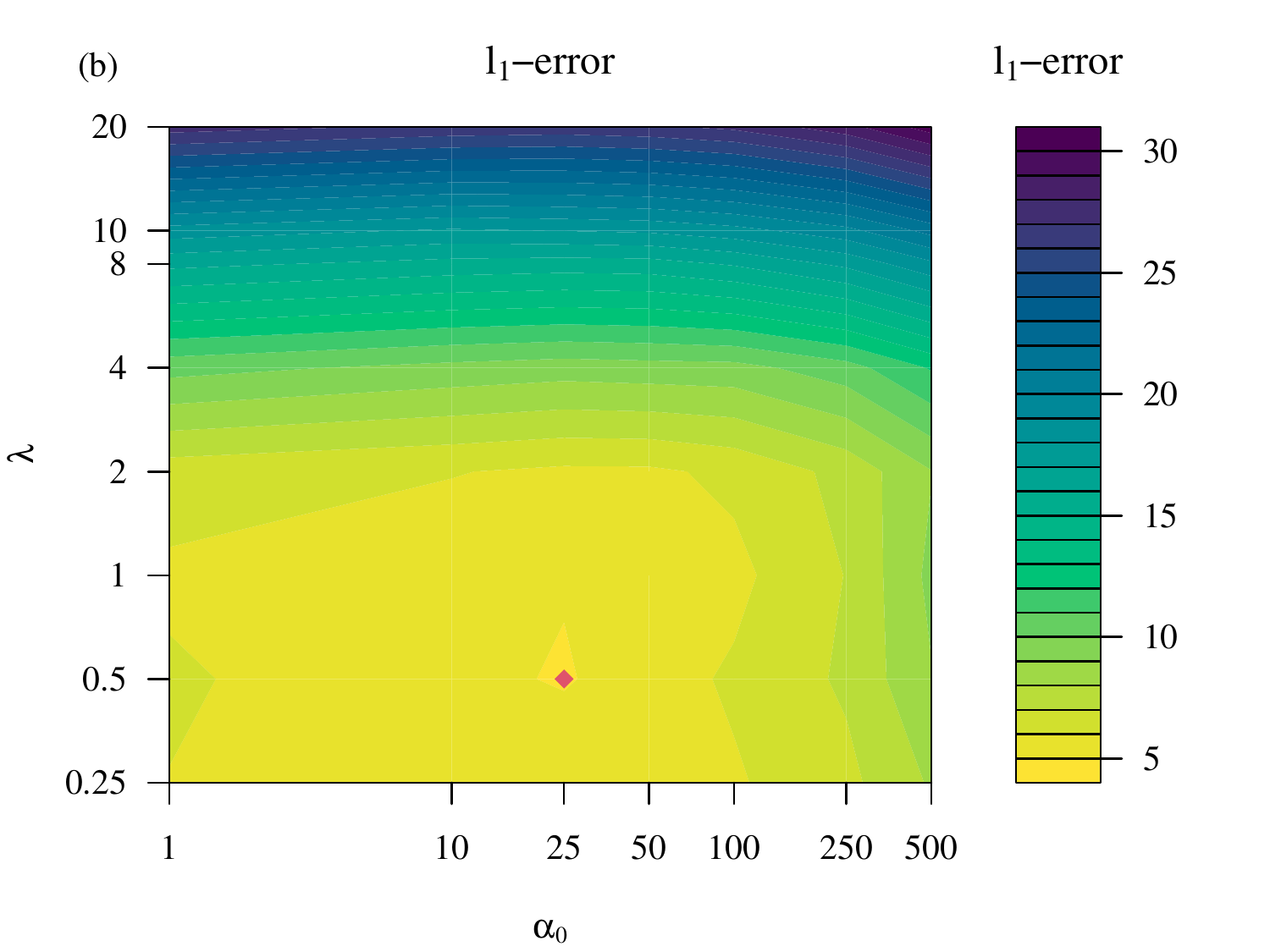}
    \includegraphics[width=.41\textwidth]{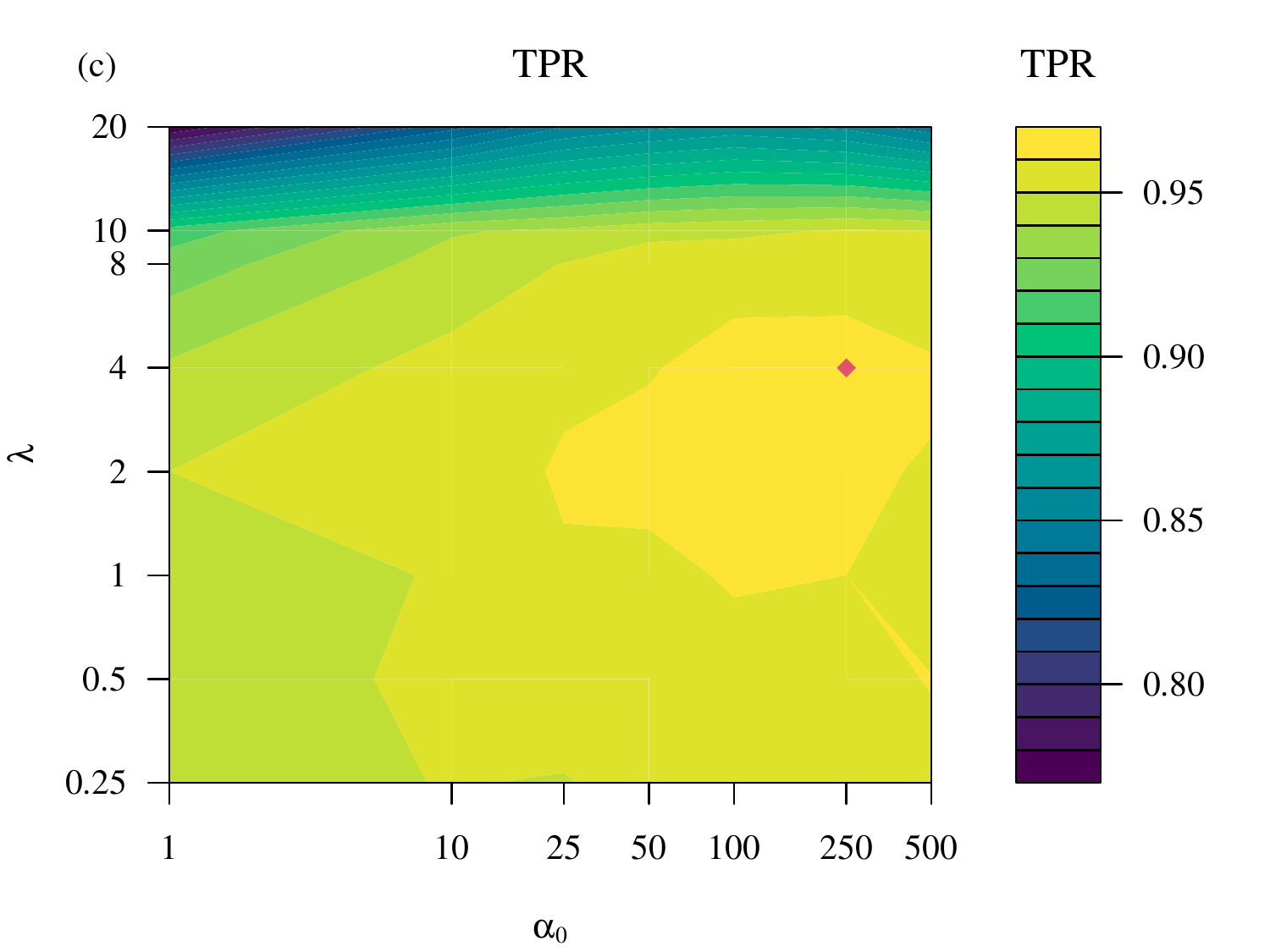}
    \includegraphics[width=.41\textwidth]{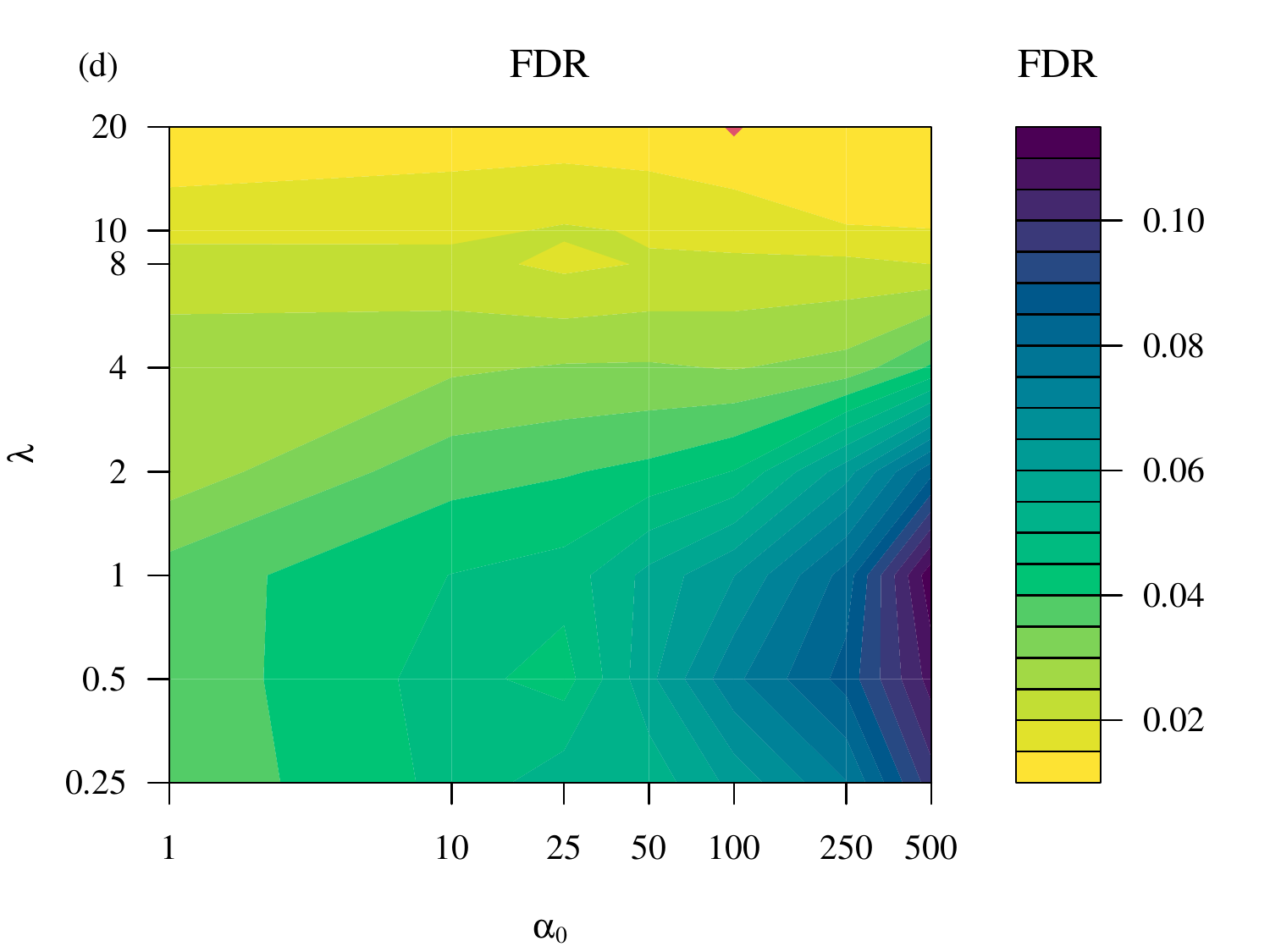}
    \includegraphics[width=.41\textwidth]{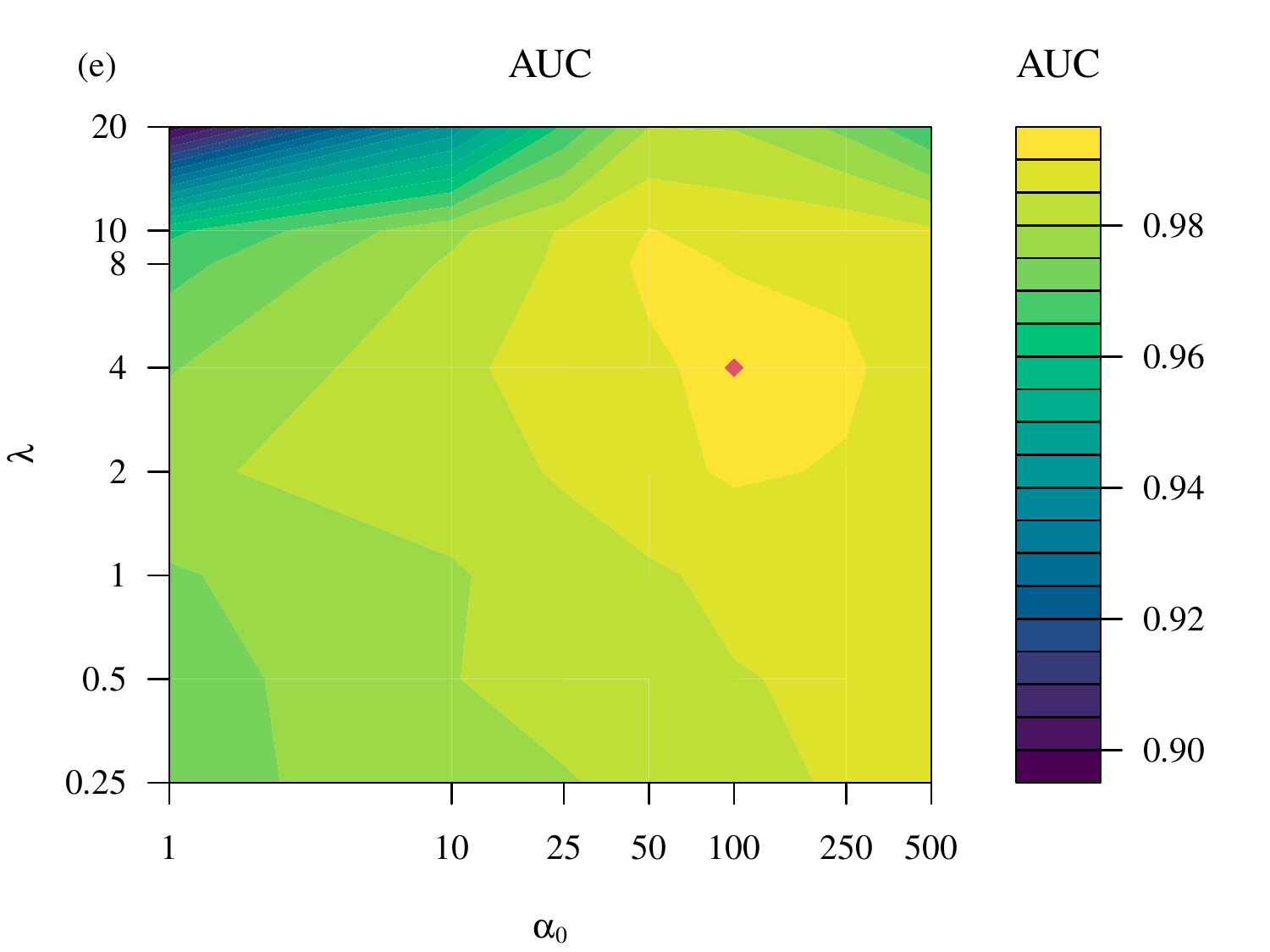}
    \includegraphics[width=.41\textwidth]{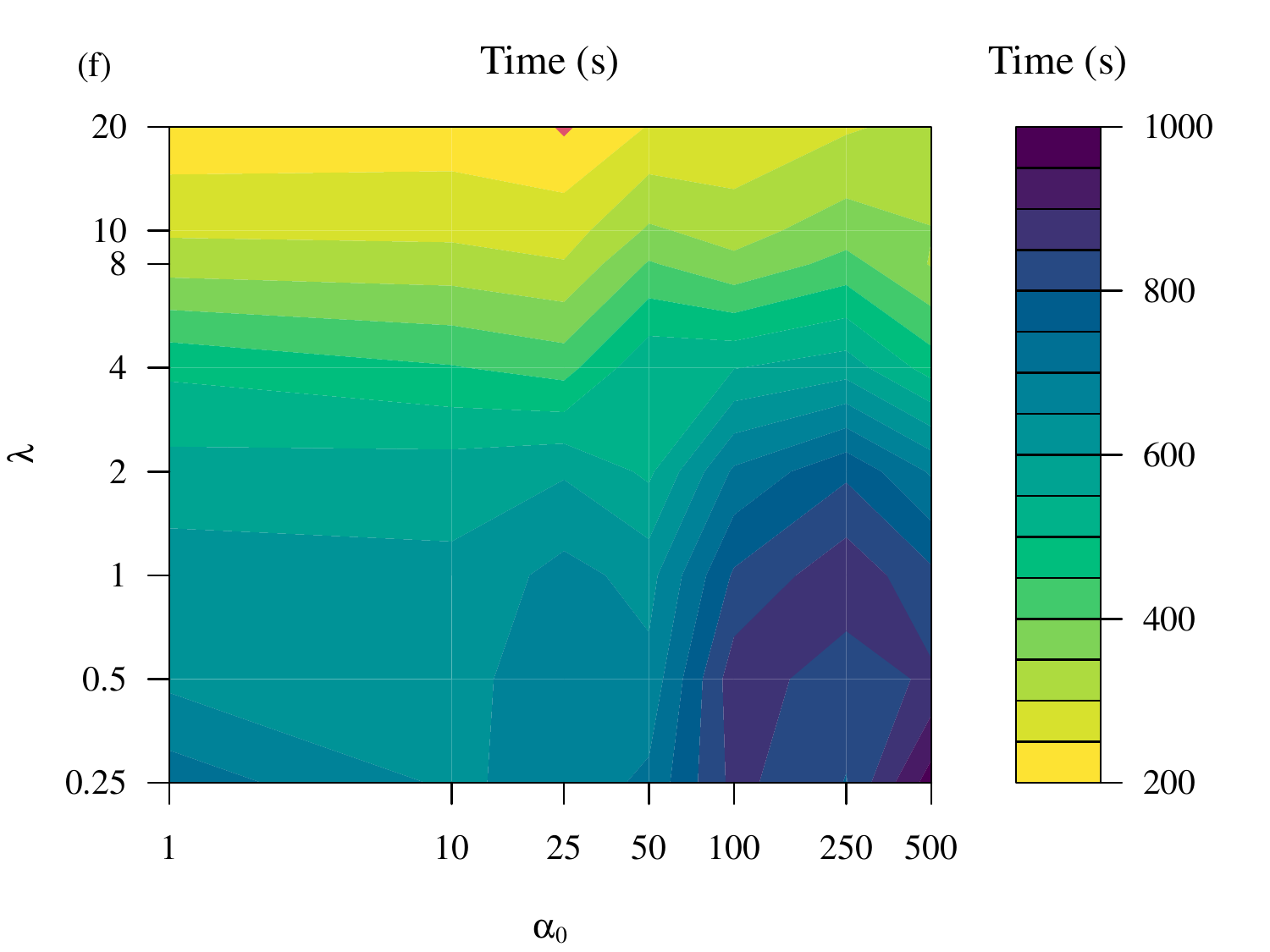}
    \includegraphics[width=.41\textwidth]{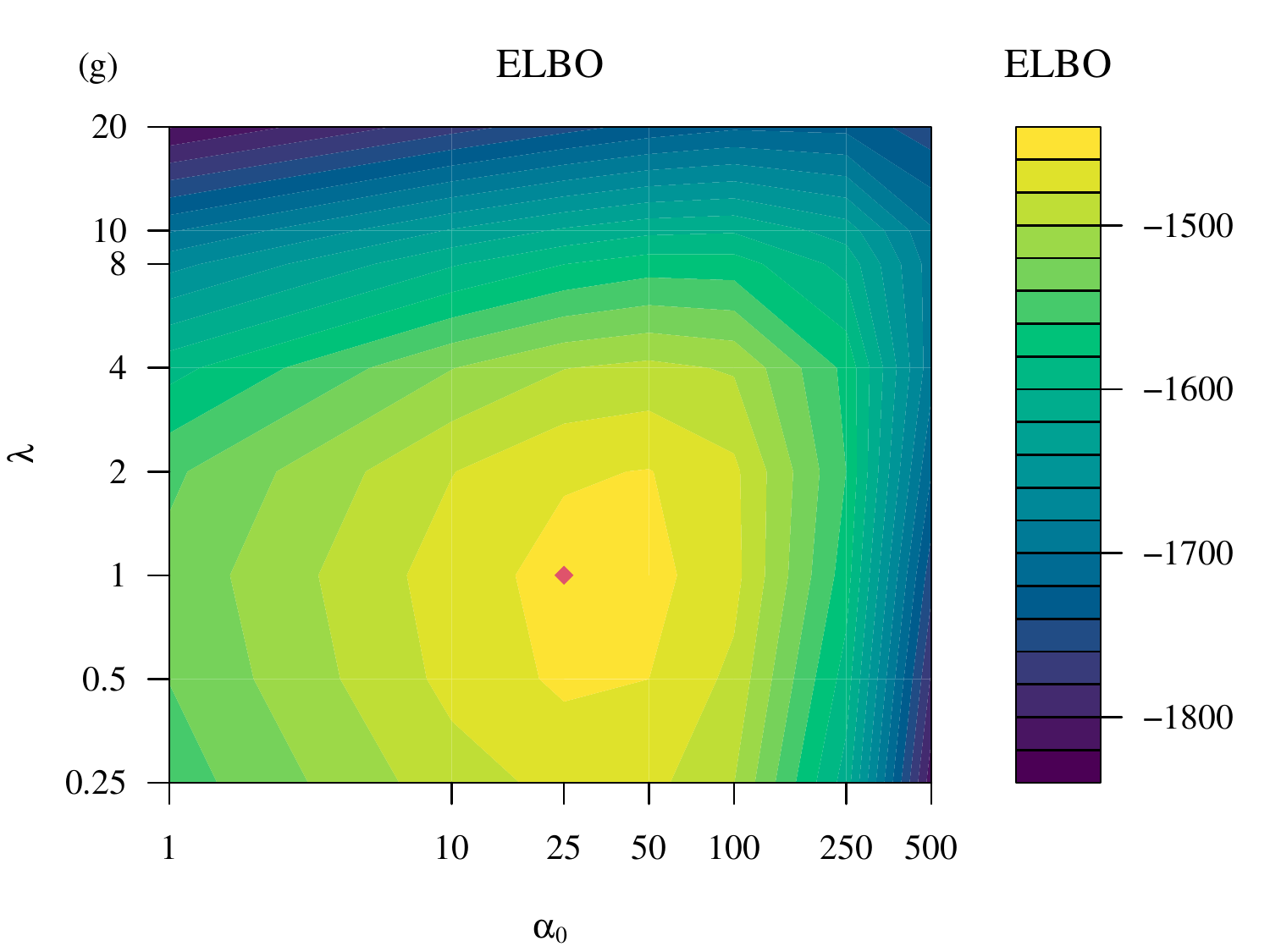}
    \includegraphics[width=.41\textwidth]{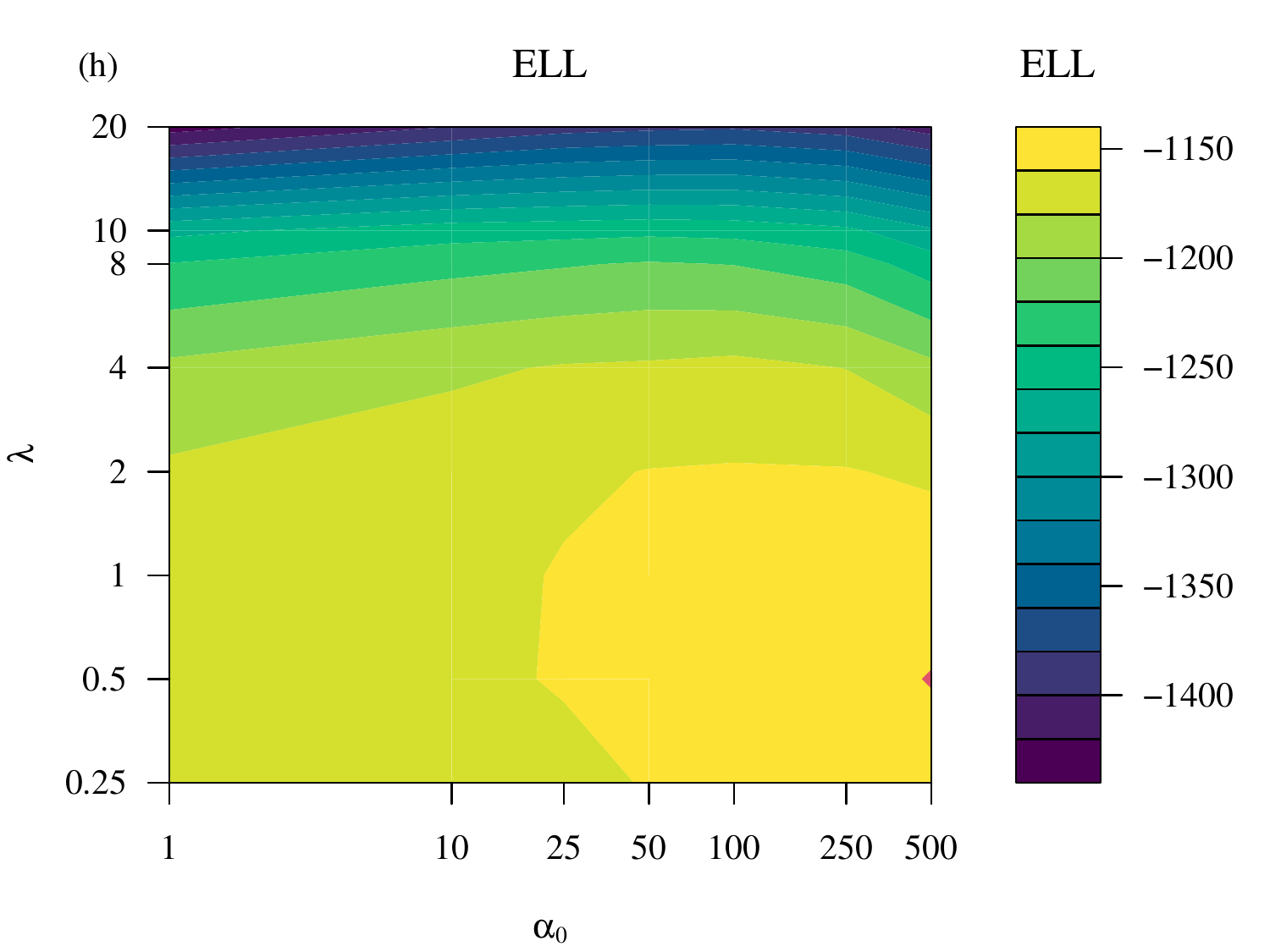}
    \includegraphics[width=.41\textwidth]{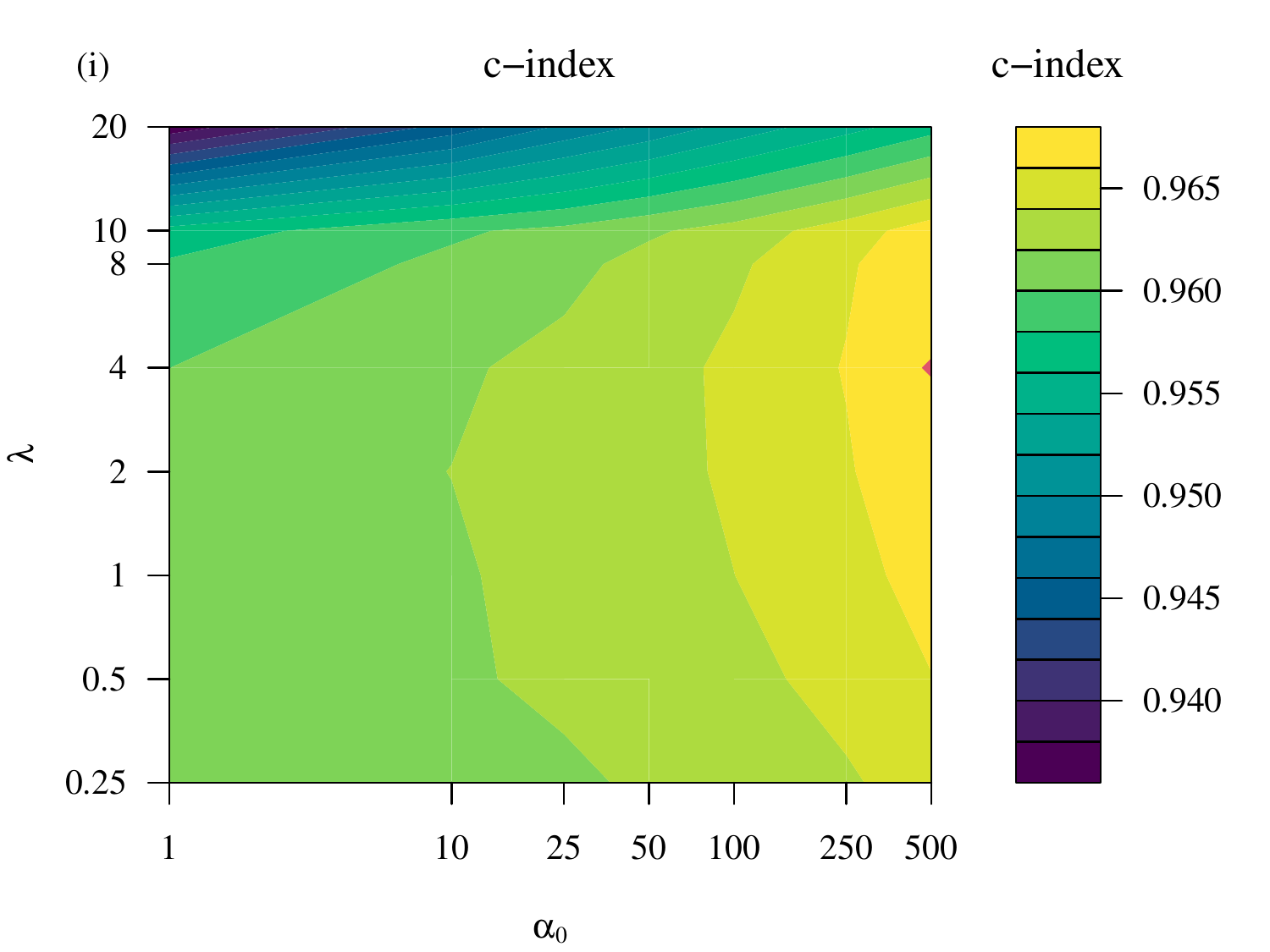}
    \caption{\textbf{Setting 3}: sensitivity with respect to $\lambda$ and $a_0$.}
    \label{fig:sensitivity_alpha_lambda_3}
\end{figure}}

\sidewaysfigure{\begin{figure}[htp]
    \centering
    \includegraphics[width=.41\textwidth]{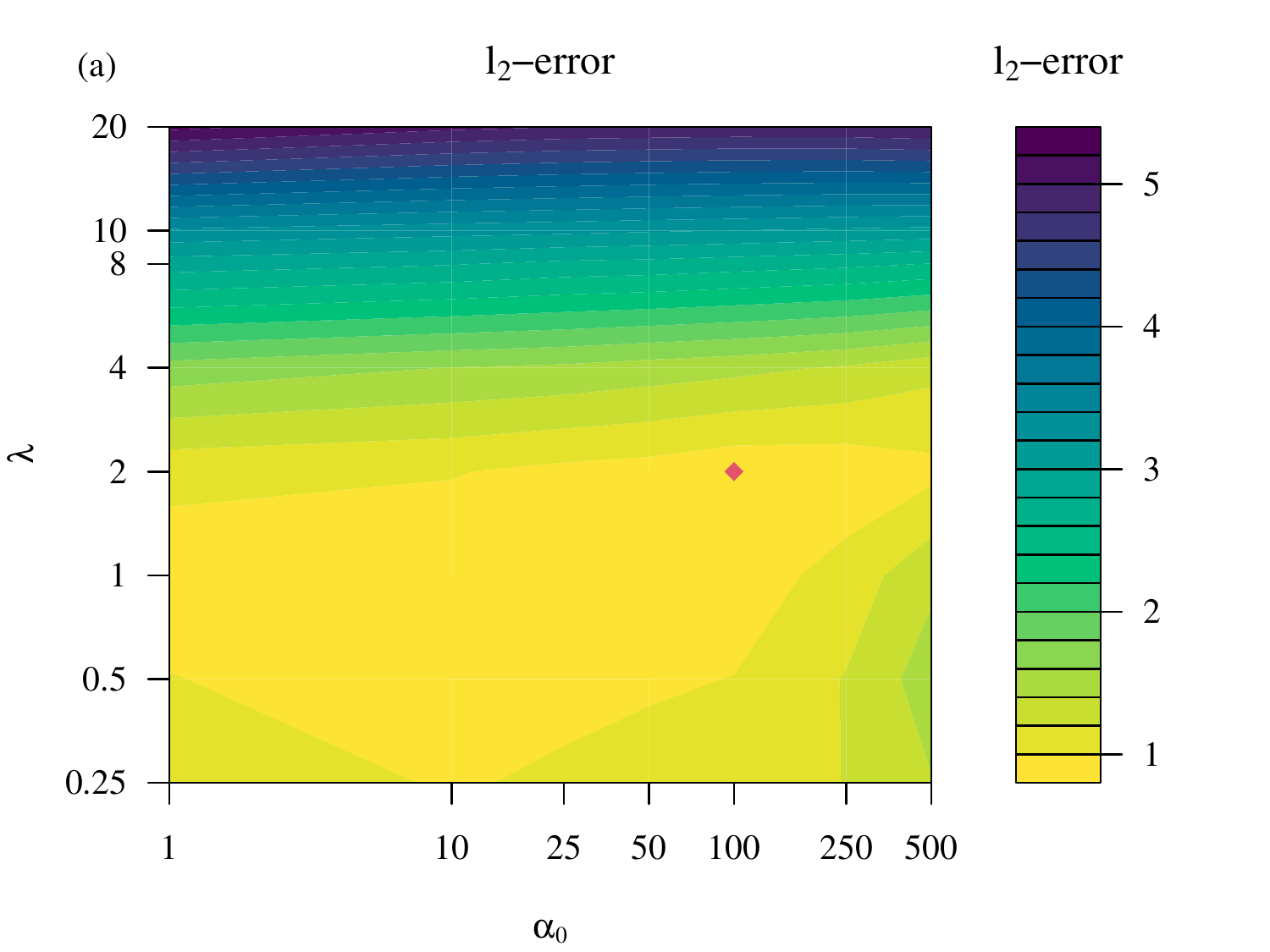}
    \includegraphics[width=.41\textwidth]{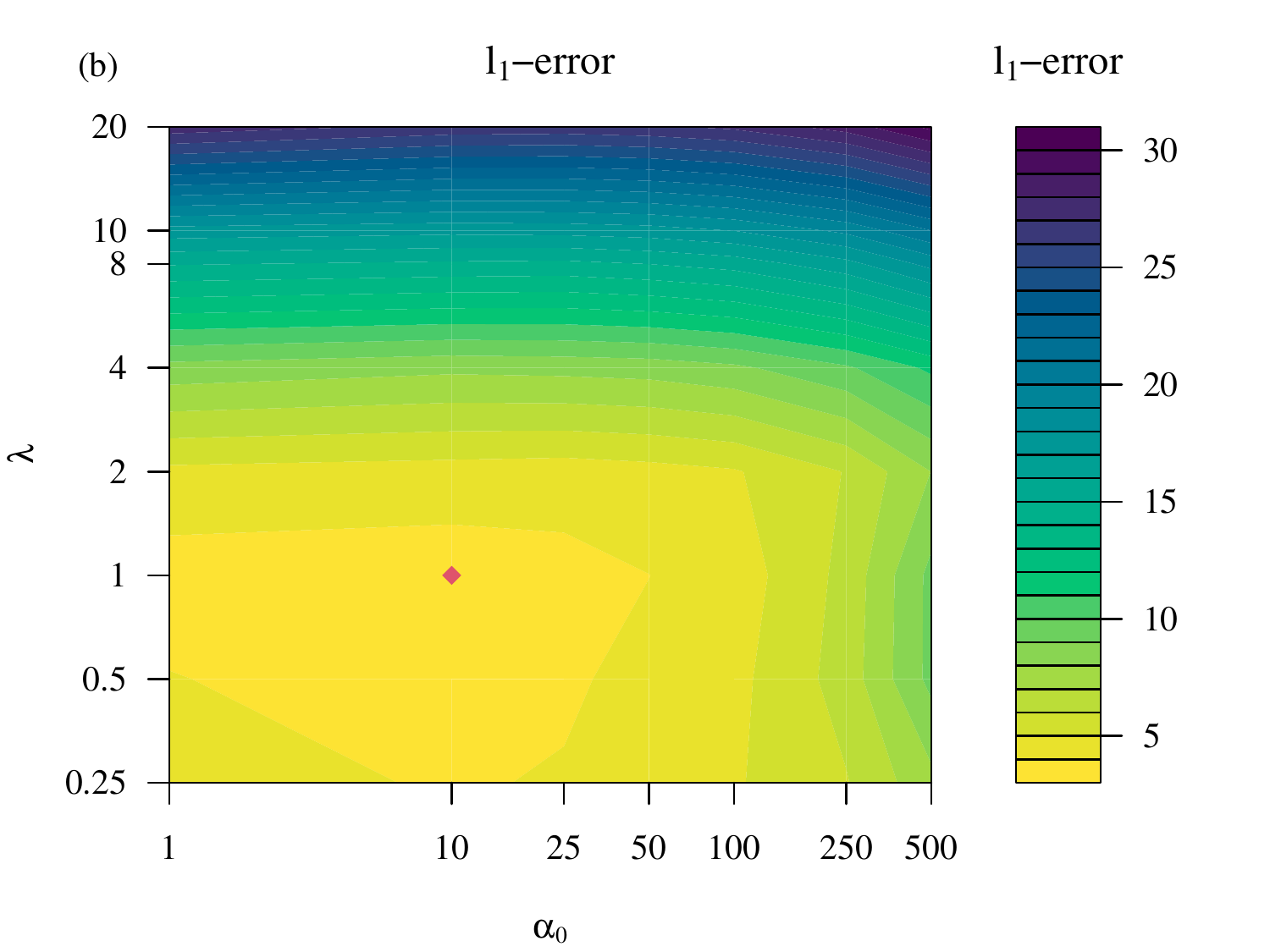}
    \includegraphics[width=.41\textwidth]{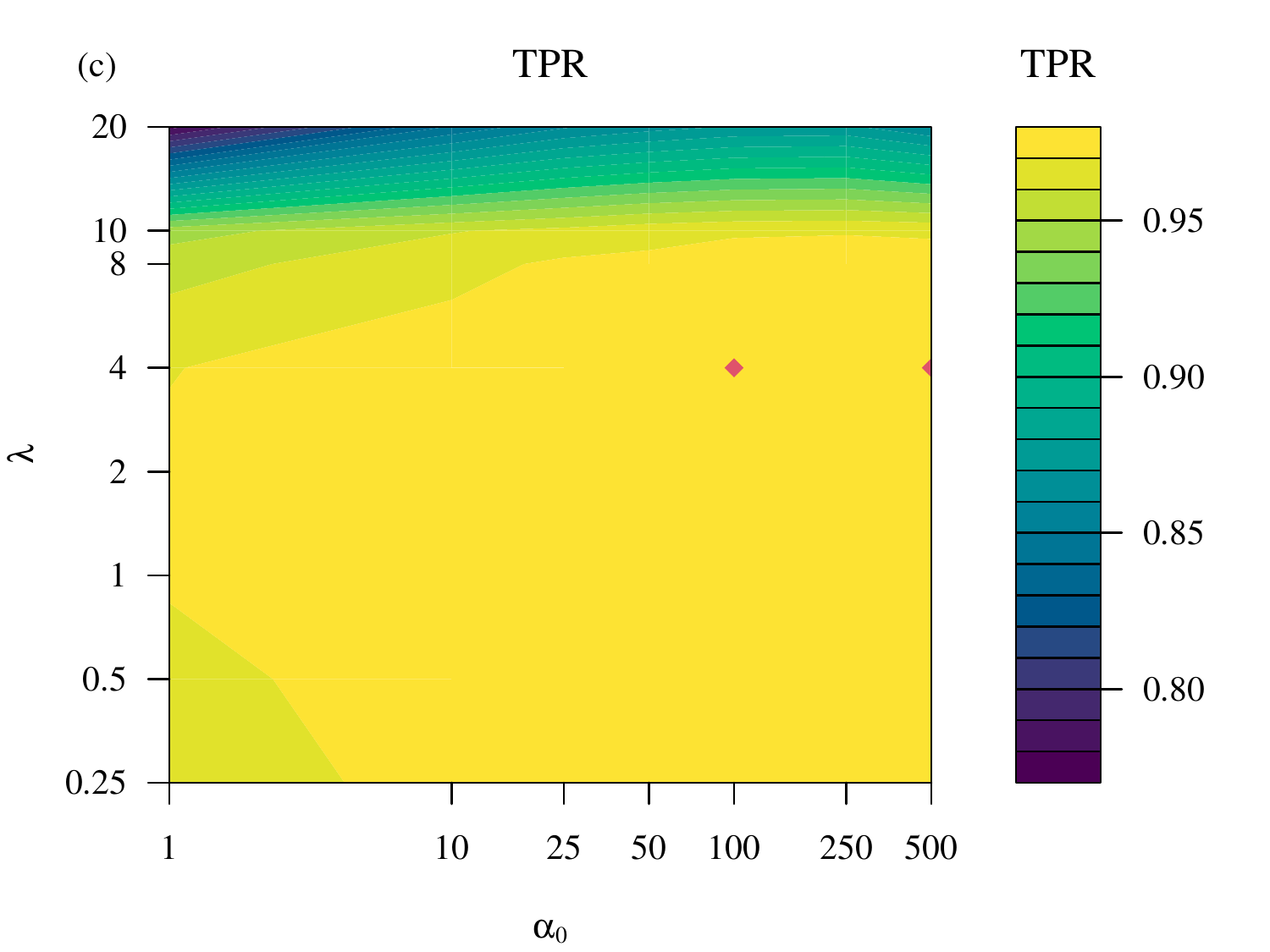}
    \includegraphics[width=.41\textwidth]{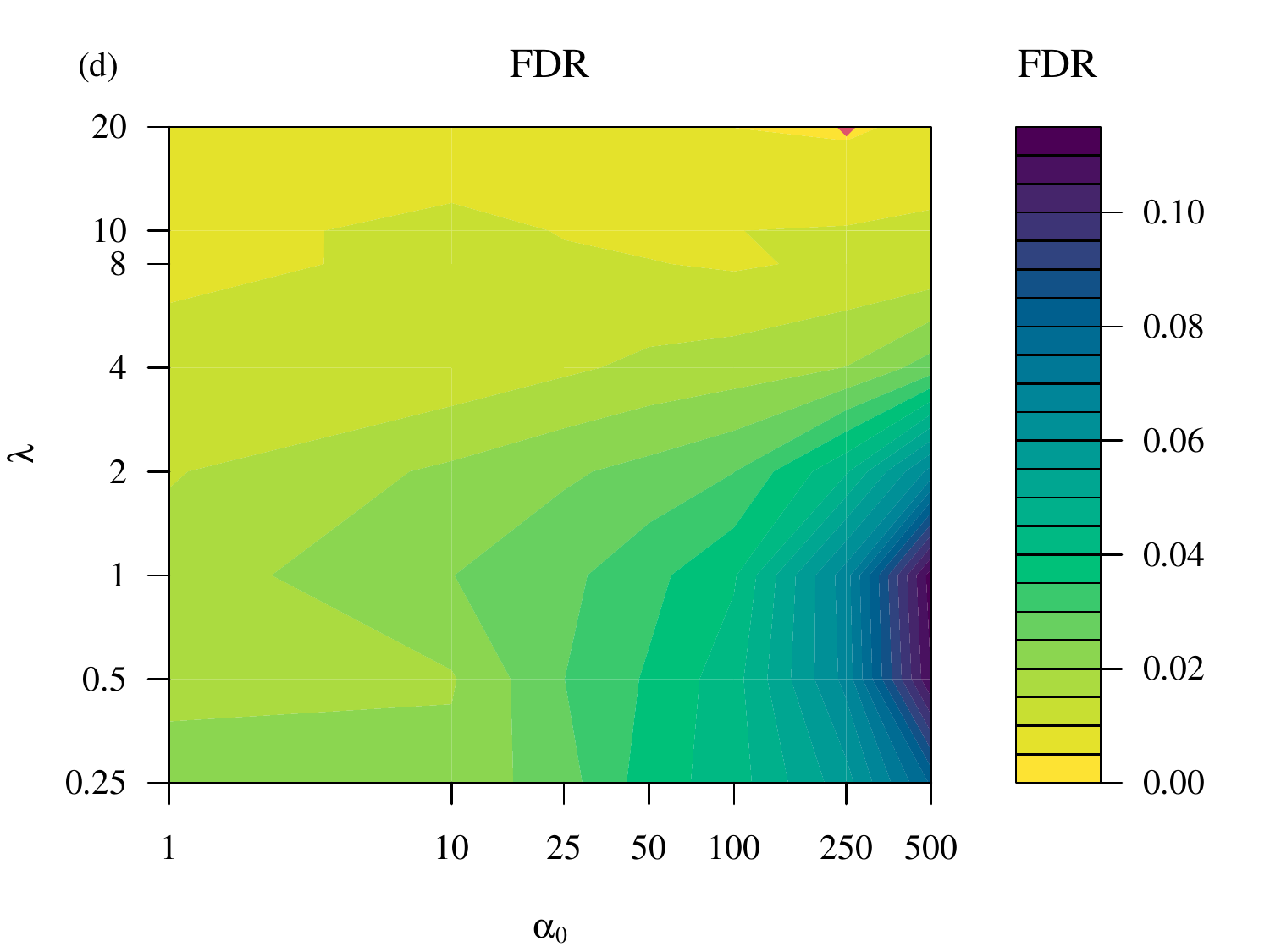}
    \includegraphics[width=.41\textwidth]{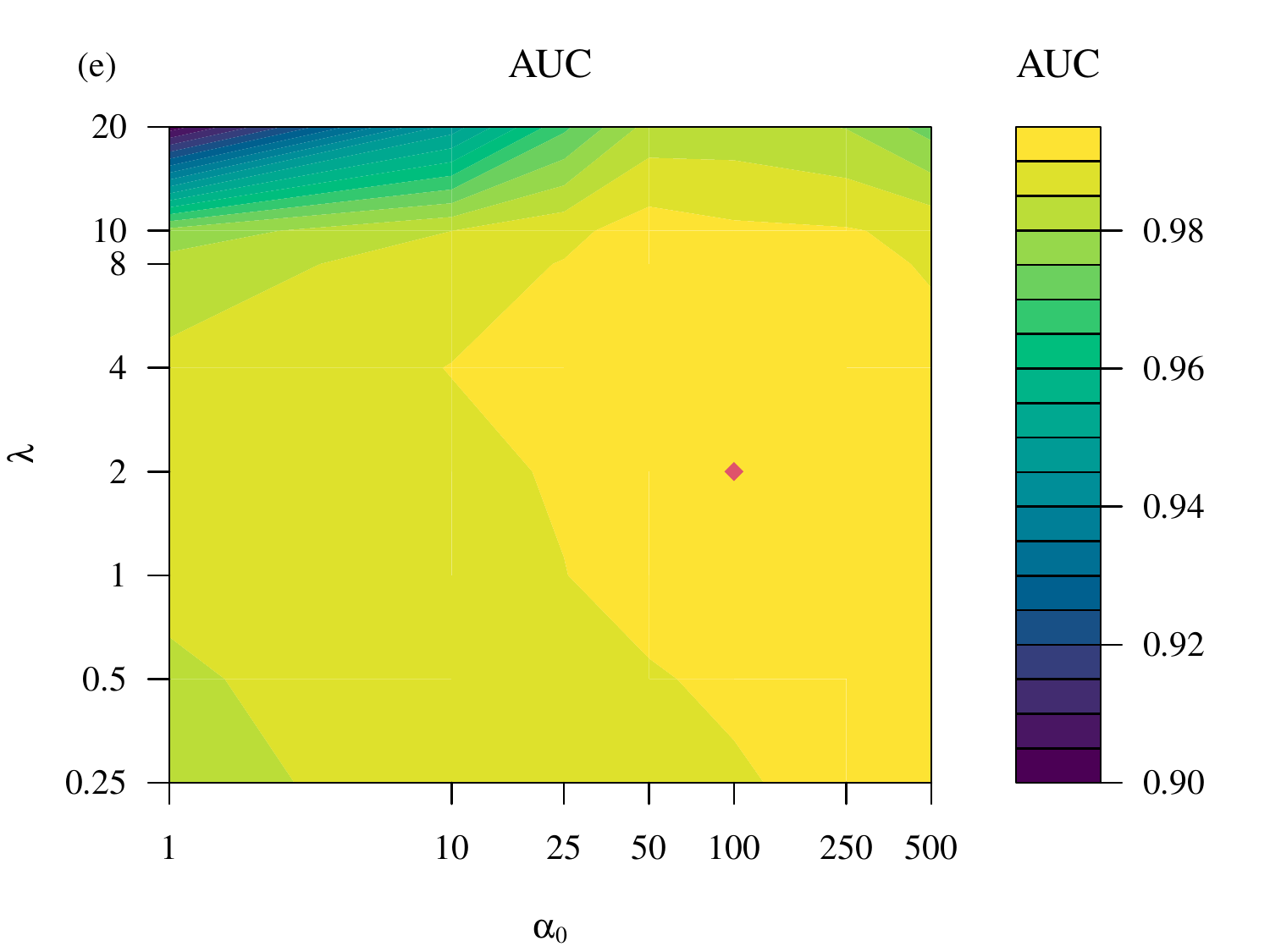}
    \includegraphics[width=.41\textwidth]{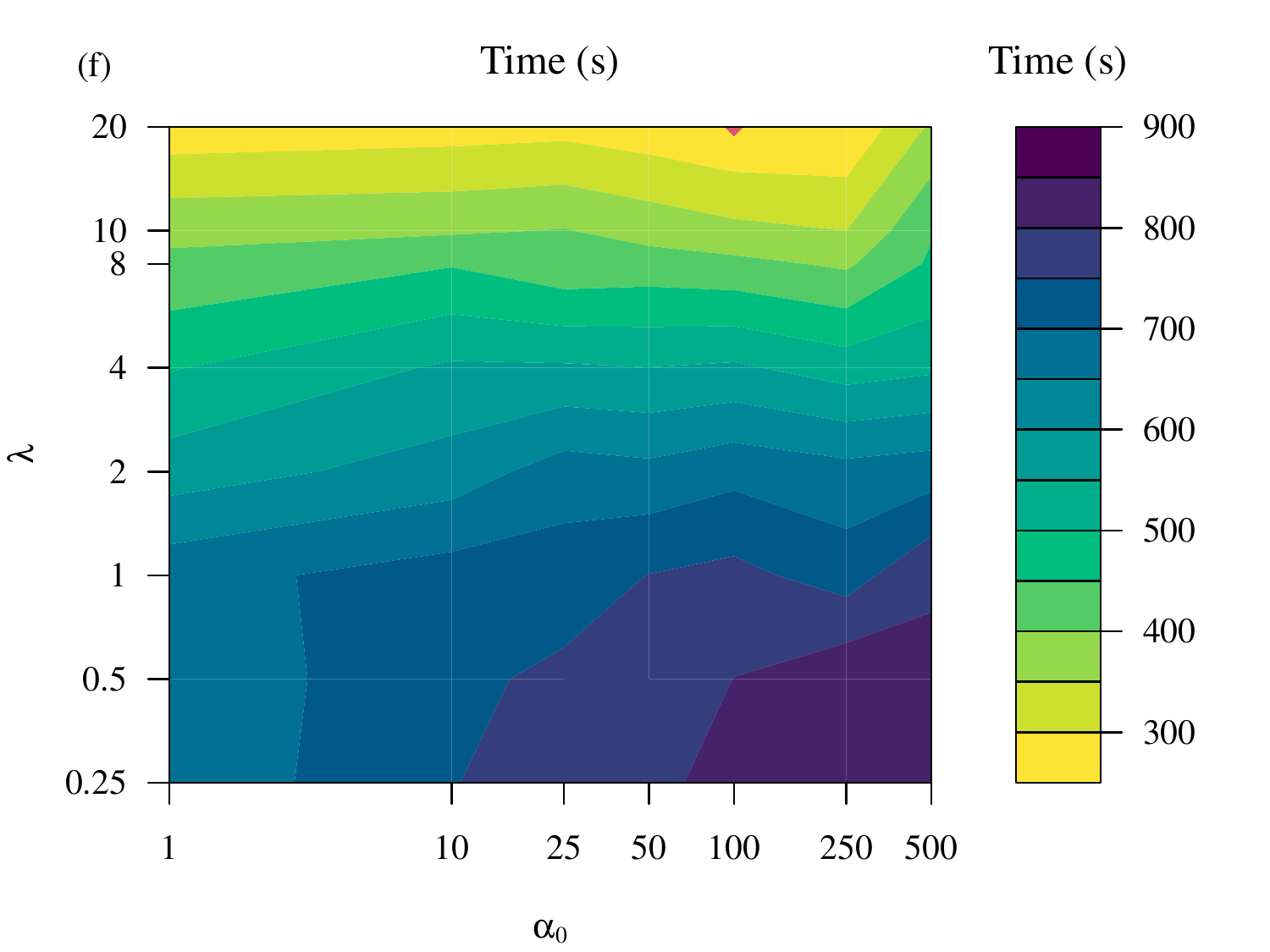}
    \includegraphics[width=.41\textwidth]{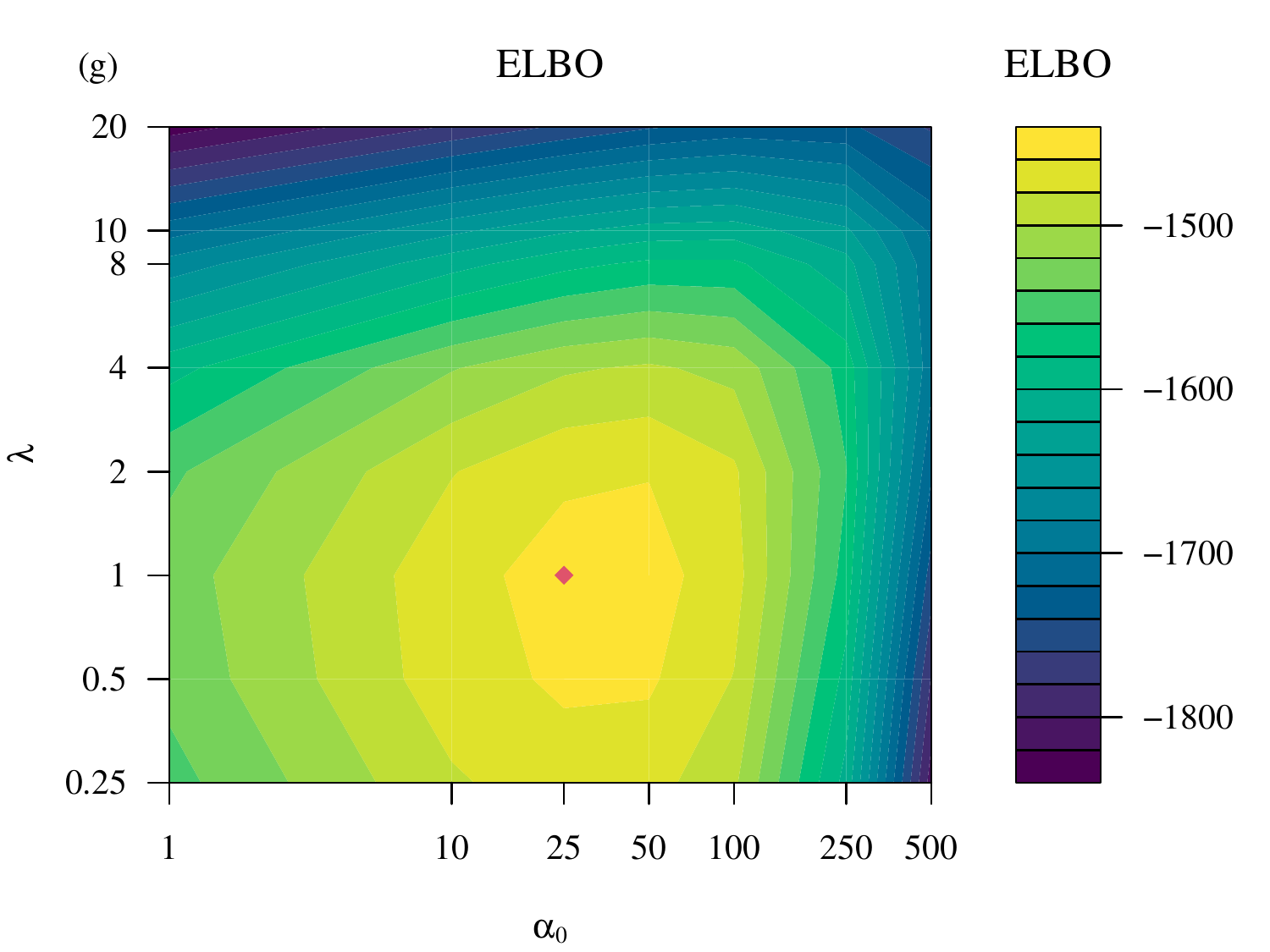}
    \includegraphics[width=.41\textwidth]{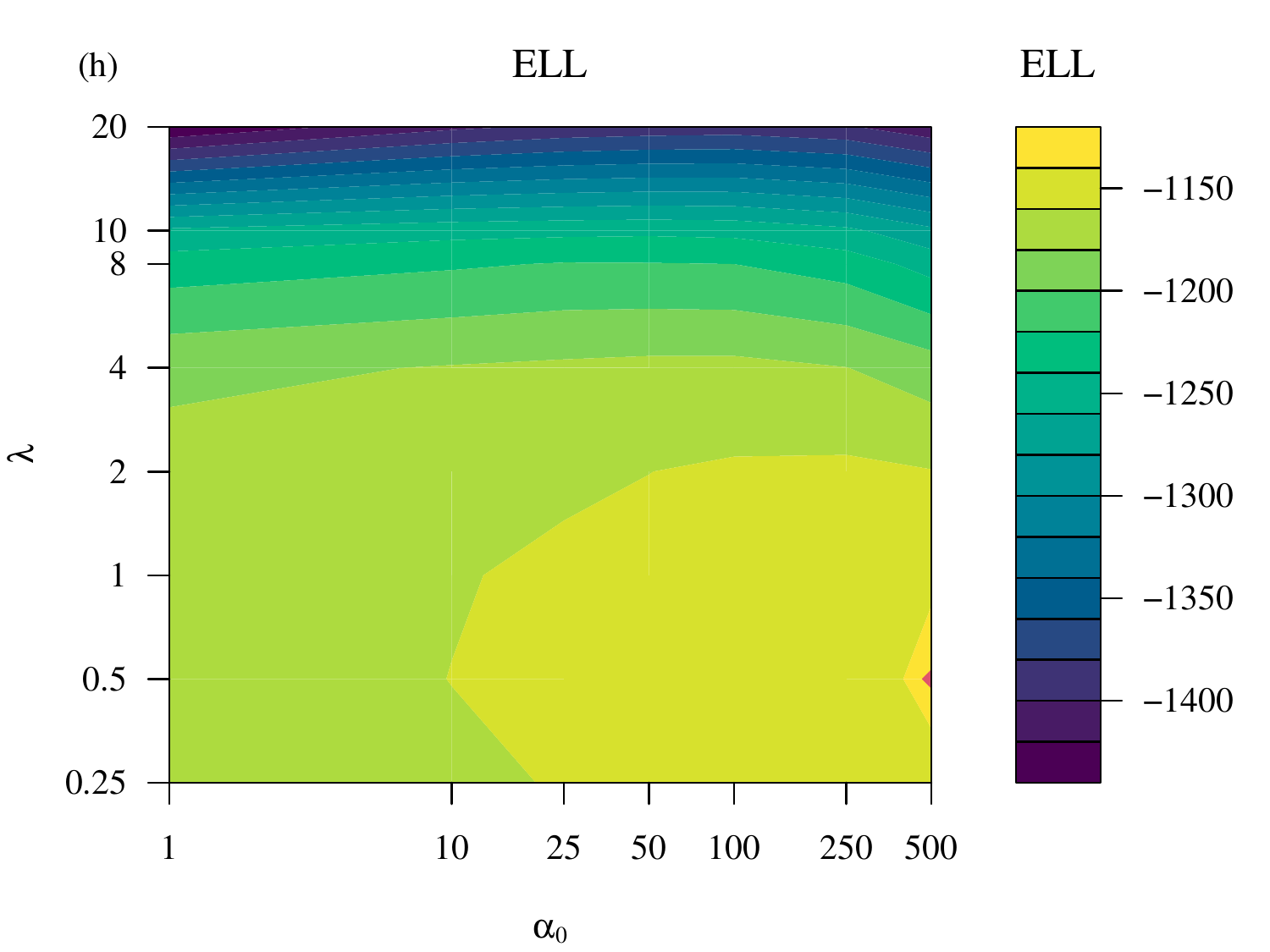}
    \includegraphics[width=.41\textwidth]{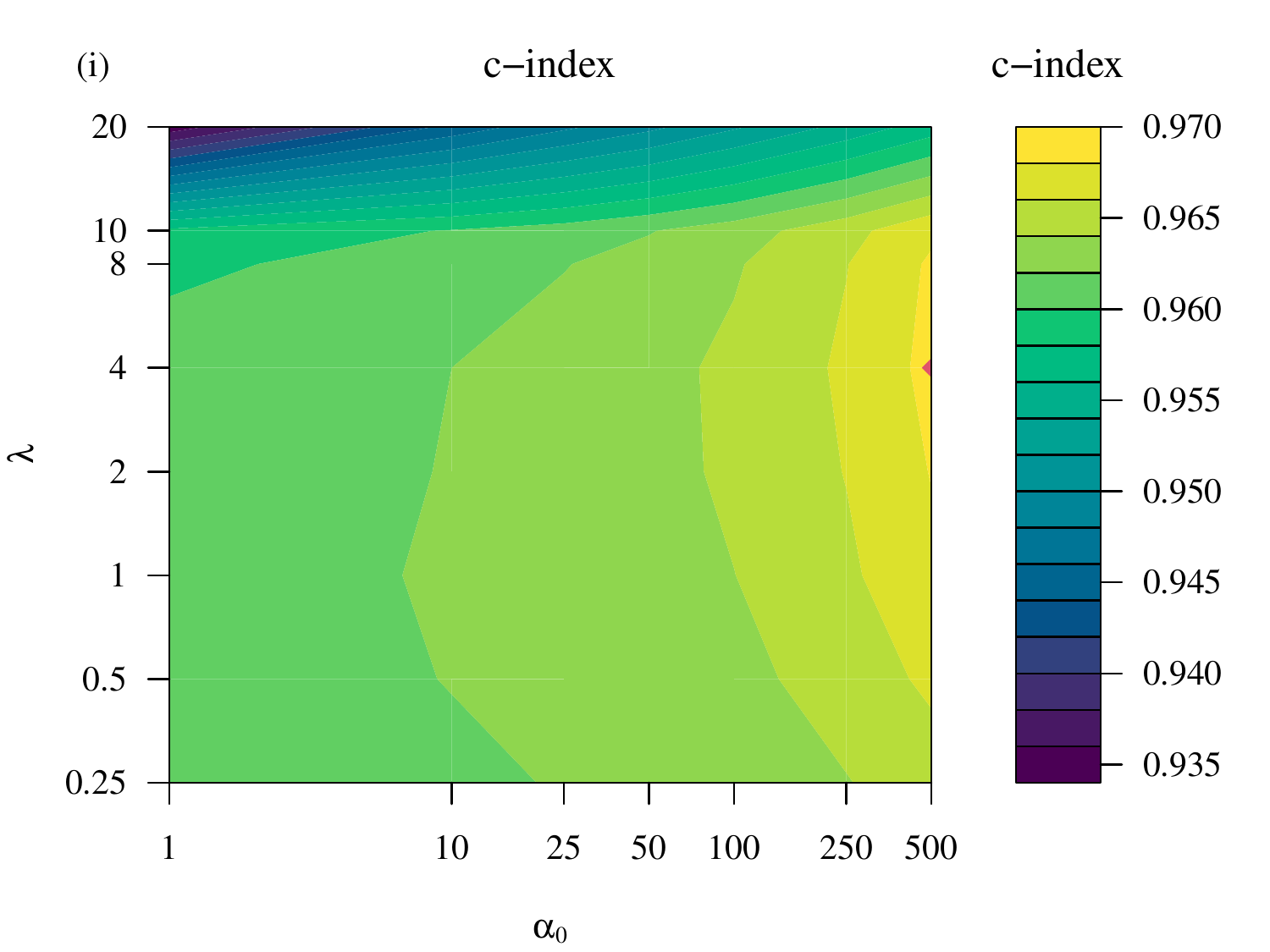}
    \caption{\textbf{Setting 4}: sensitivity with respect to $\lambda$ and $a_0$.}
    \label{fig:sensitivity_alpha_lambda_4}
\end{figure}}

% ------------------------------------------------------------------------------
\newpage
% ------------------------------------------------------------------------------
\section{Application to real data}

\subsection{Availability of datasets}

The datasets can be freely downloaded from:
\begin{itemize}
    \item Ovarian cancer dataset: \url{https://xenabrowser.net/datapages/?cohort=GDC%20TCGA%20Ovarian%20Cancer%20(OV)&removeHub=https%3A%2F%2Fxena.treehouse.gi.ucsc.edu%3A443}
    \item Breast cancer dataset: \url{https://xenabrowser.net/datapages/?cohort=Breast%20Cancer%20(Yau%202010)&removeHub=https%3A%2F%2Fxena.treehouse.gi.ucsc.edu%3A443}
\end{itemize}

\subsection{Ovarian cancer dataset}

We examine the low and high risk groups constructed using the prognostic index for the ovarian cancer dataset. To construct the groups, we selected the model fit with $\lambda=0.5$ for the first fold, which has the validation c-index of $0.57$, the lowest value obtained across the different values of $\lambda$ and folds. We chose this value of $\lambda$ and fold to demonstrate how comparison of groups can highlight outliers and indicate when splitting based on the prognostic index may not be appropriate.

\begin{figure}[htp]
    \centering
    \includegraphics[width=.8\textwidth]{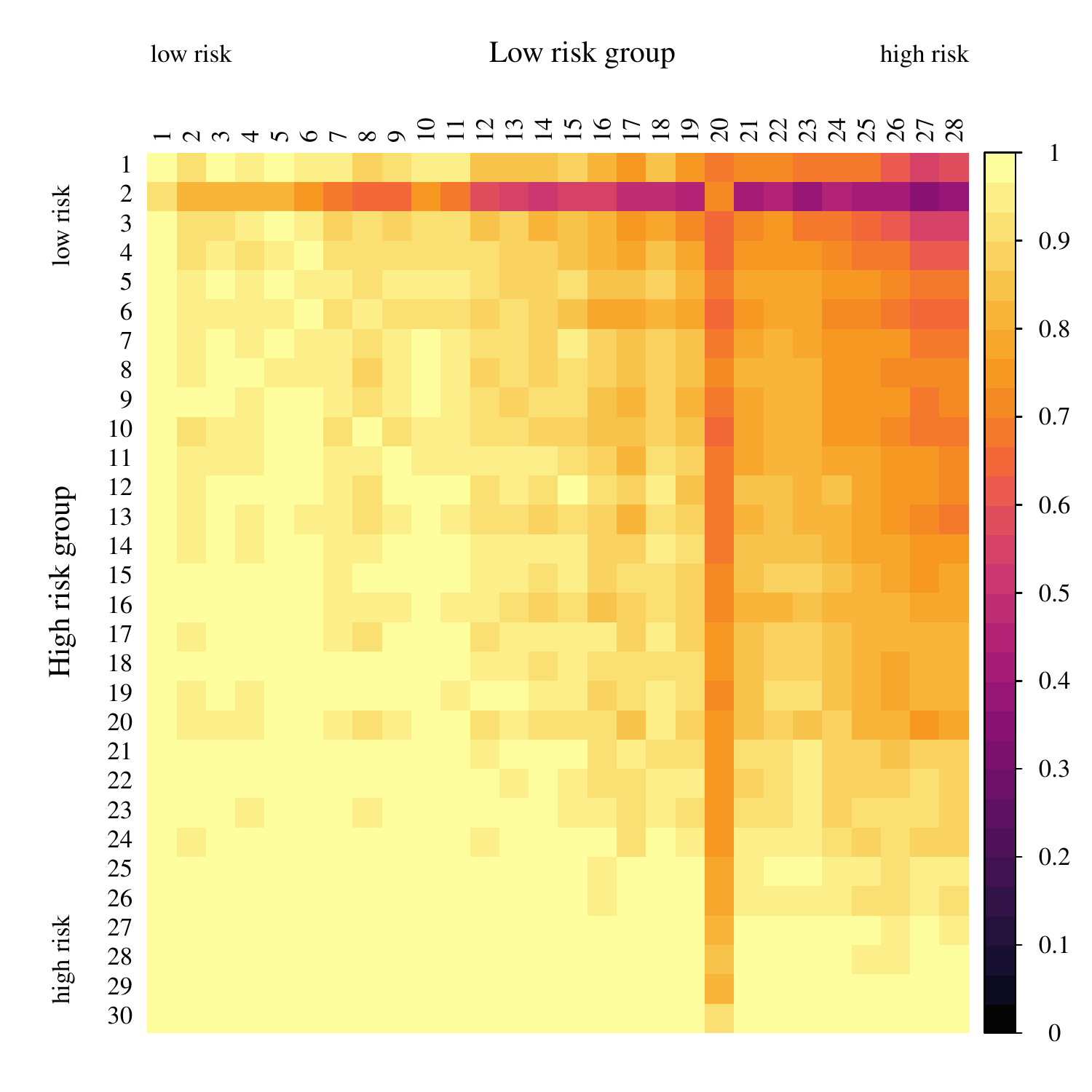}
    \caption{Comparison of validation risk groups for the TCGA ovarian cancer dataset. Rows correspond to patients in the high-risk group and columns to patients in the low-risk group. Both rows and columns are sorted from lowest to highest risk.}
    \label{fig:tcga_risk}
\end{figure}

As before, to construct the groups, the validation set was split based on the median value of the prognostic index computed on the training set. The pairwise posterior probability that a patient within the high risk group is at greater risk than a patient within the low risk group is presented in \Cref{fig:tcga_risk}. Immediately we notice two patients within each group that may be outliers: patient 2 within the high-risk group and patient 20 within the low-risk group. In the case of patient 2 within the high-risk group, examining the row from left to right we notice the posterior probability is decreasing from around 0.8 to 0.4, meaning that on furthest end, the high-risk patients in the low-risk groups are at greater risk than patient 2. Regarding patient 20 from the low risk group, we notice the posterior probability across the column is lower in comparison to the other patients within the group. Furthermore the mean posterior probability across the column is $0.73$, meaning this may be a borderline case between being in the high and low risk group.

\subsection{Breast cancer dataset}

\begin{figure}[htp]
    \centering
    \includegraphics[width=.7\textwidth]{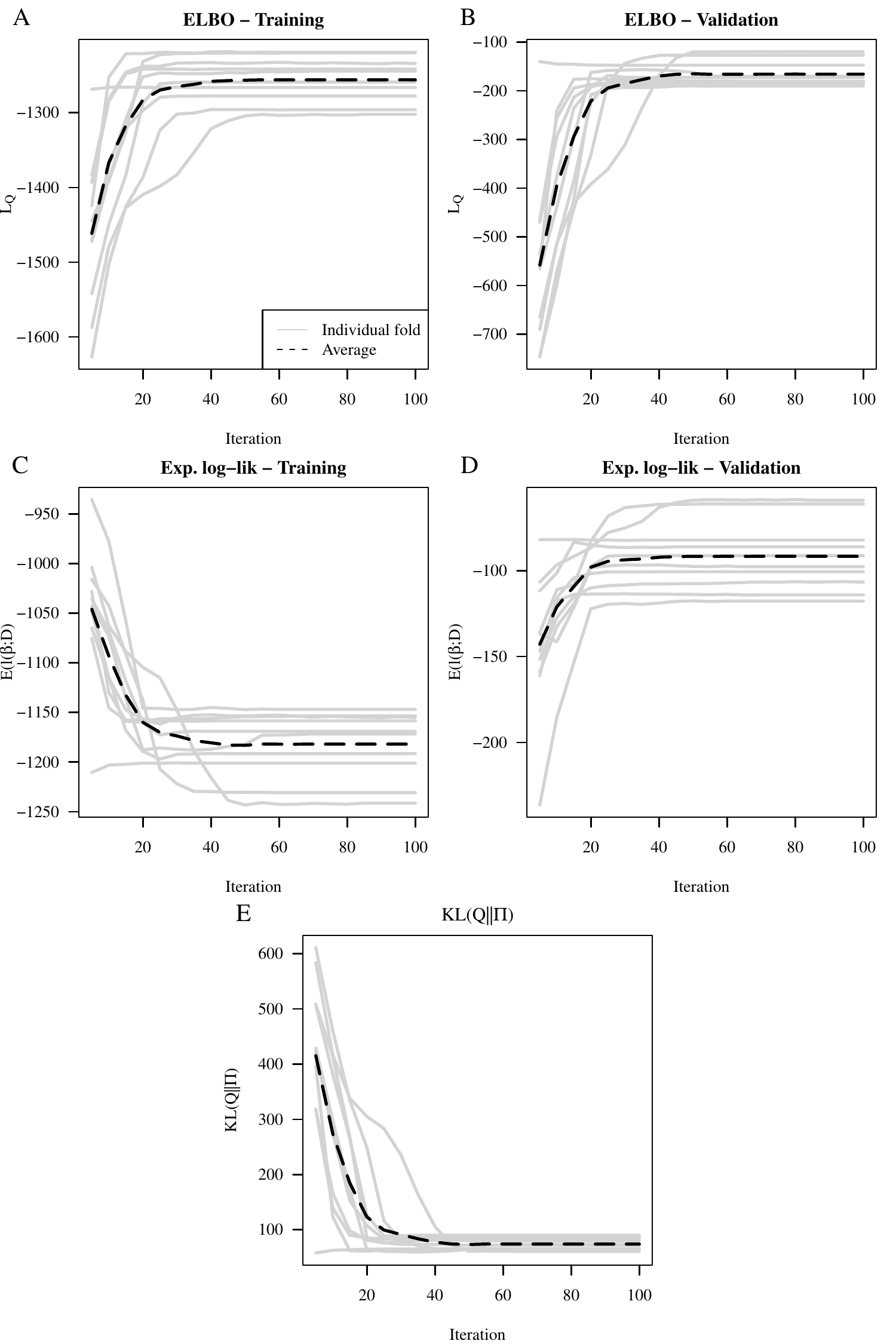}
    \caption{Breast cancer dataset convergence diagnostics for model is fit with $\lambda=2.5$, Presented is the: ELBO, ELL and KL for the training and validation set. Each fold is presented in (solid) grey, and the mean over the 10 folds presented in (dashed) black. As with the TCGA, we notice as we iterate the model fit improves, fitting better to the unseen validation set.}
    \label{fig:yau_convergence}
\end{figure}

Convergence diagnostics for the model fit to the breast cancer dataset are presented in \Cref{fig:yau_convergence} for the ovarian and breast cancer datasets respectively. As with the TCGA data we notice the ELBO is increasing as we iterate the algorithm, suggesting the model fit is improving. Furthermore, we notice that the KL($Q\| \Pi$) is decreasing, suggesting that sparsity is induced. Coupled with the fact the validation expected log-likelihood is increasing, this suggests fewer spurious variables are being selected and the model is fitting better to unseen validation set.

\subsection{Model fits}

Model fit for different values of $\lambda$ for the two datasets are presented in \Cref{tab:tcga_models} and \Cref{tab:yau_models}. Reported is the mean and standard deviation across 10 training and validation folds of the: ELBO, ELL, KL, c-index ($\widehat{k}$) and the number of parameters for which $\gamma_j > 0.5$. Examining the different metrics, we notice the model is not particularly sensitivity to the value of $\lambda$.

\sidewaystable{
\begin{table}[htp]
    \centering
    \resizebox{1.5\textwidth}{!}{ %
{\setlength{\tabcolsep}{1.7em} 
\begin{tabular}{| l | c c c c c | c c c c |}
    \hline \hline
    & \multicolumn{5}{c |}{Training} & 
    \multicolumn{4}{c|}{Validation} \\
    $\lambda$ & ELBO & ELL & KL & $\widehat{k}$ & 
    $ \#\{\gamma > 0.5\}$ & ELBO & ELL & KL & $\widehat{k}$ \\
    \hline
    0.05 & -1735.7 (21.2) & -1657.5 (26.1) & 78.2 (12.2) & 0.628 (0.018) & 2.0 (1.4) & -189.3 (20.1) & -111.1 (12.3) & 78.2 (12.2) & 0.558 (0.044) \\
0.10 & -1733.7 (20.6) & -1647.0 (31.9) & 86.8 (13.7) & 0.646 (0.018) & 3.1 (1.7) & -197.8 (24.5) & -111.0 (12.3) & 86.8 (13.7) & 0.572 (0.041) \\
0.25 & -1795.2 (208.5) & -1604.1 (129.5) & 191.1 (333.6) & 0.687 (0.085) & 15.9 (40.5) & -322.2 (398.1) & -131.1 (65.3) & 191.1 (333.6) & 0.581 (0.042) \\
0.50 & -1725.8 (21.3) & -1635.1 (30.9) & 90.7 (15.1) & 0.681 (0.021) & 4.3 (2.1) & -202.3 (25.1) & -111.7 (13.0) & 90.7 (15.1) & 0.588 (0.030) \\
0.75 & -1721.8 (22.4) & -1631.3 (31.2) & 90.5 (13.4) & 0.695 (0.015) & 4.5 (2.0) & -202.8 (24.5) & -112.3 (14.1) & 90.5 (13.4) & 0.590 (0.050) \\
1.00 & -1719.2 (20.6) & -1620.4 (37.5) & 98.8 (26.8) & 0.701 (0.023) & 6.3 (3.9) & -211.6 (37.3) & -112.8 (13.6) & 98.8 (26.8) & 0.585 (0.034) \\
1.25 & -1713.7 (22.9) & -1622.7 (27.3) & 91.0 (14.1) & 0.703 (0.016) & 5.5 (2.2) & -204.7 (21.0) & -113.7 (14.0) & 91.0 (14.1) & 0.594 (0.037) \\
1.50 & -1712.8 (22.9) & -1620.3 (28.4) & 92.4 (15.0) & 0.706 (0.015) & 6.1 (2.4) & -205.4 (24.8) & -113.0 (14.5) & 92.4 (15.0) & 0.587 (0.043) \\
1.75 & -1709.9 (20.8) & -1625.1 (29.9) & 84.9 (12.0) & 0.704 (0.014) & 5.2 (1.9) & -197.7 (24.0) & -112.8 (14.5) & 84.9 (12.0) & 0.605 (0.044) \\
2.00 & -1707.6 (21.7) & -1618.2 (28.5) & 89.4 (13.1) & 0.717 (0.021) & 6.1 (2.0) & -202.8 (22.7) & -113.4 (14.1) & 89.4 (13.1) & 0.589 (0.038) \\
2.50 & -1704.5 (22.9) & -1620.3 (28.7) & 84.2 (9.1) & 0.717 (0.011) & 5.8 (1.5) & -198.1 (22.7) & -113.9 (16.1) & 84.2 (9.1) & 0.610 (0.054) \\
3.00 & -1702.5 (21.7) & -1623.2 (31.2) & 79.3 (10.6) & 0.715 (0.014) & 5.5 (1.8) & -191.2 (23.7) & -111.9 (13.7) & 79.3 (10.6) & 0.606 (0.034) \\
4.00 & -1695.7 (21.9) & -1626.0 (28.8) & 69.6 (10.7) & 0.713 (0.011) & 4.8 (1.8) & -181.4 (20.5) & -111.7 (13.2) & 69.6 (10.7) & 0.602 (0.021) \\
5.00 & -1693.9 (21.4) & -1629.1 (25.9) & 64.7 (8.7) & 0.713 (0.012) & 4.4 (1.3) & -176.3 (19.4) & -111.5 (13.6) & 64.7 (8.7) & 0.605 (0.045) \\
    \hline \hline
\end{tabular}
}
}

    \caption{\textbf{Ovarian cancer dataset}, model fit for different values of $\lambda$}
    \label{tab:tcga_models}
\end{table}
}

\sidewaystable{
\begin{table}[H]
    \centering
    \resizebox{1.5\textwidth}{!}{ %
{\setlength{\tabcolsep}{1.7em} 
\begin{tabular}{| l | c c c c c | c c c c |}
    \hline \hline
    & \multicolumn{5}{c |}{Training} & 
    \multicolumn{4}{c|}{Validation} \\
    $\lambda$ & ELBO & ELL & KL & $\widehat{k}$ & 
    $ \#\{\gamma > 0.5\}$ & ELBO & ELL & KL & $\widehat{k}$ \\
    \hline
    0.05 & -1284.1 (29.8) & -1216.9 (28.9) & 67.1 (1.6) & 0.693 (0.009) & 2.0 (0.0) & -155.1 (18.1) & -87.9 (19.0) & 67.1 (1.6) & 0.625 (0.039) \\
0.10 & -1282.2 (29.8) & -1210.6 (30.9) & 71.6 (7.3) & 0.706 (0.010) & 2.6 (1.0) & -161.1 (22.9) & -89.5 (20.2) & 71.6 (7.3) & 0.615 (0.071) \\
0.25 & -1278.9 (29.5) & -1199.7 (31.2) & 79.3 (12.3) & 0.729 (0.019) & 4.1 (1.8) & -170.6 (24.7) & -91.4 (19.7) & 79.3 (12.3) & 0.598 (0.071) \\
0.50 & -1274.6 (29.1) & -1189.3 (34.9) & 85.3 (14.9) & 0.746 (0.013) & 5.4 (2.1) & -176.8 (29.6) & -91.6 (20.8) & 85.3 (14.9) & 0.615 (0.075) \\
0.75 & -1270.2 (29.2) & -1189.0 (29.8) & 81.2 (9.4) & 0.752 (0.011) & 5.1 (1.6) & -174.1 (22.9) & -92.9 (20.4) & 81.2 (9.4) & 0.593 (0.081) \\
1.00 & -1268.4 (30.5) & -1187.8 (28.5) & 80.6 (12.0) & 0.760 (0.012) & 5.2 (2.1) & -173.8 (22.5) & -93.2 (21.0) & 80.6 (12.0) & 0.603 (0.083) \\
1.25 & -1264.1 (29.2) & -1184.0 (30.8) & 80.0 (11.3) & 0.768 (0.009) & 5.4 (2.0) & -173.8 (24.2) & -93.8 (20.3) & 80.0 (11.3) & 0.590 (0.075) \\
1.50 & -1262.9 (29.6) & -1181.5 (32.4) & 81.3 (11.0) & 0.772 (0.010) & 5.8 (1.8) & -174.8 (25.4) & -93.4 (20.8) & 81.3 (11.0) & 0.596 (0.064) \\
1.75 & -1260.9 (29.4) & -1179.0 (32.6) & 81.9 (15.9) & 0.776 (0.012) & 6.2 (2.7) & -175.8 (27.6) & -93.9 (20.6) & 81.9 (15.9) & 0.597 (0.064) \\
2.00 & -1259.1 (29.5) & -1180.6 (32.6) & 78.5 (11.0) & 0.775 (0.009) & 5.7 (1.8) & -172.6 (25.0) & -94.1 (20.5) & 78.5 (11.0) & 0.584 (0.077) \\
2.50 & -1255.9 (29.5) & -1181.9 (33.3) & 74.0 (10.1) & 0.777 (0.009) & 5.3 (1.6) & -165.6 (25.5) & -91.6 (20.1) & 74.0 (10.1) & 0.626 (0.075) \\
3.00 & -1252.8 (29.5) & -1184.1 (30.1) & 68.7 (8.7) & 0.781 (0.005) & 4.8 (1.6) & -161.6 (22.0) & -92.9 (19.6) & 68.7 (8.7) & 0.612 (0.072) \\
4.00 & -1249.3 (29.7) & -1183.1 (31.1) & 66.2 (7.7) & 0.785 (0.004) & 4.9 (1.4) & -158.6 (23.1) & -92.4 (20.2) & 66.2 (7.7) & 0.619 (0.072) \\
5.00 & -1246.6 (28.7) & -1185.8 (31.4) & 60.8 (9.4) & 0.787 (0.008) & 4.5 (1.8) & -153.2 (24.8) & -92.4 (20.2) & 60.8 (9.4) & 0.610 (0.069) \\
    \hline \hline
\end{tabular}
}
}

    \caption{\textbf{Breast cancer dataset}, model fit for different values of $\lambda$}
    \label{tab:yau_models}
\end{table}
}

\end{document}